\documentclass[traditabstract]{aa}
\usepackage[pdftex]{graphicx}
\usepackage[varg]{txfonts}
\usepackage[english]{babel}
\usepackage{natbib}
\usepackage{pdfpages}
\usepackage[caption=false]{subfig}

\begin{document}
\title{A low-luminosity type-1 QSO sample}
\subtitle{I. Overluminous host spheroidals or undermassive black holes?\thanks{Based on observations with ESO-NTT, proposal no. 083.B-0739}}

\author{G. Busch\inst{1}, 
   	J. Zuther\inst{1},
   	M. Valencia-S.\inst{1},
   	L. Moser\inst{1},
	S. Fischer\inst{1,2},
        A. Eckart\inst{1,3},
        J. Scharw\"achter\inst{4},
        D. A. Gadotti\inst{5},
        \and L. Wisotzki\inst{6}
        }

\institute{I. Physikalisches Institut, Universit\"at zu K\"oln,
           Z\"ulpicher Str. 77, 50937 K\"oln, Germany \\
           \email{busch@ph1.uni-koeln.de}
	\and
	Deutsches Zentrum f\"ur Luft- und Raumfahrt (DLR), K\"onigswinterer Str. 522-524, 53227 Bonn, Germany
         \and
         Max-Planck-Institut f\"ur Radioastronomie,
         Auf dem H\"ugel 69, 53121 Bonn, Germany
         \and
         Observatoire de Paris, LERMA, 61 Av. de l'Observatoire, 75014 Paris, France
         \and
         European Southern Observatory (ESO), Casilla 19001, Santiago 19, Chile 
         \and
         Leibniz-Institut f\"ur Astrophysik Potsdam (AIP), An der Sternwarte 16, 14482 Potsdam, Germany 
             }

\date{Received 14 August 2013 / Accepted 29 October 2013}

\abstract{
Recognizing the properties of the host galaxies of quasi-stellar objects (QSOs) is essential for understanding the suspected coevolution of central supermassive black holes (BHs) and their host galaxies.
Low-luminosity type-1 QSOs (LLQSOs) are ideal targets because of their small cosmological distance, which allows a for detailed structural analysis.
We selected a subsample of the Hamburg/ESO survey for bright UV-excess QSOs that contains only the 99 nearest QSOs with redshift $z\leq 0.06$. From this low-luminosity type-1 QSO sample, we observed 20 galaxies and performed aperture photometry and bulge-disk-decomposition with BUDDA on near-infrared $J$-, $H$-, and $K$-band images to separate disk, bulge, bar, and nuclear component.
From the photometric decomposition of these 20 objects and visual inspection of images of another 26, we find that $\sim 50$\% of the hosts are disk galaxies and most of them (86\%) are barred.
Stellar masses, calculated from parametric models based on inactive galaxy colors, range from $2\times 10^9 M_\odot$ to $2\times 10^{11} M_\odot$ with an average mass of $7\times 10^{10} M_\odot$.
Black hole masses measured from single-epoch spectroscopy range from $1\times 10^6 M_\odot$ to $5\times 10^8 M_\odot$ with a median mass of $3\times 10^7 M_\odot$.
In comparison with higher-luminosity QSO samples, LLQSOs tend to have lower stellar and BH masses. Moreover, in the effective radius vs. mean surface-brightness projection of the fundamental plane, they lie in the transition area between luminous QSOs and normal galaxies. This can be seen as additional evidence that they populate a region intermediate between the local Seyfert population and luminous QSOs at higher redshift. This region has not been well studied so far.
Eleven low-luminosity type-1 QSOs, for which we have reliable morphological decompositions and BH mass estimations, lie below the published BH mass vs. bulge luminosity relations for inactive galaxies. This can partially be explained if one assumes that the bulges of active galaxies contain much younger stellar populations than the bulges of inactive galaxies. Another possibility would be that their BHs are undermassive. This might indicate that the growth of the host spheroid precedes that of the BH.
}   
   

\titlerunning{LLQSO sample - NIR study of AGN host galaxies}
\authorrunning{Busch et al.}

\maketitle

\section{Introduction}
\label{sec:intro}
There is wide agreement that every galaxy, or at least those with a considerable bulge component, hosts a central supermassive black hole \citep[SMBH; e.g.][]{1995ARA&A..33..581K, 1998IAUS..184..451R, 1999ASSL..234..157H}. Over the past decades, masses of SMBHs have been measured by dynamical modeling and spatially resolved kinematics as well as by reverberation mapping and single-epoch spectroscopy. A tight correlation between black hole mass $M_\mathrm{BH}$ and the velocity dispersion $\sigma$ of the bulge component of the host galaxy has been established \citep{2000ApJ...539L...9F,2000ApJ...539L..13G}. Similar correlations with bulge luminosity and mass have been found \citep[e.g.][]{1998AJ....115.2285M,2003ApJ...589L..21M,2004ApJ...604L..89H}. Others, with the bulge S\'ersic index (definition in Sec. \ref{sec:decomposition}) for instance, are still under discussion \citep{2007ApJ...655...77G,2012MNRAS.419.2264V}. 

Tight correlations between central SMBHs and the host galaxy (resp. their bulges) have been interpreted as manifestations of an SMBH-host galaxy coevolution. \cite{1988ApJ...325...74S} established a merger-driven scenario from ultraluminous infrared galaxies (ULIRGs) to quasi-stellar objects (QSOs) and quiescent dead ellipticals. However, in the past years, the detection of SMBHs in bulgeless galaxies and the finding that a considerable amount of local spiral galaxies host a disk-like pseudobulge instead of a classical bulge \citep[e.g.][]{2004ARA&A..42..603K,2009MNRAS.393.1531G}, shows that the picture of galaxy evolution is not yet complete.

\begin{figure}
\centering
\includegraphics[width=\columnwidth]{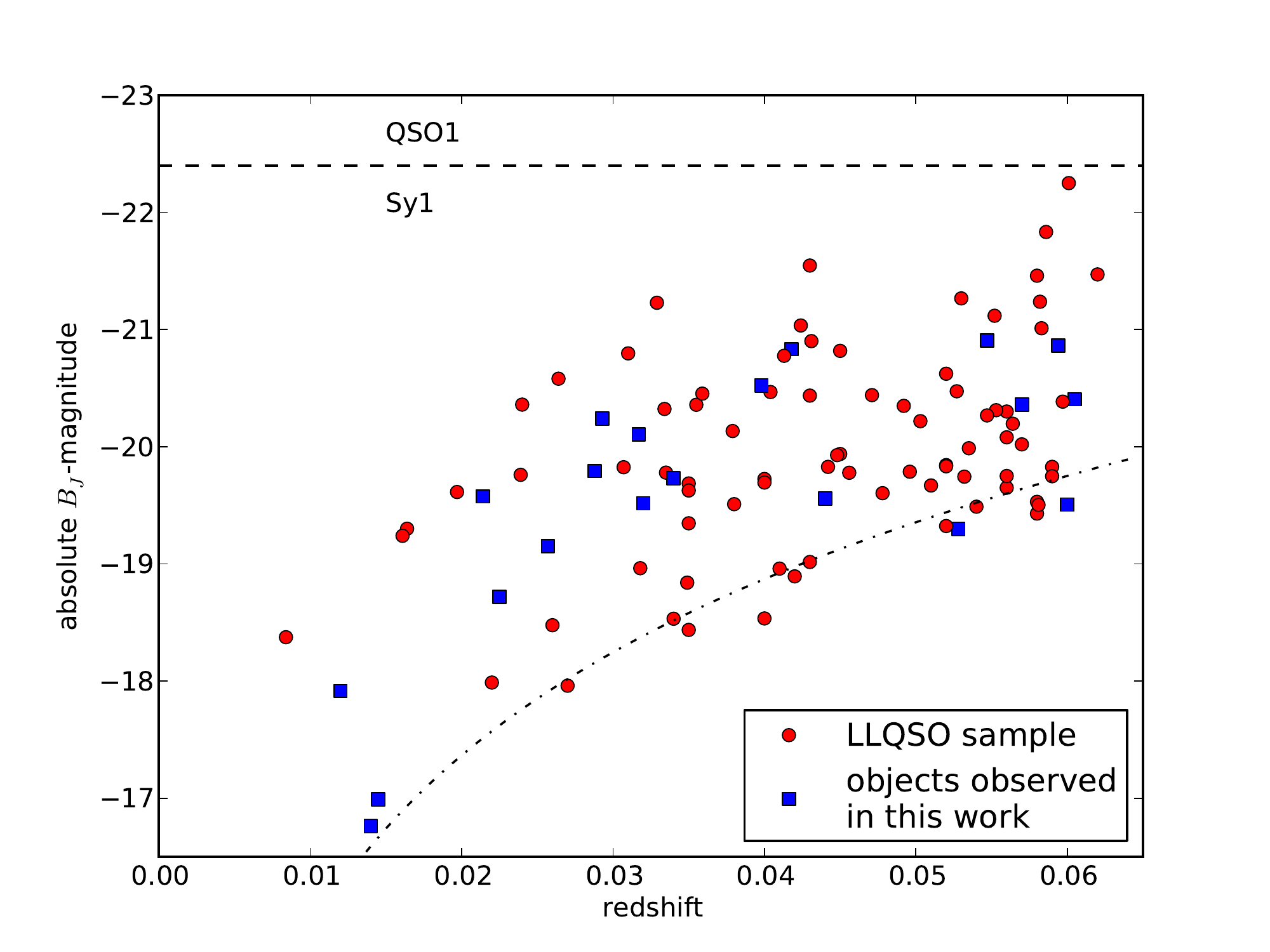}
\caption{Redshift-magnitude diagram of the 99 low-luminosity type-1 QSOs. The galaxies that have been observed and analyzed in this work are marked with squares. The limiting magnitude ($B_J \leq 17.3$) of the HES is marked by the dashed-dotted line.}
\label{fig:redsh-magn}
\end{figure}

In the local Universe ($z\lesssim 0.2$), studies of thousands of objects have shown that in sources with active galactic nuclei (AGN), the star formation and SMBH accretion rates are related to each other and also to the age of the stellar population. The history of SMBH growth (corresponding to the quasar activity) and star formation in the Universe is similar \citep[e.g.,][and references therein]{1998ApJ...498..106M,2004ApJ...613..109H,2006ApJ...651..142H,2009ApJ...696..396S,2010MNRAS.401.2531A}. Several other findings \citep[see review of][]{2013ARA&A..51..511K} support the assumption that AGN feedback is crucial for the evolution of galaxies, at least at past epochs. However, in this context, the role of AGN is still unclear. Thus, a key question is whether hosts of AGN are different from those of inactive galaxies. Furthermore, it is unclear whether they follow the same black hole - bulge relations, or if any deviations from these correlations are indicative of an early evolutionary stage or of an entirely different evolution process driven by secular evolution or major mergers \citep[e.g.,][]{2009ApJ...698..812G,2011Natur.469..374K,2012ApJ...746..113G,2013ApJ...768...76S,2013ARA&A..51..511K}.

Many studies have focused on nearby Seyfert galaxies with very moderate nuclear emission on the one hand, and on the analysis of global properties of large samples of powerful QSOs at higher redshift on the other hand. To fill the gap between these two approaches, we analyzed a sample of \emph{low-luminosity type-1 QSOs} at low redshift. This sample contains the closest known bright AGN that are still close enough for spatially resolved observations and a detailed structural analysis of the host galaxies. These galaxies constitute the important link between the local, but less luminous AGN and the powerful QSOs at higher redshifts ($z\geq 0.5$). The detailed study of the hosts of type-1 QSOs is challenging because of the high contrast between the bright nuclear point source and the faint host. 

The Hamburg/ESO Survey \citep[HES,][]{2000A&A...358...77W} is a wide-angle survey for optically bright QSOs and other rare objects in the southern hemisphere, covering a formal area of $\approx 9500$ deg$^2$ on the sky. A multitude of selection criteria is applied to select type-1\footnote{AGN with broad permitted lines in the optical spectra are referred to as ``type-1''.} AGN up to $z\approx 3.2$. The limiting magnitude is $B_J < 17.3$ with a dispersion of 0.5 mag between individual fields. The HES provides a methodologically complete sample of the local AGN population, that is, all AGN selected by the given criteria are included.

To select only the nearest AGN, we set a redshift limit of $z= 0.06$. This particular value was chosen to ensure the presence of the CO(2-0) band head in near-infrared (NIR) $K$-band spectra. This stellar absorption feature can be used to derive the stellar kinematics of the host galaxy or to analyze stellar populations \citep{1995PASP..107...68G,2006A&A...452..827F}. These selection criteria result in a sample of 99 galaxies that we call the \emph{low luminosity type-1 QSO sample} (LLQSO) throughout the text \citep[for more information see][]{2007A&A...470..571B}. A redshift-magnitude diagram is shown in Fig. \ref{fig:redsh-magn}. The redshifts are taken from the HES. The magnitudes are nuclear $B_J$ magnitudes (corresponding to the seeing disk) taken from the HES. The absolute magnitude range is $-22.3 < M_{B_J} < -16.8$ with a median value of $-19.8$. The median redshift is $z=0.043$. The redshift-magnitude diagram shows that the selected galaxies lie below the commonly used division line of $M_B=-21.5+5\log h_0$ between QSOs and Seyfert galaxies. Several galaxies from the LLQSO sample have been observed in molecular gas \citep[\element{CO}][]{2007A&A...470..571B}, atomic hydrogen \citep[\ion{H}{i}][]{2009A&A...507..757K}, and \element{H}$_2$\element{O}-maser-emission \citep{2012MNRAS.420.2263K}. \cite{2006A&A...452..827F} studied nine galaxies in the NIR, focusing on spectroscopy.

In this work, we focus on structural and morphological properties of 20 galaxies (marked with squares in Fig. \ref{fig:redsh-magn}, object information in Table \ref{tab:objects}) that were randomly selected from the LLQSO sample. Our aim is to compare the properties of the hosts of these AGN with those of brighter QSOs and those of inactive galaxies. We analyze images in the NIR $J$-, $H$-, and $K$-band because these are less affected by dust extinction \citep[$A_H/A_V \approx 0.175$,][]{1985ApJ...288..618R}. Furthermore, they are less affected by recent star formation and mainly trace intermediate-to-old (age $\gtrsim 10^8\,{\rm yr}$) red stars that dominate the stellar mass. Additionally, the contrast between the nuclear point source and the host galaxy has a minimum in the NIR \citep{1997quho.conf...45M}. In Sect. \ref{sec:observations}, we report on the details of the observations as well as on the data reduction and calibration.

We performed a detailed structural analysis with bulge-disk-bar-AGN decomposition using \textsc{Budda} and a photometric analysis of high-quality NIR images. The methods used for decomposition and photometry are described in Sect. \ref{sec:results}. We estimated host stellar masses and bulge luminosities and compare them with other samples in the literature. In agreement with previous studies of type-1 AGN in the optical \citep[e.g.,][]{2004ApJ...615..652N,2008ApJ...687..767K,2011ApJ...726...59B}, the LLQSOs examined here do not follow the published black hole mass ($M_\mathrm{BH}$) vs. bulge luminosity ($L_\mathrm{bulge}$) relations of inactive galaxies. In Sect. \ref{sec:discussion}, we discuss possible reasons for the observed discrepancy. Summary and conclusions are presented in Sect. \ref{sec:summary}. Throughout this paper, we use a standard cosmological model with $H_0 =70\ \mathrm{km}\ \mathrm{s}^{-1}\ \mathrm{Mpc}^{-1}$, $\Omega_m=0.3$ and $\Omega_\Lambda=0.7$. 

\begin{table*}
\centering
\caption{Object information for the observed galaxies.}
\label{tab:objects}
\begin{tabular}{cccccccccc} \hline \hline
ID & Name & RA & Dec & $z$ & $B_J$ & $M_{B_J}$ & Instrument & Seeing & $\sigma_\mathrm{sky}$ \\ 
 & & (J2000) & (J2000) & & (mag) & (mag) & & (arcsec) & (mag arcsec$^{-2}$) \\ \hline

05 & HE0036-5133 & $00~39~15.80$ & $-51~17~02.04$ & 0.0288 &  15.71 & -19.79 & SofI & 0.9 - 0.9 - 0.8 & 23.93 - 23.12 - 22.74 \\
08 & HE0045-2145 & $00~47~41.30$ & $-21~29~26.88$ & 0.0214 &  15.27 & -19.58 & SofI & 1.0 - 1.0 - 1.2 & 23.76 - 23.16 - 23.12 \\
11 & HE0103-5842 & $01~05~16.99$ & $-58~26~16.08$ & 0.0257 &  16.10 & -19.15 & SofI & 1.4 - 1.5 - 1.4 & 23.60 - 22.89 - 22.68 \\
16 & HE0119-0118 & $01~21~59.81$ & $-01~02~25.01$ & 0.0547 &  16.03 & -20.91 & SofI & 1.0 - 1.2 - 1.0 & 23.94 - 23.28 - 23.16 \\
24 & HE0224-2834 & $02~26~25.70$ & $-28~20~58.92$ & 0.0605 &  16.76 & -20.40 & SofI & 1.3 - 1.1 - 1.0 & 23.16 - 23.14 - 23.35 \\
29 & HE0253-1641 & $02~56~02.59$ & $-16~29~16.08$ & 0.0320 &  16.22 & -19.52 & SofI & 1.0 - 1.0 - 1.0 & 23.86 - 23.32 - 23.74 \\
69 & HE1248-1356 & $12~51~32.40$ & $-14~13~17.04$ & 0.0145 &  17.00 & -16.99 & LUCI & 1.0 - 0.8 - 0.8 & 22.59 - 21.11 - 20.98 \\
70 & HE1256-1805 & $12~58~42.96$ & $-18~21~36.00$ & 0.0140 &  17.15 & -16.76 & LUCI & 0.9 - 0.9 - 0.9 & 22.49 - 21.75 - 21.50 \\
71 & HE1310-1051 & $13~13~05.76$ & $-11~07~41.88$ & 0.0340 &  16.14 & -19.73 & LUCI & 1.0 - 1.1 - 0.8 & 22.96 - 21.54 - 21.07 \\
74 & HE1330-1013 & $13~32~39.12$ & $-10~28~53.04$ & 0.0225 &  16.24 & -18.72 & LUCI & 0.9 - 0.9 - 0.9 & 21.75 - 21.45 - 20.42 \\
75 & HE1338-1423 & $13~41~12.96$ & $-14~38~39.84$ & 0.0418 &  15.50 & -20.83 & LUCI & 0.9 - 1.0 - 0.6 & 22.84 - 20.74 - 20.45 \\
77 & HE1348-1758 & $13~51~29.52$ & $-18~13~46.92$ & 0.0120 &  15.66 & -17.91 & LUCI & 1.0 - 0.9 - 0.8 & 21.96 - 20.09 - 20.95 \\
79 & HE1417-0909 & $14~20~06.24$ & $-09~23~12.98$ & 0.0440 &  16.89 & -19.56 & LUCI & 0.8 - 0.7 - 0.8 & 22.04 - 20.11 - 20.71 \\
80 & HE2112-5926 & $21~15~51.60$ & $-59~13~54.12$ & 0.0317 &  15.61\tablefootmark{a} & -20.11\tablefootmark{a} & SofI & 1.4 - 1.4 - 0.7 & 24.10 - 23.23 - 22.78 \\
81 & HE2128-0221 & $21~30~49.92$ & $-02~08~15.00$ & 0.0528 &  17.56 & -19.30 & SofI & 0.8 - 0.7 - 0.7 & 23.98 - 23.06 - 22.95 \\
82 & HE2129-3356 & $21~32~02.16$ & $-33~42~54.00$ & 0.0293 &  15.30 & -20.24 & SofI & 1.4 - 1.2 - 1.0 & 24.07 - 23.46 - 22.79 \\
83 & HE2204-3249 & $22~07~45.12$ & $-32~35~02.04$ & 0.0594 &  16.26 & -20.86 & SofI & 1.4 - 1.1 - 1.6 & 24.27 - 23.58 - 22.75 \\
84 & HE2211-3903 & $22~14~42.00$ & $-38~48~24.12$ & 0.0398 &  15.70 & -20.52 & SofI & 1.2 - 1.3 - 1.1 & 23.40 - 22.69 - 22.69 \\
85 & HE2221-0221 & $22~23~49.68$ & $-02~06~13.00$ & 0.0570 &  16.67 & -20.36 & SofI & 1.3 - 1.0 - 0.9 & 23.75 - 23.19 - 22.91 \\
89 & HE2236-3621 & $22~39~05.28$ & $-36~05~53.16$ & 0.0600 &  17.64 & -19.51 & SofI & 1.2 - 1.4 - 1.2 & 24.43 - 23.80 - 22.82 \\

\hline
\end{tabular}
\tablefoot{
The table presents the ID in our sample, the corresponding Hamburg/ESO survey name, the coordinates, redshift, apparent and absolute $B_J$ nuclear magnitude, the used instrument, seeing, and the sky deviation in the $J$-, $H$-, and $K$-band.\\
\tablefoottext{a}{magnitude from NED}
}
\end{table*}

\section{Observation, reduction, and calibration}
\label{sec:observations}
In this study we used a set of thirteen galaxies drawn from the \emph{low-luminosity type-1 QSO sample} that has been observed in the $J$-, $H$-, and $K_s$-band with the Son of ISAAC (SofI) infrared spectrograph and imaging camera on the New Technology Telescope (NTT, La Silla, Chile) during September 2009. The $1024\times 1024$ Hawaii HgCdTe array provides a pixel scale of $0.288\arcsec/\mathrm{pixel}$ with a field of view of $4.9 \times 4.9$ arcmin$^2$.

Seven additional galaxies have been observed in the $J$-, $H$-, and $K_s$-band with the LBT NIR Spectrograph Utility with Camera and Integral-Field Unit for Extragalactic Research (LUCI) at the Large Binocular Telescope (LBT, Mt. Graham, USA) during May 2011 in seeing-limited mode. The $2048\times 2048$ Rockwell Hawaii-2 HdCdTe array provides a pixel scale of $0.12\arcsec/\mathrm{pixel}$ with a field of view of $4\times 4$ arcmin$^2$.

All images were obtained in jitter mode. Initially, flat-fielding (twilight-flat) and bad-pixel-correction were applied. The sky background was then removed by subtracting consecutive frames from each other. After aligning the resulting frames, a median frame was created. The images were flux calibrated using data from the 2 Micron All Sky Survey (2MASS) for foreground stars. Comparing the results of different stars, we can estimate that the calibration error is not higher than 10\%.

We determined the seeing by fitting Gaussians to unresolved objects (stars) in the images. We present the average full width at half maximum (FWHM) of the fitted stars as measure of the seeing in Table \ref{tab:objects}. The seeing was around $1 \arcsec$ in most cases. To determine the depth of the observations, we measured the standard deviation $\sigma_\mathrm{sky}$ of the sky-subtracted background. We then calculated the  mean $1\sigma$ sky deviation surface brightness
\begin{equation}
\sigma_\mathrm{sky} = -2.5 \log(\sigma[\mathrm{counts}]) + \mathrm{ZP} + 5 \log(\mathrm{pixelscale}).
\end{equation}
The calculated values are presented in Table \ref{tab:objects} as well and indicate a high sensitivity.

\section{Results}
\label{sec:results}
We describe the methods used for the surface-brightness decomposition, the tests we developed to establish the accuracy of such procedure and the derived results. Some relevant information about the studied sources is also summarized.

\subsection{Decomposition}
\label{sec:decomposition}
We used \textsc{Budda}\footnote{http://www.sc.eso.org/\~{}dgadotti/budda.html} (Bulge/Disk Decomposition Analysis) to perform the two-dimensional decomposition of the near-infrared galaxy images. \textsc{Budda} was developed by de Souza, Gadotti and dos Anjos and has been made publicly available \citep{2004ApJS..153..411D,2008MNRAS.384..420G}.

We assumed the disk component to be fitted best by an exponential function \citep{1970ApJ...160..811F}
\begin{equation}
\mu_\mathrm{disk} (r) = \mu_0 + 1.086 \frac{r}{h_r}.
\end{equation}
Here, $\mu_0$ denotes the central surface brightness of the disk component and $h_r$ the characteristic scale length.

Furthermore, we used the S\'{e}rsic-function \citep{1968adga.book.....S}
\begin{equation}
\mu (r) = \mu_e + c_n \left[ \left( \frac{r}{r_e}\right)^{1/n} -1 \right]
\end{equation}
to fit bulges and bars. In the past, the bulge component has often been modeled with a S\'ersic-profile with an index fixed to $n= 4$ (the so-called de Vaucouleurs-profile, see \cite{1948AnAp...11..247D}). However, it has been shown that letting the S\'ersic index of the bulge vary leads to better results \citep[e.g.][]{2009MNRAS.393.1531G}. For $n=1$, the profile reduces to an exponential function. Even bars can be modeled with a S\'{e}rsic-index between $\approx 0.5$ and $\approx 1.0$ \citep{2008MNRAS.384..420G}. In the formula, $r_e$ denotes the effective radius, that is, the radius that contains half of the fitted components' flux, $\mu_e$ is the effective surface brightness, that is, the surface brightness at the effective radius. $n$ is the S\'ersic index and $c_n$ a parameter that depends on $n$: $c_n= 2.5 (0.868 n -0.142)$ \citep{1993MNRAS.265.1013C}.

The components were fitted with generalized ellipses
\begin{equation}
\left( \frac{|x|}{a}\right)^c + \left( \frac{|y|}{b} \right)^c = r^c.
\end{equation}
A fixed parameter $c=2$ corresponds to simple ellipses. This setup was used for the bulges and disks for which the code only fitted the ellipticity and the position angle. For a bar, the parameter was let free, mostly resulting in $c>2$, that is, boxy ellipses.

The AGN was modeled using a circular Moffat distribution with FWHM fixed as the same as the PSF. The latter was measured in the image and kept fixed. Only the peak intensity of the AGN was fitted as a free parameter.

The results of the decomposition in the $J$-, $H$-, and $K$-band are presented in Tables \ref{tab:jband}, \ref{tab:hband} and \ref{tab:kband}. They give the structural parameters of bulge and disk as well as the luminosity fractions of the components. Figures with images of all decomposed galaxies and radial profiles are provided in the appendix. Figure \ref{fig:decomp_ex} shows an example of the surface-brightness fit for the galaxy HE 2211--3903: the original image, the \textsc{Budda} model image and a residual image, obtained by dividing the galaxy image by the the model. Additionally, Fig. \ref{fig:decomp_ex} shows the radial profile of the surface brightness of the galaxy, together with profiles of the individual components and the total model, as well as radial plots of the difference between the surface brightness of the galaxy and the model, ellipticity and position angle. All radial plots are based on ellipse fits with the \textsc{Ellipse} task of \textsc{Iraf}.

The residual images and radial profiles were used to check the quality of the models obtained with \textsc{Budda}. Furthermore, they can be used to identify otherwise hidden substructures and irregularities such as spiral arms (e.g. HE1310--1051), inner rings (e.g. HE2211--3903) or dust lanes (HE2204--3249). \cite{1995ApJ...441...96M} considered a galaxy to have a bar if the ellipticity has a bump while the position angle stays constant and there is a shoulder in the intensity profile at the same position. We identify this behavior for instance in HE2211--3903. However, these authors reported that there are also galaxies that clearly show a bar in images but do not show these features in the radial profiles. 

\begin{figure*}
\sidecaption
\includegraphics[width=12cm]{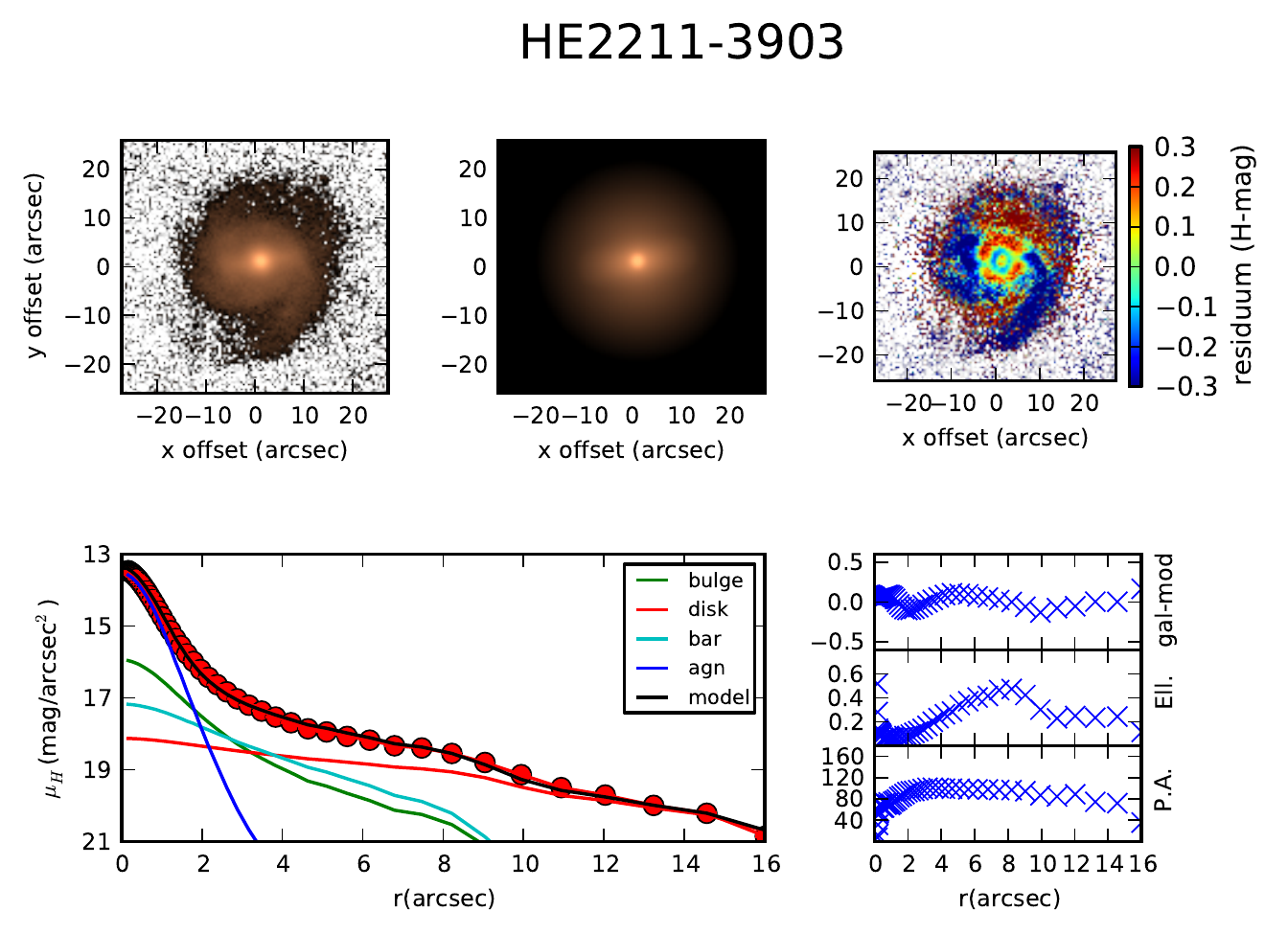}
\caption{Example of decomposition with \textsc{Budda}. We show (from left to right) the original $H$-band image, the model and the residuum (galaxy/model). Blue indicates regions where the model is fainter than the galaxy, red indicates regions where the model is brighter than the galaxy. In the lower row, we show an elliptically averaged radial profile of the galaxy, the single components, and the whole model. In the ellipse fits of the model, ellipticity and position angle are fixed to those of the original image. Furthermore, we show the difference between galaxy and model, ellipticity and position angle.}
\label{fig:decomp_ex}
\end{figure*}

\subsection{Reliability of \textsc{Budda} fits}
\label{sec:reliability}

We used \textsc{Budda} to fit the described brightness profiles (see Sect. \ref{sec:decomposition}) to the galaxy images. As results we obtained the structural parameters and images of the model components. From these data, we calculated the luminosities of the components. We are mainly interested in two parameters:
\begin{itemize}
\item The flux of the host galaxy to estimate NIR colors (Sect. \ref{sec:method_photometry}) and the stellar mass (Sect. \ref{sec:stellmasses}).
\item The flux of the bulge component (equal to the host galaxy flux in case of elliptical galaxies) to estimate the black hole mass and/or study the black hole/bulge relationship.
\end{itemize}

\begin{figure*}
\centering
\includegraphics[width=0.45\linewidth]{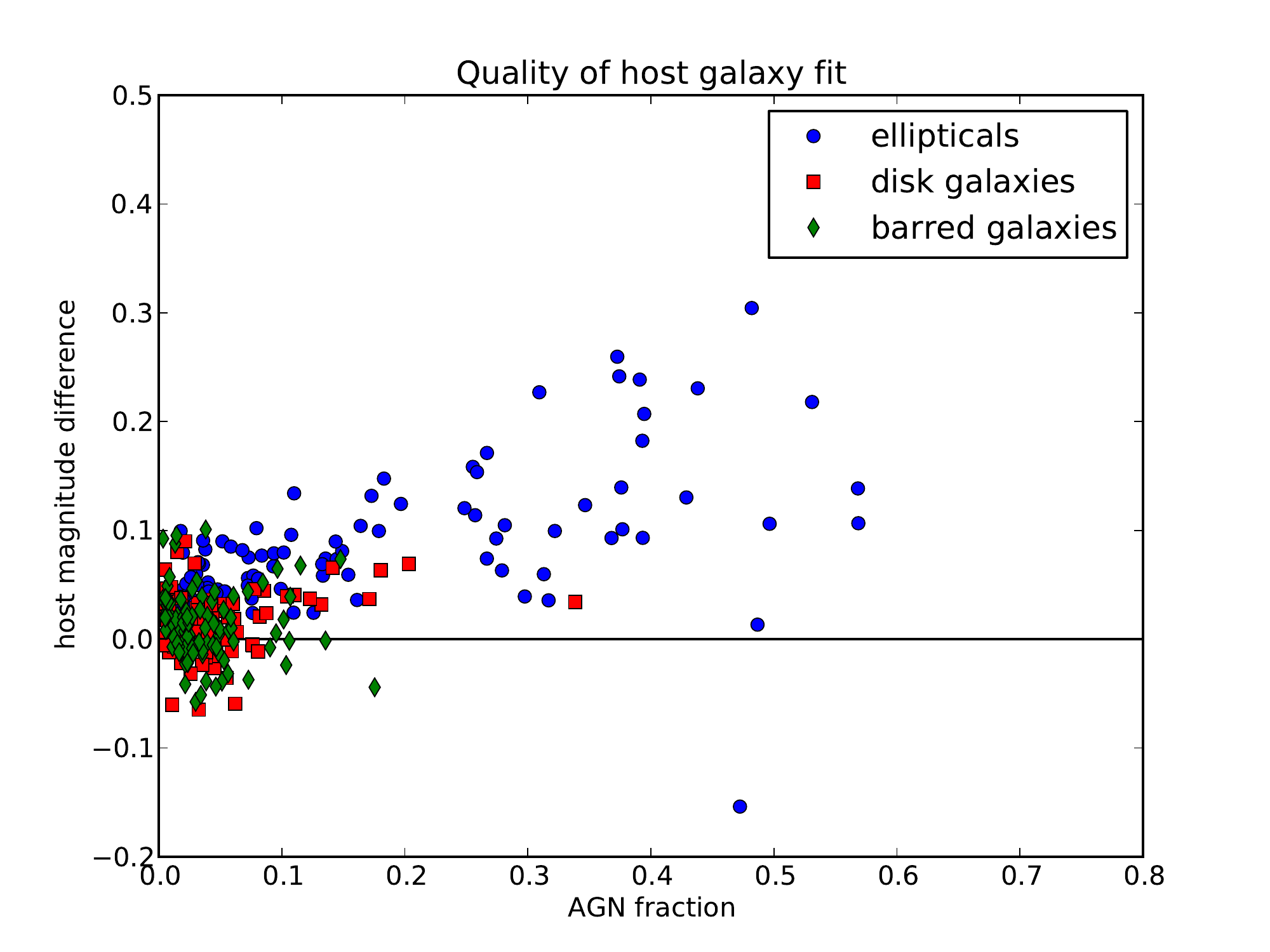}
\includegraphics[width=0.45\linewidth]{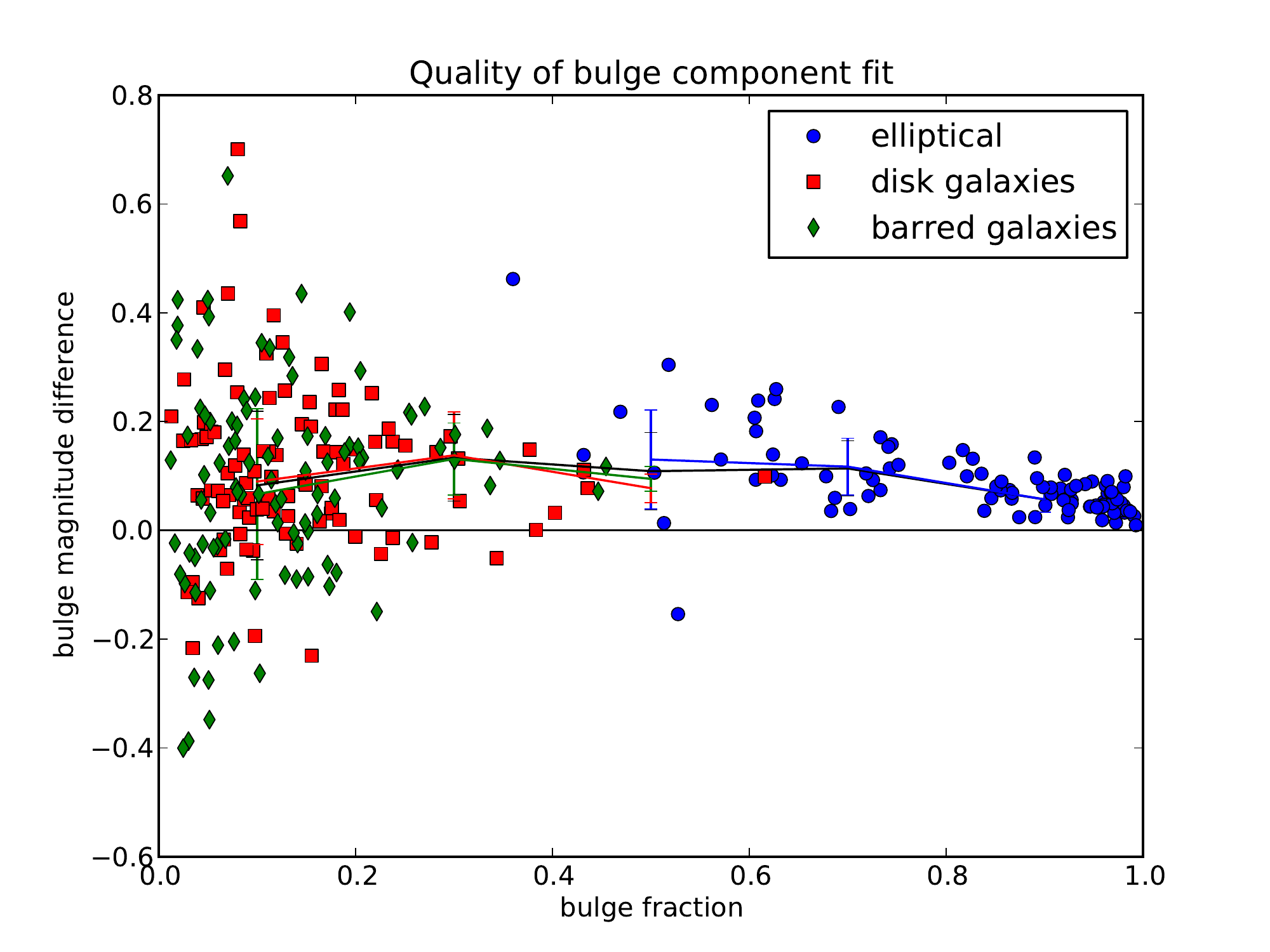}
\caption{Quality of the AGN subtraction with \textsc{Budda}: Difference between host galaxy in synthetic image and \textsc{Budda} fit in magnitudes (left), same for bulge component (right), as function of the fitted AGN (left) or bulge (right) luminosity fraction. Positive errors indicate that there is more flux in the model than in the original component, e.g. that the host galaxy or bulge resp. is overestimated, squares indicate disk-dominated galaxies, circles bulge-dominated galaxies, and diamonds barred galaxies. In the right panel, we mark trends by the mean and standard deviation of the data points in bins of $0.2$ of the AGN/bulge fraction.}
\label{fig:qualbudda}
\end{figure*}

\begin{figure*}
\sidecaption
\includegraphics[width=12cm]{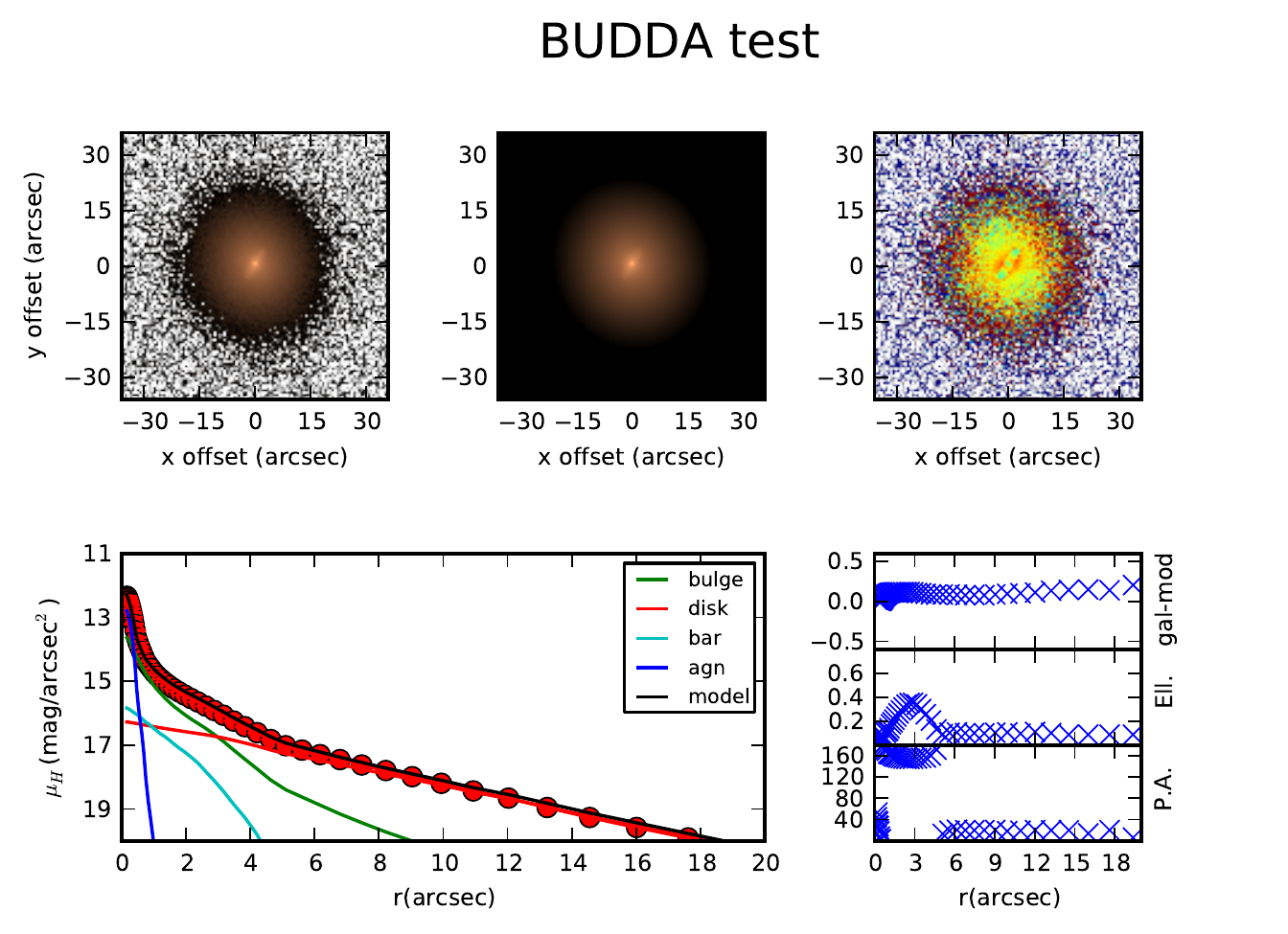}
\caption{Synthetic galaxy produced with \textsc{Budda Bmod} (see Sect. \ref{sec:reliability}) and its \textsc{Budda} model in analogy to Fig. \ref{fig:decomp_ex}.}
\label{fig:testgal}
\end{figure*}

To do this, we need to assess how well \textsc{Budda} reproduces these parameters. Therefore, we created artificial galaxy images with the \textsc{Bmod} (build model)-function of \textsc{Budda}. In particular, we produced 99 galaxies with bulge and AGN (displayed with circles in Fig. \ref{fig:qualbudda}), 99 galaxies with bulge, disk, and AGN component (displayed with squares), and 99 galaxies with bulge, disk, bar, and AGN (displayed with diamonds). To produce the artificial galaxies, we assumed the profiles described in Sect. \ref{sec:decomposition}. Although the bar structure seems to be much more complex \citep[e.g.][]{2011MNRAS.412.2211G}, in a first approach, we simulated bars by using a S\'ersic profile with $n\approx 0.5 ... 1$. In all images, we added Gaussian noise. Then, we fitted the synthetic images with \textsc{Budda}, analogously to the science data. Figure \ref{fig:testgal} shows an example of an artificial barred galaxy and the \textsc{Budda} fit results. 

Figure \ref{fig:qualbudda} (left) shows the deviation between host galaxy magnitude in the artificial image and the \textsc{Budda} fit as a function of the (fitted) AGN fraction. The quality of the fit is better for low AGN fractions. For AGN fractions below 30\%, which is a good assumption in most cases, the difference is below 0.2 mag, which is acceptable. In cases of higher AGN fractions, the residuum should be analyzed carefully. We note that even in these cases the error in the measured properties of the AGN can be as low as 5\%, and this feature arises because of the extraordinary brightness of the AGN. In inactive galaxies, this feature is absent, as previous tests have shown \citep{2004ApJS..153..411D,2009MNRAS.393.1531G}. We also investigated the quality of the bulge fits. Figure \ref{fig:qualbudda} (right) shows the deviation between bulge magnitude in the artificial image and the \textsc{Budda} fit as a function of the (fitted) bulge fraction. The difference in the bulge magnitude becomes lower with increasing bulge fraction. It is below 0.2 mag in most cases. Often, the fit is even better. In general, \textsc{Budda} tends to slightly overestimate ($\approx$0.1 mag) the bulge component.

Since an accurate sky-background subtraction can be crucial for the reliability of the decomposition, especially in the NIR, we measured the sky background very thoroughly: following the practice of \cite{2012MNRAS.419.2264V}, we measured the sky background by manually placing some 50 $10\times10\, \mathrm{pixel}$ boxes around the galaxy and calculating the median value in each of these boxes. We then took the mean of these values as sky background and the standard deviation as measure of the spatial variability of the sky $\sigma_\mathrm{sky}$. In general the sky background is close to zero, that is, it has been subtracted very well in the reduction process. However, if the measured sky background was not zero, as was the case in some images, we subtracted it subsequently. A background gradient was not observed in either case. 

\begin{figure*}
\includegraphics[width=0.57\linewidth]{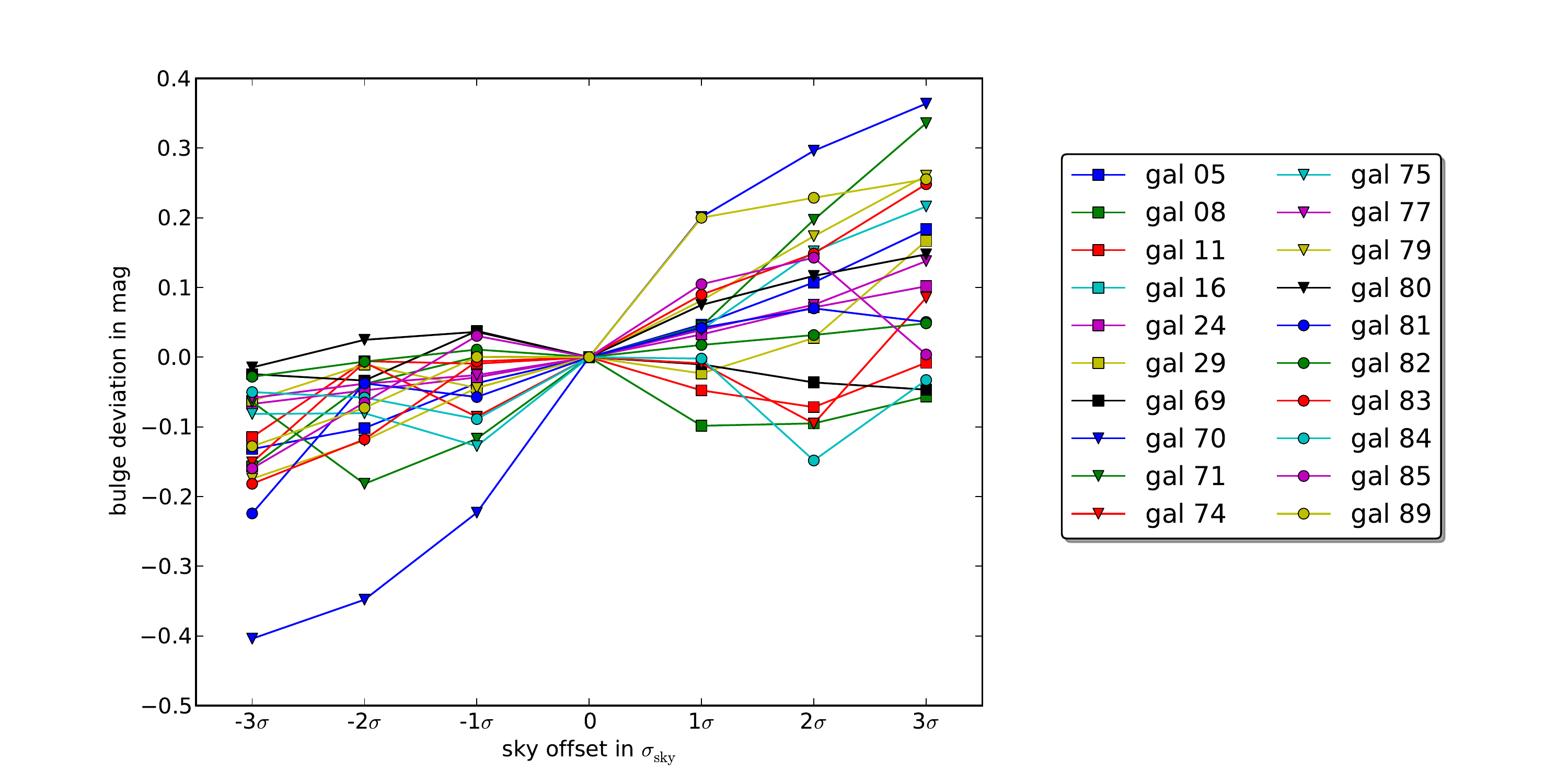}
\includegraphics[width=0.38\linewidth]{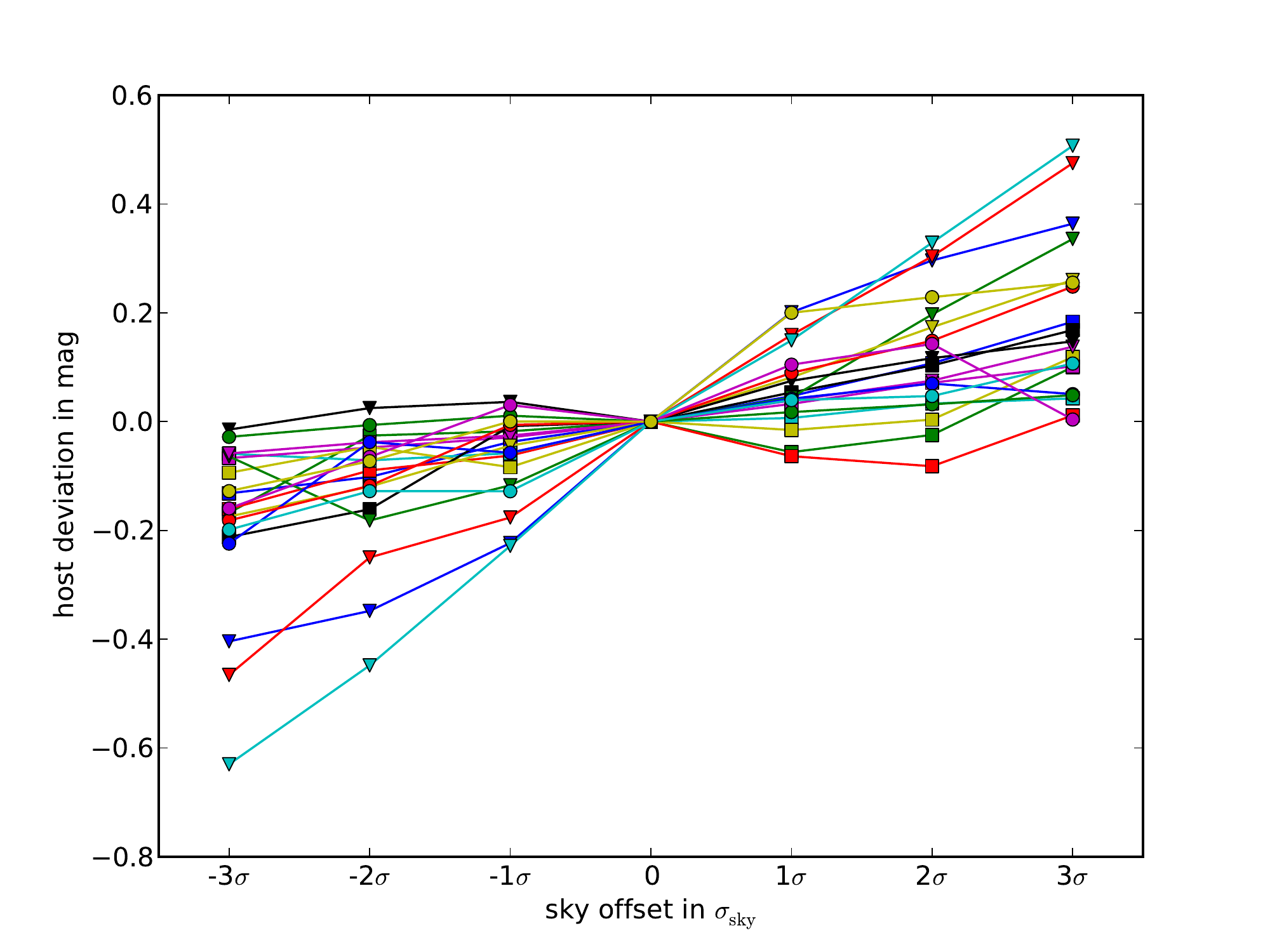}
\caption{Influence of the sky-background subtraction on the \textsc{Budda} fit in the $K$-band: We added a sky offset of $(-3,-2,-1,1,2,3)\times \sigma_\mathrm{sky}$ to the sky subtracted images and repeated the fit. A positive sky offset means that the sky background was underestimated and therefore not completely subtracted. The plots show the hereby introduced deviation of the bulge (left) and host (right) magnitude. Positive magnitudes indicate an overestimation of the bulge/host, while negative magnitudes indicate underestimation.}
\label{fig:skybackgr}
\end{figure*}

To estimate the influence of the background subtraction on the decomposition, we added sky offsets to the images ($(-3,-2,-1,1,2,3)\times \sigma_\mathrm{sky}$) and fitted them again. Figure \ref{fig:skybackgr} shows the difference in the bulge (left) and host (right) mangnitude compared with the fit without sky offset, exemplarily in $K$-band. In general, an underestimation of the sky background (resulting in a positive offset in the background-subtracted image) leads to an overestimation of the bulge and host flux and vice versa. However, in most cases the error is below the assumed errors introduced by calibration and decomposition.  But in cases with low S/N (e.g., gal 70), the introduced errors can be large and an accurate sky subtraction is important.

\begin{table*}
\centering
\caption{Parameters of the galaxies in the $J$-band.} 
\label{tab:jband}
\begin{tabular*}{\textwidth}{@{\extracolsep{\fill}}cc|ccccccccc} \hline \hline
ID & Name & $\mu_0$ & $h$ & $\mu_e$ & $r_e$ & $n$ & B/T & D/T & Bar/T & AGN/T \\
(1) & (2) & (3) & (4) & (5) & (6) & (7) & (8) & (9) & (10) & (11) \\ \hline

05 & HE0036-5133 & - & - & 18.1 & 2.4 & 2.5 & 0.683 & - & - & 0.317 \\
08 & HE0045-2145 & 19.2 & 7.6 & 18.5 & 1.7 & 2.7 & 0.137 & 0.524 & 0.239 & 0.1 \\
11 & HE0103-5842 & 17.7 & 7.1 & 17.9 & 2.6 & 2.9 & 0.472 & 0.395 & - & 0.133 \\
16 & HE0119-0118 & 18.2 & 4.0 & - & - & - & - & 0.63 & 0.221 & 0.149 \\
24 & HE0224-2834 & - & - & 19.9 & 4.4 & 4.9 & 0.767 & - & - & 0.233 \\
29 & HE0253-1641 & 18.8 & 4.4 & 18.1 & 1.1 & 2.3 & 0.143 & 0.473 & 0.177 & 0.208 \\
69 & HE1248-1356 & 17.2 & 7.9 & 17.4 & 1.7 & 2.2 & 0.153 & 0.813 & - & 0.033 \\
70 & HE1256-1805 & - & - & 21.1 & 5.3 & 3.0 & 0.866 & - & - & 0.134 \\
71 & HE1310-1051 & - & - & 19.5 & 4.0 & 2.5 & 0.691 & - & - & 0.309 \\
74 & HE1330-1013 & 19.5 & 14.8 & 19.4 & 2.1 & 1.7 & 0.06 & 0.67 & 0.209 & 0.061 \\
75 & HE1338-1423 & 19.6 & 12.3 & 18.9 & 3.9 & 4.4 & 0.512 & 0.313 & 0.068 & 0.107 \\
77 & HE1348-1758 & - & - & 19.0 & 5.0 & 3.7 & 0.792 & - & - & 0.208 \\
79 & HE1417-0909 & - & - & 19.8 & 3.8 & 3.4 & 0.773 & - & - & 0.227 \\
80 & HE2112-5926 & - & - & 19.2 & 4.5 & 3.3 & 0.903 & - & - & 0.097 \\
81 & HE2128-0221 & - & - & 18.3 & 2.0 & 3.0 & 0.86 & - & - & 0.14 \\
82 & HE2129-3356 & - & - & 19.1 & 3.8 & 3.2 & 0.943 & - & - & 0.057 \\
83 & HE2204-3249 & - & - & 18.6 & 3.6 & 3.3 & 0.867 & - & - & 0.133 \\
84 & HE2211-3903 & 18.7 & 6.7 & 17.9 & 1.4 & 2.5 & 0.159 & 0.531 & 0.124 & 0.186 \\
85 & HE2221-0221 & - & - & 18.4 & 1.9 & 3.9 & 0.7 & - & - & 0.3 \\
89 & HE2236-3621 & - & - & 20.7 & 2.6 & 3.6 & 0.667 & - & - & 0.333 \\

\hline
\end{tabular*}
\tablefoot{
Parameters of bulge, disk, bar, and AGN. Columns (1) and (2) give ID and name of the galaxy. Columns (3) and (4) give the central surface brightness (in mag/arcsec$^2$) and the scale length (in arcsec) of the disk. Columns (5), (6), and (7) give the effective surface brightness and effective radius (again in mag/arcsec$^2$ and arcsec) as well as the S\'{e}rsic index of the bulge. Columns (8) to (11) give the luminosity fractions of bulge, disk, bar, and AGN compared to the total luminosity.
}
\end{table*}

\begin{table*}
\begin{center}
\caption{Parameters of the galaxies in the $H$-band. For explanations, see the notes in Table \ref{tab:jband}.} 
\label{tab:hband}
\begin{tabular*}{\textwidth}{@{\extracolsep{\fill}}cc|ccccccccc} \hline \hline
ID & Name & $\mu_0$ & $h$ & $\mu_e$ & $r_e$ & $n$ & B/T & D/T & Bar/T & AGN/T \\
(1) & (2) & (3) & (4) & (5) & (6) & (7) & (8) & (9) & (10) & (11) \\ \hline

05 & HE0036-5133 & - & - & 17.3 & 2.3 & 1.7 & 0.546 & - & - & 0.454 \\
08 & HE0045-2145 & 18.8 & 7.6 & 17.9 & 2.6 & 2.7 & 0.318 & 0.44 & 0.158 & 0.084 \\
11 & HE0103-5842 & 17.2 & 7.5 & 17.4 & 3.0 & 2.8 & 0.488 & 0.349 & - & 0.163 \\
16 & HE0119-0118 & 17.7 & 5.3 & - & - & - & - & 0.656 & 0.125 & 0.219 \\
24 & HE0224-2834 & - & - & 18.6 & 3.6 & 4.1 & 0.756 & - & - & 0.244 \\
29 & HE0253-1641 & 18.4 & 5.2 & 17.6 & 1.3 & 2.5 & 0.182 & 0.43 & 0.176 & 0.212 \\
69 & HE1248-1356 & 16.4 & 7.9 & 16.4 & 1.6 & 3.0 & 0.167 & 0.792 & - & 0.042 \\
70 & HE1256-1805 & - & - & 20.5 & 5.1 & 3.2 & 0.841 & - & - & 0.159 \\
71 & HE1310-1051 & - & - & 18.8 & 3.7 & 3.0 & 0.717 & - & - & 0.283 \\
74 & HE1330-1013 & 18.3 & 12.7 & 18.7 & 2.1 & 1.6 & 0.053 & 0.706 & 0.165 & 0.076 \\
75 & HE1338-1423 & 18.4 & 11.1 & 18.2 & 3.7 & 4.7 & 0.334 & 0.44 & 0.096 & 0.13 \\
77 & HE1348-1758 & - & - & 18.5 & 5.1 & 5.0 & 0.726 & - & - & 0.274 \\
79 & HE1417-0909 & - & - & 19.4 & 4.8 & 4.1 & 0.751 & - & - & 0.249 \\
80 & HE2112-5926 & - & - & 18.3 & 4.0 & 4.3 & 0.918 & - & - & 0.082 \\
81 & HE2128-0221 & - & - & 17.7 & 2.0 & 2.8 & 0.82 & - & - & 0.18 \\
82 & HE2129-3356 & - & - & 18.6 & 4.4 & 2.9 & 0.866 & - & - & 0.134 \\
83 & HE2204-3249 & - & - & 17.4 & 2.9 & 3.0 & 0.891 & - & - & 0.109 \\
84 & HE2211-3903 & 18.0 & 6.1 & 17.4 & 1.9 & 2.2 & 0.183 & 0.427 & 0.142 & 0.248 \\
85 & HE2221-0221 & - & - & 18.3 & 2.4 & 3.2 & 0.477 & - & - & 0.523 \\
89 & HE2236-3621 & - & - & 20.9 & 3.6 & 3.5 & 0.593 & - & - & 0.407 \\

\hline
\end{tabular*}
\end{center}
\end{table*}

\begin{table*}
\begin{center}
\caption{Parameters of the galaxies in the $K$-band. For explanations, see the notes in Table \ref{tab:jband}} 
\label{tab:kband}
\begin{tabular*}{\textwidth}{@{\extracolsep{\fill}}cc|ccccccccc} \hline \hline
ID & Name & $\mu_0$ & $h$ & $\mu_e$ & $r_e$ & $n$ & B/T & D/T & Bar/T & AGN/T \\
(1) & (2) & (3) & (4) & (5) & (6) & (7) & (8) & (9) & (10) & (11) \\ \hline

05 & HE2236-3621 & - & - & 16.8 & 2.1 & 1.7 & 0.422 & - & - & 0.578 \\
08 & HE2236-3621 & 18.3 & 7.3 & 17.4 & 2.7 & 3.0 & 0.33 & 0.389 & 0.176 & 0.104 \\
11 & HE2236-3621 & 16.8 & 7.1 & 16.8 & 2.8 & 3.0 & 0.476 & 0.315 & - & 0.209 \\
16 & HE2236-3621 & 16.6 & 3.0 & - & - & - & - & 0.541 & 0.139 & 0.32 \\
24 & HE2236-3621 & - & - & 18.2 & 3.4 & 4.0 & 0.601 & - & - & 0.399 \\
29 & HE2236-3621 & 17.9 & 4.7 & 17.2 & 1.3 & 2.4 & 0.168 & 0.38 & 0.146 & 0.305 \\
69 & HE2236-3621 & 16.5 & 9.3 & 16.4 & 2.2 & 2.9 & 0.266 & 0.683 & - & 0.051 \\
70 & HE2236-3621 & - & - & 20.2 & 7.7 & 1.8 & 0.839 & - & - & 0.161 \\
71 & HE2236-3621 & - & - & 18.4 & 4.0 & 4.3 & 0.622 & - & - & 0.378 \\
74 & HE2236-3621 & 18.6 & 24.4 & 18.4 & 2.1 & 2.8 & 0.041 & 0.76 & 0.108 & 0.091 \\
75 & HE2236-3621 & 17.9 & 11.5 & 17.8 & 3.5 & 4.0 & 0.209 & 0.497 & 0.069 & 0.226 \\
77 & HE2236-3621 & - & - & 17.9 & 4.3 & 4.0 & 0.637 & - & - & 0.363 \\
79 & HE2236-3621 & - & - & 18.9 & 4.5 & 3.1 & 0.648 & - & - & 0.352 \\
80 & HE2236-3621 & - & - & 17.7 & 3.6 & 3.3 & 0.805 & - & - & 0.195 \\
81 & HE2236-3621 & - & - & 17.3 & 1.9 & 2.6 & 0.758 & - & - & 0.242 \\
82 & HE2236-3621 & - & - & 17.8 & 3.4 & 3.4 & 0.984 & - & - & 0.016 \\
83 & HE2236-3621 & - & - & 17.5 & 3.4 & 3.0 & 0.783 & - & - & 0.217 \\
84 & HE2236-3621 & 17.9 & 6.4 & 17.0 & 1.9 & 2.9 & 0.196 & 0.282 & 0.089 & 0.432 \\
85 & HE2236-3621 & - & - & 17.5 & 2.2 & 3.0 & 0.328 & - & - & 0.672 \\
89 & HE2236-3621 & - & - & 20.3 & 3.6 & 3.2 & 0.602 & - & - & 0.398 \\

\hline
\end{tabular*}
\end{center}
\end{table*}

\subsection{Photometry}
\label{sec:method_photometry}
Near-infrared colors provide information on extinction and can help to distinguish whether the nuclear or the stellar component dominates the luminosity of the galaxy. Several components contribute to the NIR flux of an active galaxy and determine the colors: a stellar component, a non-stellar continuum source, an extinction component, and a component corresponding to the hot dust emission of the obscuring torus. \cite{1982MNRAS.199..943H} calculated the colors of a zero-redshift quasar (i.e. an object entirely dominated by the nonstellar continuum emission instead of stellar emission) to be $J-H=0.95$, $H-K=1.15$. The colors of ordinary galaxies are $J-H=0.78$, $H-K=0.22$ \citep{1984MNRAS.211..461G}.

We measured the flux of the galaxies in three apertures after smoothing the images in the $J$-, $H$-, and $K$-band to a common seeing. We began with an aperture whose radius corresponded to the seeing FWHM and therefore mainly contained the nucleus. The second aperture has a radius of 4''. The third given value corresponds to the magnitude of the \textsc{Budda} model and was supposed to contain the complete object. All apertures were centered on the nucleus. 

We also measured the NIR colors of the host galaxy. We used the \textsc{Budda} model of the stellar components. Spiral arms and other residual structures that are not modeled with \textsc{Budda} do not significantly affect our measurements. \cite{2008MNRAS.384..420G} pointed out that averaged over the complete annuli the deviation is almost zero. Therefore, we cannot deliver spatially resolved color information with our models. However, as we showed in Section \ref{sec:reliability}, total fluxes are reliable. Therefore, the colors for the whole galaxy are expected to be reliable as well.

\begin{figure*}
\centering
\includegraphics[width=\linewidth]{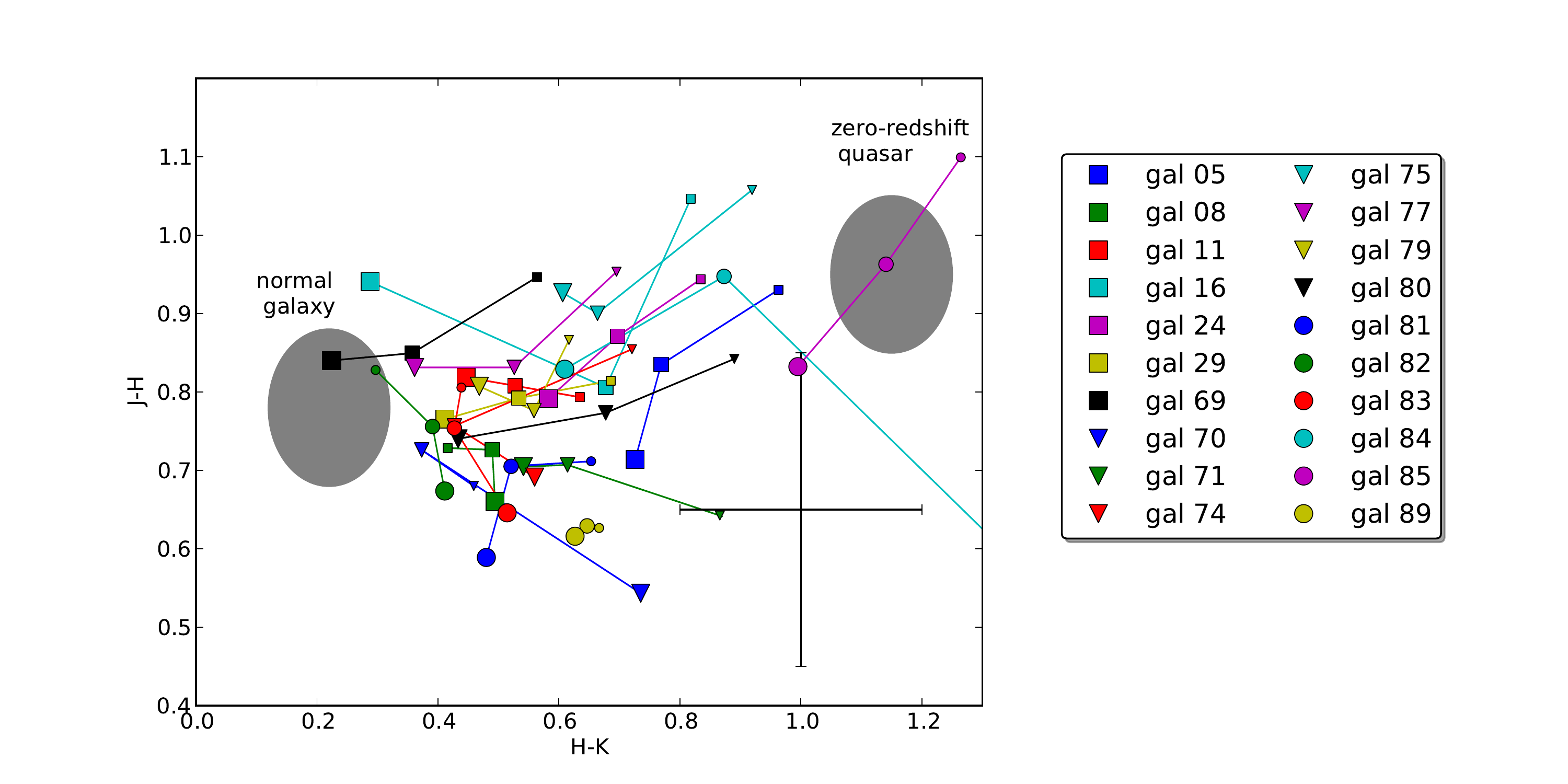}
\caption{NIR two-color diagram of the 20 observed galaxies. The three data points mark the different apertures (from large to small symbols: $14\arcsec$, $8\arcsec$, and $1.6\arcsec$ diameter). The error bar in the lower right corner indicates the typical error for individual measurements. The expected locations of normal (i.e. inactive) galaxies and zero-redshift quasars are indicated by ellipses. }
\label{fig:ccdiagram}
\end{figure*}

As expected from former studies \citep[e.g.][]{2006A&A...452..827F}, the measured colors lie well between the positions of inactive galaxies and zero redshift quasars that can be seen in the color-color diagram in Fig. \ref{fig:ccdiagram}. Small apertures centered on the nucleus are redder than large apertures that contain the whole galaxy, including blue light from the host galaxy. However, some galaxies show an unexpected behavior, that is, the color of the nucleus is bluer than the color of the complete galaxy. These peculiarities coincide with irregularities in the decomposition (e.g. galaxy 82 or 89). The mean colors of the AGN-subtracted galaxies are $J-H=0.68$ and $H-K=0.40$. If we neglect galaxies with the above-mentioned irregularities and poor S/N, we obtain $J-H=0.73$ and $H-K=0.35$.

Until now, $K$-corrections have not been applied because they are strongly dependent on the shape of the SEDs, which are not known in our case. If we assume a typical stellar population of quiescent galaxies and use the corrections given by \cite{1999A&A...351..869F}, $K_{J-H}=-0.2 z$ and $K_{H-K}=2.7 z$, we obtain mean host galaxy colors of $J-H=0.74$ and $H-K=0.25$. The standard deviations of the sample are below $\sigma = 0.15$ in all cases. However, the measurement uncertainties are much higher. Taking into account decomposition (see Sect. \ref{sec:reliability}) and calibration (see Sect. \ref{sec:observations}) errors, the uncertainties are around 0.2 mag. Taking into account these uncertainties, the colors are consistent with host galaxy colors in the literature, supporting the assumption that quasar host galaxies are not much different from ``normal'', inactive galaxies.

We also measured the colors of only the AGN component model. The median colors are $J-H=0.92$ and $H-K=0.80$, which overlaps with the colors of the zero-redshift quasar. However, in this case, the standard deviations ($\sigma_{J-H}=0.30$ and $\sigma_{H-K}=0.66$) as well as the measurement uncertainties are much higher.

\begin{table}
\begin{center}
\caption{NIR colors of the observed galaxies.}
\label{tab:colors} 
\begin{tabular}{cc|ccccc} \hline \hline
ID & Name & Aperture & $H$ & $M_H$ & $J-H$ & $H-K$ \\ \hline

05 & HE0036-5133 &
nucl. & 13.24 & -22.27 & 0.93 & 0.96 \\
& & 
interm. & 12.57 & -22.93 & 0.84 & 0.77 \\
& & 
compl. & 12.41 & -23.09 & 0.71 & 0.73 \\
& & 
host & 13.07 & -22.43 & 0.47 & 0.45 \\
\hline
08 & HE0045-2145 &
nucl. & 13.59 & -21.25 & 0.73 & 0.42 \\
& & 
interm. & 12.53 & -22.31 & 0.73 & 0.49 \\
& & 
compl. & 11.59 & -23.26 & 0.66 & 0.49 \\
& & 
host & 11.68 & -23.17 & 0.68 & 0.47 \\
\hline
11 & HE0103-5842 &
nucl. & 12.55 & -22.7 & 0.79 & 0.63 \\
& & 
interm. & 11.8 & -23.46 & 0.81 & 0.53 \\
& & 
compl. & 11.0 & -24.25 & 0.82 & 0.45 \\
& & 
host & 11.19 & -24.06 & 0.78 & 0.38 \\
\hline
16 & HE0119-0118 &
nucl. & 13.65 & -23.29 & 1.05 & 0.82 \\
& & 
interm. & 12.8 & -24.14 & 0.81 & 0.68 \\
& & 
compl. & 11.9 & -25.04 & 0.94 & 0.29 \\
& & 
host & 12.17 & -24.77 & 0.85 & 0.14 \\
\hline
24 & HE0224-2834 &
nucl. & 13.6 & -23.56 & 0.94 & 0.83 \\
& & 
interm. & 13.01 & -24.16 & 0.87 & 0.7 \\
& & 
compl. & 12.6 & -24.57 & 0.79 & 0.58 \\
& & 
host & 12.9 & -24.26 & 0.78 & 0.33 \\
\hline
29 & HE0253-1641 &
nucl. & 13.44 & -22.3 & 0.81 & 0.69 \\
& & 
interm. & 12.64 & -23.1 & 0.79 & 0.53 \\
& & 
compl. & 12.0 & -23.74 & 0.77 & 0.41 \\
& & 
host & 12.26 & -23.48 & 0.76 & 0.27 \\
\hline
69 & HE1248-1356 &
nucl. & 13.25 & -20.74 & 0.95 & 0.56 \\
& & 
interm. & 11.91 & -22.08 & 0.85 & 0.36 \\
& & 
compl. & 10.53 & -23.46 & 0.84 & 0.22 \\
& & 
host & 10.58 & -23.41 & 0.83 & 0.21 \\
\hline
70 & HE1256-1805 &
nucl. & 15.49 & -18.42 & 0.68 & 0.46 \\
& & 
interm. & 14.39 & -19.53 & 0.73 & 0.37 \\
& & 
compl. & 13.88 & -20.03 & 0.54 & 0.74 \\
& & 
host & 14.07 & -19.84 & 0.51 & 0.73 \\
\hline
71 & HE1310-1051 &
nucl. & 13.72 & -22.16 & 0.64 & 0.87 \\
& & 
interm. & 12.95 & -22.92 & 0.71 & 0.61 \\
& & 
compl. & 12.54 & -23.34 & 0.7 & 0.54 \\
& & 
host & 12.9 & -22.97 & 0.74 & 0.39 \\
\hline
74 & HE1330-1013 &
nucl. & 13.75 & -21.21 & 0.85 & 0.72 \\
& & 
interm. & 12.56 & -22.4 & 0.76 & 0.43 \\
& & 
compl. & 11.01 & -23.94 & 0.69 & 0.56 \\
& & 
host & 11.1 & -23.86 & 0.67 & 0.54 \\
\hline
\end{tabular} 
 \end{center} 
 \end{table} 
 \begin{table} 
 \begin{center} 
 \ContinuedFloat 
 \caption{continued} 
 \begin{tabular}{cc|ccccc} \hline \hline 
 ID & Name & Aperture & $H$ & $M_H$ & $J-H$ & $H-K$ \\ \hline
75 & HE1338-1423 &
nucl. & 12.93 & -23.41 & 1.06 & 0.92 \\
& & 
interm. & 12.02 & -24.31 & 0.9 & 0.66 \\
& & 
compl. & 11.14 & -25.2 & 0.93 & 0.61 \\
& & 
host & 11.29 & -25.05 & 0.9 & 0.48 \\
\hline
77 & HE1348-1758 &
nucl. & 12.69 & -20.89 & 0.95 & 0.7 \\
& & 
interm. & 11.96 & -21.62 & 0.83 & 0.53 \\
& & 
compl. & 11.34 & -22.23 & 0.83 & 0.36 \\
& & 
host & 11.69 & -21.88 & 0.74 & 0.22 \\
\hline
79 & HE1417-0909 &
nucl. & 14.53 & -21.92 & 0.87 & 0.62 \\
& & 
interm. & 13.55 & -22.9 & 0.78 & 0.56 \\
& & 
compl. & 13.28 & -23.17 & 0.81 & 0.47 \\
& & 
host & 13.59 & -22.86 & 0.78 & 0.31 \\
\hline
80 & HE2112-5926 &
nucl. & 13.6 & -22.12 & 0.84 & 0.89 \\
& & 
interm. & 12.71 & -23.01 & 0.77 & 0.68 \\
& & 
compl. & 11.98 & -23.74 & 0.74 & 0.43 \\
& & 
host & 12.07 & -23.65 & 0.76 & 0.29 \\
\hline
81 & HE2128-0221 &
nucl. & 14.8 & -22.05 & 0.71 & 0.65 \\
& & 
interm. & 13.81 & -23.04 & 0.71 & 0.52 \\
& & 
compl. & 13.56 & -23.3 & 0.59 & 0.48 \\
& & 
host & 13.77 & -23.08 & 0.54 & 0.39 \\
\hline
82 & HE2129-3356 &
nucl. & 14.07 & -21.47 & 0.83 & 0.3 \\
& & 
interm. & 13.2 & -22.34 & 0.76 & 0.39 \\
& & 
compl. & 12.65 & -22.89 & 0.67 & 0.41 \\
& & 
host & 12.8 & -22.74 & 0.58 & 0.55 \\
\hline
83 & HE2204-3249 &
nucl. & 13.42 & -23.7 & 0.81 & 0.44 \\
& & 
interm. & 12.74 & -24.38 & 0.75 & 0.43 \\
& & 
compl. & 12.26 & -24.86 & 0.65 & 0.51 \\
& & 
host & 12.38 & -24.74 & 0.68 & 0.37 \\
\hline
84 & HE2211-3903 &
nucl. & 13.53 & -22.69 & 0.09 & 2.02 \\
& & 
interm. & 11.91 & -24.31 & 0.95 & 0.87 \\
& & 
compl. & 11.16 & -25.07 & 0.83 & 0.61 \\
& & 
host & 11.46 & -24.76 & 0.74 & 0.3 \\
\hline
85 & HE2221-0221 &
nucl. & 13.14 & -23.89 & 1.1 & 1.26 \\
& & 
interm. & 12.71 & -24.32 & 0.96 & 1.14 \\
& & 
compl. & 12.47 & -24.56 & 0.83 & 1.0 \\
& & 
host & 13.28 & -23.75 & 0.42 & 0.59 \\
\hline
89 & HE2236-3621 &
nucl. & 15.34 & -21.81 & 0.63 & 0.67 \\
& & 
interm. & 14.86 & -22.28 & 0.63 & 0.65 \\
& & 
compl. & 14.49 & -22.65 & 0.62 & 0.63 \\
& & 
host & 15.06 & -22.08 & 0.49 & 0.64 \\
\hline

\end{tabular}

\end{center}
\end{table}

\subsection{Notes on individual objects}
\paragraph{05 HE0036-5133}
The galaxy is an elongated elliptical ($\epsilon=0.4$). The position in the color-color-diagram is consistent with a high AGN fraction. Only fits with AGN and one Sersic-component are successful. However, the low resulting Sersic-index of about 2, the bump in the ellipticity and the residuum can be seen as indicators for a hidden substructure. Fischer (2008, PhD-thesis) reported indications for a nuclear bar and spiral arms. These features cannot be seen with our resolution, but could be consistent with our results. A more detailed study is needed.

\paragraph{08 HE0045-2145}
Almost face-on spiral galaxy with a bar and a disk scale-length of 3.3 kpc. There is possibly an inner ring that can only be seen in the residuum. The nuclear colors are bluer than the colors of the complete galaxy. Additionally, the fitted AGN fraction is very low. Optical spectra of this source (in preparation) suggest that the classification as Seyfert 1 is incorrect. 

\paragraph{11 HE0103-5842}
At first sight a barred galaxy. More precise inspection reveals a spiral structure inside the bar-like feature. This could be interpreted as a highly inclined spiral structure. In the residuum, a smaller-scaled elongated structure can be seen (possibly a bar?). Furthermore, additional large-scale structure - first interpreted as spiral arms - is seen. Its nature is unclear. Therefore, we classified the galaxy as irregular. A detailed study of the dynamics could be interesting.

\paragraph{16 HE0119-0118}
The galaxy has a redshift of almost 0.06 and thus a small angular diameter. A very prominent bar and spiral arms make a good disk-fit impossible. In the residuum, indications for an inner ring can be seen. A bulge component could not be fitted. Therefore, we excluded this galaxy when estimating BH masses from the bulge luminosity.

\paragraph{24 HE0224-2834}
This galaxy - the object with the highest redshift in our sample - is currently merging and highly perturbed. Because of the strong perturbations we only fit a general Sersic-profile, resulting in a Sersic index of $n\approx4.5$ and an effective radius of $r_e = 4.4$ kpc.

\paragraph{29 HE0253-1641}
A face-on barred spiral galaxy. The bulge component is very compact with an effective radius of $r_e=0.8$ kpc. 

\paragraph{69 HE1248-1356}
A highly inclined ($i\approx 69^\circ$) spiral galaxy at low redshift of $z=0.0145$. The disk has a scale length of $h_r = 2.5$ kpc and very prominent spiral arms. We see no indication for a substructure. However, because of the high inclination, the detection of a bar would be difficult. The AGN fraction is very low. 

\paragraph{70 HE1256-1805}
A circular elliptical. The galaxy is quite dim and a bright foreground star was in the field of view, resulting in a poor S/N. The nuclear colors indicate a low AGN fraction. This is consistent with the decomposition result (AGN fraction below 20\%). The colors in larger apertures have large errors and thus are less reliable. The decomposition does not reveal any irregularities. We fit a Sersic profile, resulting in an effective radius of $r_e=1.7$ kpc and a Sersic index of $n\approx 2-3$.

\paragraph{71 HE1310-1051}
The galaxy has no prominent features that deviate from the spherical symmetry in our NIR images. We fit an AGN component and a bulge component with effective radius $r_e=2.6$ kpc and Sersic index varying between $n=2.5$ in the $J$-band and $n=4.3$ in the $K$-band. The subtraction of the model reveals an arm-like structure that could be identified as spiral arms or tidal tails. In HST images \citep[][as PG 1310-108]{1998ApJS..117...25M}, the arm is more prominent.

\paragraph{74 HE1330-1013}
The galaxy has a disk scale length of $h_r=8$ kpc and shows a prominent elongated bar-like structure surrounded by a ring. On one side, a tail can be seen. Its origin cannot be explained with imaging data. A more detailed study to reveal the dynamical properties would be interesting. The bulge component has a low S\'ersic-index value and lies slightly offset from the Kormendy-relation (Fig. \ref{fig:fp}). This could be seen as indication of a pseudo-bulge \citep{2009MNRAS.393.1531G}.

\paragraph{75 HE1338-1423}
The galaxy consists of a disk with an inclination $i\approx 54^\circ$ and a scale length of $h_r=9.6$ kpc. Additionally, we found a spheroidal component with ellipticity $\epsilon\approx 0.2$ and effective radius $3.1$ kpc, that is, we found a slightly smaller bulge and disk than in the fit of \cite{2004MNRAS.352..399J}. This could be an effect of including a bar in our fit. However, we found deviations from a purely elliptical shape, which might be indicative of a more complex structure, for instance, a warped disk.

\paragraph{77 HE1348-1758}
A very circular elliptical. There are no signs for irregularities or perturbations. No candidates for companion galaxies are found in the field of view.

\paragraph{79 HE1417-0909}
In the $K$-band image, we were unable to avoid residua originating from the data reduction. Therefore, the colors and fits are not reliable. We found an elliptical with $\epsilon=0.4$ and no other irregularities. There are several neighboring galaxies at the same redshift in the region (distance $\approx$ 0.5-1\arcmin, redshift from NED), which could be interacting with HE1417--0909. Another cluster of galaxies at the same redshift can be found at $\approx$ 6\arcmin  distance.

\paragraph{80 HE2112-5926}
A very circular elliptical with no sign for perturbations. The center as well as the AGN-subtracted host galaxy are very red. However, the decomposition reveals a low AGN fraction. The reddening is more likely to come from dust extinction. There is a nearby companion (spiral galaxy), which forms the interacting galaxy pair ESO 144-IG 021 with HE2112--5926.

\paragraph{81 HE2128-0221}
A very elongated elliptical with $\epsilon=0.5$ at a redshift of $z=0.0528$. We fit a Sersic profile with $r_e=2$ kpc and $n\approx 3$. The residuum implies a substructure. However, we were unable to find indications of other structures, neither in optical, nor in SDSS images.

\paragraph{82 HE2129-3356}
We found an elongated elliptical. The color analysis shows that the host is redder than the nucleus. Decomposition indicates that the AGN fraction is very low and even drops in the $K$-band. We found the galaxy to be surrounded by several objects, some of them extended. However, we have no additional information about these objects and cannot confirm interaction.

\paragraph{83 HE2204-3249}
The AGN fraction in the $H$-band is lower than expected compared with that of the $J$- and $K$-band. Additionally, the host galaxy is redder than the nucleus. We suspect that a dust lane follows the semi-major axis of the galaxy. This is supported by the residual image of our decomposition analysis, which clearly indicates a residual structure that follows the symmetry axis of the galaxy.

\paragraph{84 HE2211-3903}
A spiral galaxy with a disk scale-length of $h_r=5$ kpc and a small bulge ($r_e=1.4$ kpc). The 2D decomposition of this barred spiral galaxy reveals an inner ring and a third spiral arm \citep[see also][]{2011AJ....142...43S}.

\paragraph{85 HE2221-0221}
Very symmetric, almost circular elliptical with $r_e=2.4$ kpc. The nuclear colors are very red. This is consistent with the high AGN fraction of 65\% obtained from our fits in $K$-band.

\paragraph{89 HE2236-3621}
The decomposition of this quite dim elliptical galaxy reveals a tail-like irregularity. The residuum also shows an irregularity in the center (which cannot be explained by a PSF-mismatch etc.). The origin of the perturbation is unclear.

\section{Discussion}
\label{sec:discussion}
We discuss the results of photometry and decomposition by commenting on the morphology, estimating stellar and black hole masses, and bringing them into context with the overall AGN population.

\subsection{Morphology}
\begin{figure}
\centering
\includegraphics[width=\columnwidth]{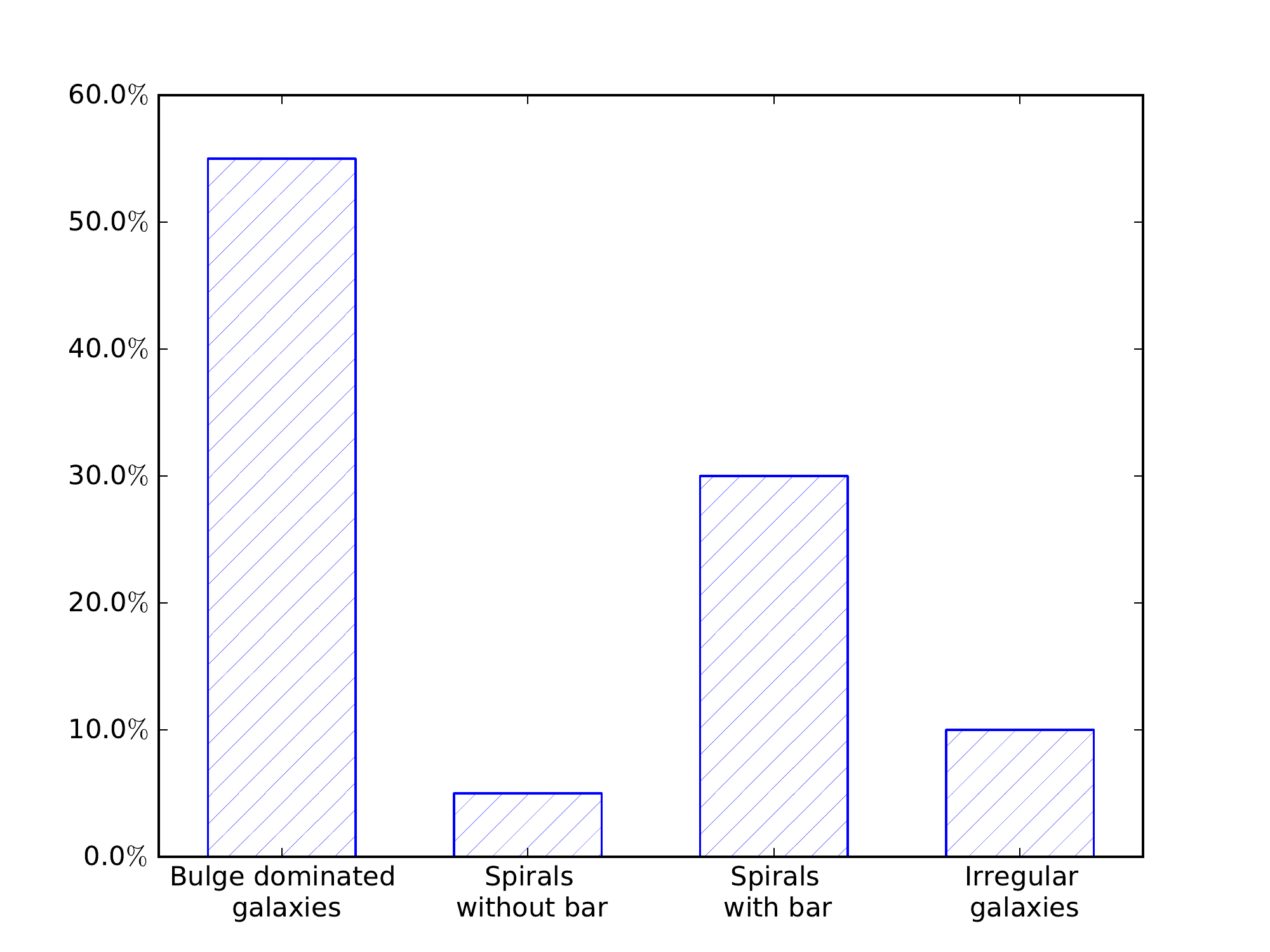}
\caption{Distribution of Hubble types of the 20 observed galaxies.}
\label{fig:morph}
\end{figure}

Of the 20 observed objects, we classified eleven as bulge dominated. In one case, we fitted only disk and bar. Eight galaxies have disk and bulge components. Seven galaxies show a prominent bar. Figure \ref{fig:morph} shows a histogram of the Hubble classification of the 20 galaxies that are examined in this study.

In this paper, we present a detailed study of the structural parameters based on decomposition. In a previous study \citep{2013arXiv1307.1049B}, however, we presented a morphological classification that was only based on visual inspection. We used images of an additional 26 galaxies from the SDSS (optical) and the ESO archive (NIR, obtained with NACO and ISAAC) to augment our morphological statistics. This resulted in 46 galaxies, almost $50\%$ of the LLQSO sample. The statistics show consistently that the fraction of barred galaxies within spiral galaxies is very high (6 out of 7 resp. 19 out of 22 (86\%) spiral galaxies are barred). Observations of the remaining 53 galaxies of the LLQSO sample might be useful to ensure the statistical significance.

\cite{2012ApJ...750..141L} analyzed a volume-limited sample of SDSS galaxies with $0.01<z<0.05$, $M_r<-20.3$ and $M_*>10^{10} M_\odot$, similar to our sample. They found a bar-fraction of 41.4\% at optical wavelengths. Other studies \citep[e.g.][]{2009ASPC..419..402H,2012ApJ...745..125L} likewise suggested that one third of the spiral galaxies is barred at optical wavelengths. However, in the NIR, the bar detection rate tends to be higher, about two thirds \citep{2004ARA&A..42..603K}. The bar fraction in the sample presented here is higher. This is an interesting fact since bars are commonly seen as a possible way to fuel central SMBHs \citep[e.g.][]{2003ASPC..290..411C}. On the other hand, it is discussed that bars are not the only way to feed the central engine. Some groups have observed a higher bar fraction in AGN galaxies than in non-AGN galaxies \citep[e.g.][]{2002ApJ...567...97L}. However, \cite{2012ApJ...750..141L} found that no significant difference in bar fraction between active and inactive galaxies can be observed if one compares samples of galaxies that are matched in mass, color, etc.

\begin{figure}
\centering
\includegraphics[width=\columnwidth]{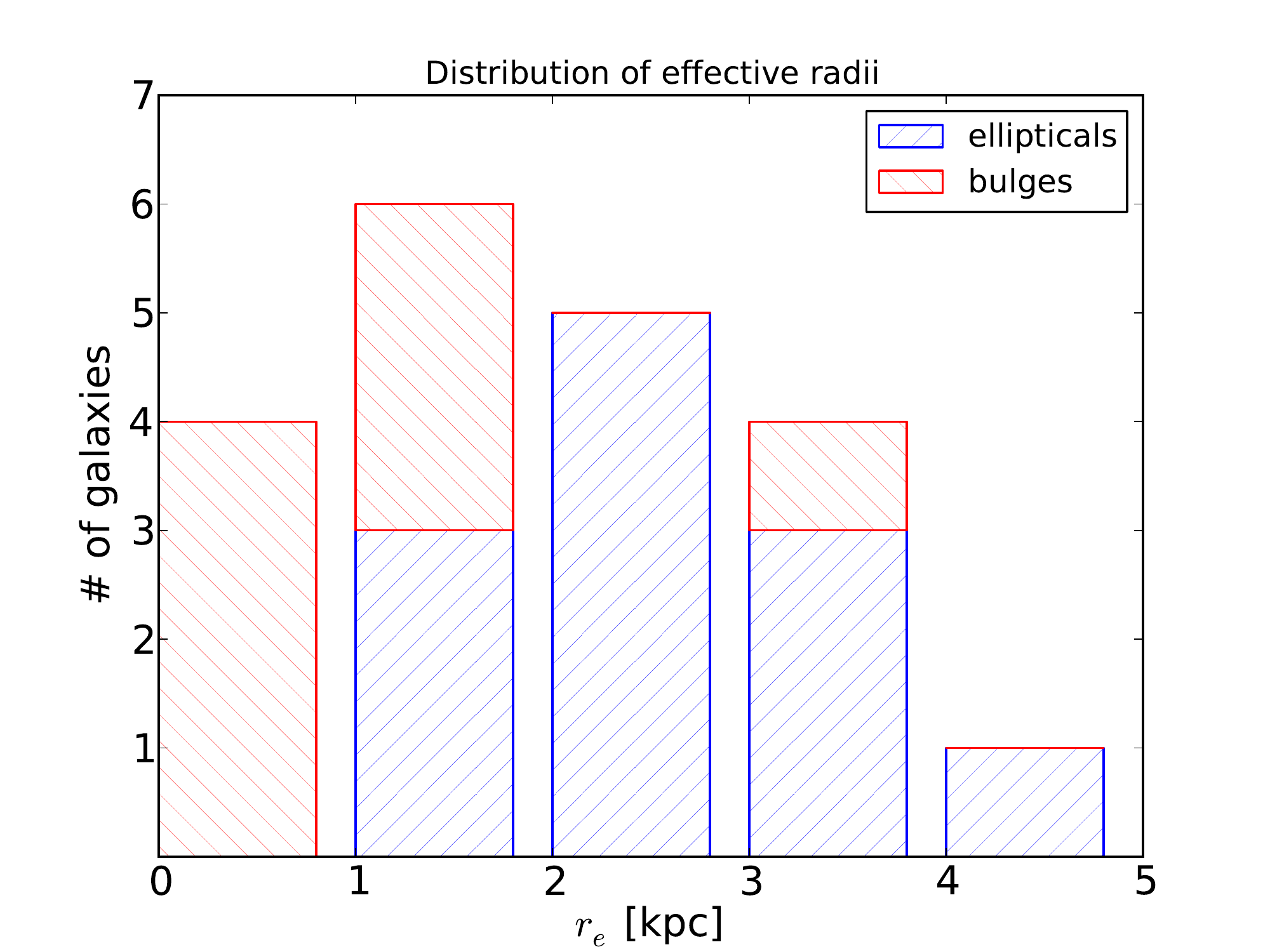}
\caption{Histograms of the effective radius of the bulge component (red) of spiral galaxies or the whole host galaxy for elliptical galaxies (blue). To calculate the physical scales, we assumed the standard cosmology described in Sect. \ref{sec:intro}.}
\label{fig:linearscale}
\end{figure}

The distribution of the linear extent of the central spheroidal components is shown in Fig. \ref{fig:linearscale}. Disk scale-lengths range from 2.5 to 9.6 kpc, with an average of 5 kpc. The effective radii of the spheroidal component of the spiral galaxies range from 0.5 to 3.0 kpc, with an average of 1.3 kpc. The effective radii of the elliptical hosts range from 1.2 to 4.5 kpc, with an average of 2.7 kpc. \cite{2007ApJ...657..102D} observed 12 local (mainly Palomar-Green) QSOs and found a mean effective radius of 3.9 kpc for QSOs and 2.2 kpc for ULIRGs, analyzed in previous studies \citep{2006ApJ...638..745D,2006ApJ...651..835D}. \cite{2004MNRAS.352..399J} observed nearby quasars compiled from the Hamburg/ESO survey and found a mean effective radius of 5.4 kpc for the elliptical hosts. In this sample with even closer galaxies ($z<0.06$), we found a mean effective radius of 2.7 kpc for elliptical galaxies of the sample. We conclude that the LLQSOs are less extended than the more powerful PG quasars and HES quasars.

To investigate the dynamical properties of the spheroids of the host galaxies in more detail, we investigated the positions of the objects in the fundamental plane \citep{1987ApJ...313...59D,1987ApJ...313...42D}. Kinematics and light distribution are both consequences of the overall dynamical state of the system. Furthermore, the light distribution depends on the mass-to-light ratio $M_*/L$ and thus on the star formation history. Massive elliptical galaxies and classical bulges follow a well-defined relation in the $R_\mathrm{eff}-\mu_\mathrm{eff}$ projection of the fundamental plane using the effective radius $r_\mathrm{eff}$ and the mean $K$-band surface brightness within the effective radius $\langle \mu_\mathrm{eff} \rangle$ from the \textsc{Budda} fits. This relation is also known as the Kormendy-relation \citep[first published by][]{1977ApJ...218..333K}. Pseudo-bulges that are built by secular evolution and therefore show different dynamics, are supposed to be found as outliers of this relation \citep{2009MNRAS.393.1531G}.

\begin{figure*}
\sidecaption
\includegraphics[width=12cm]{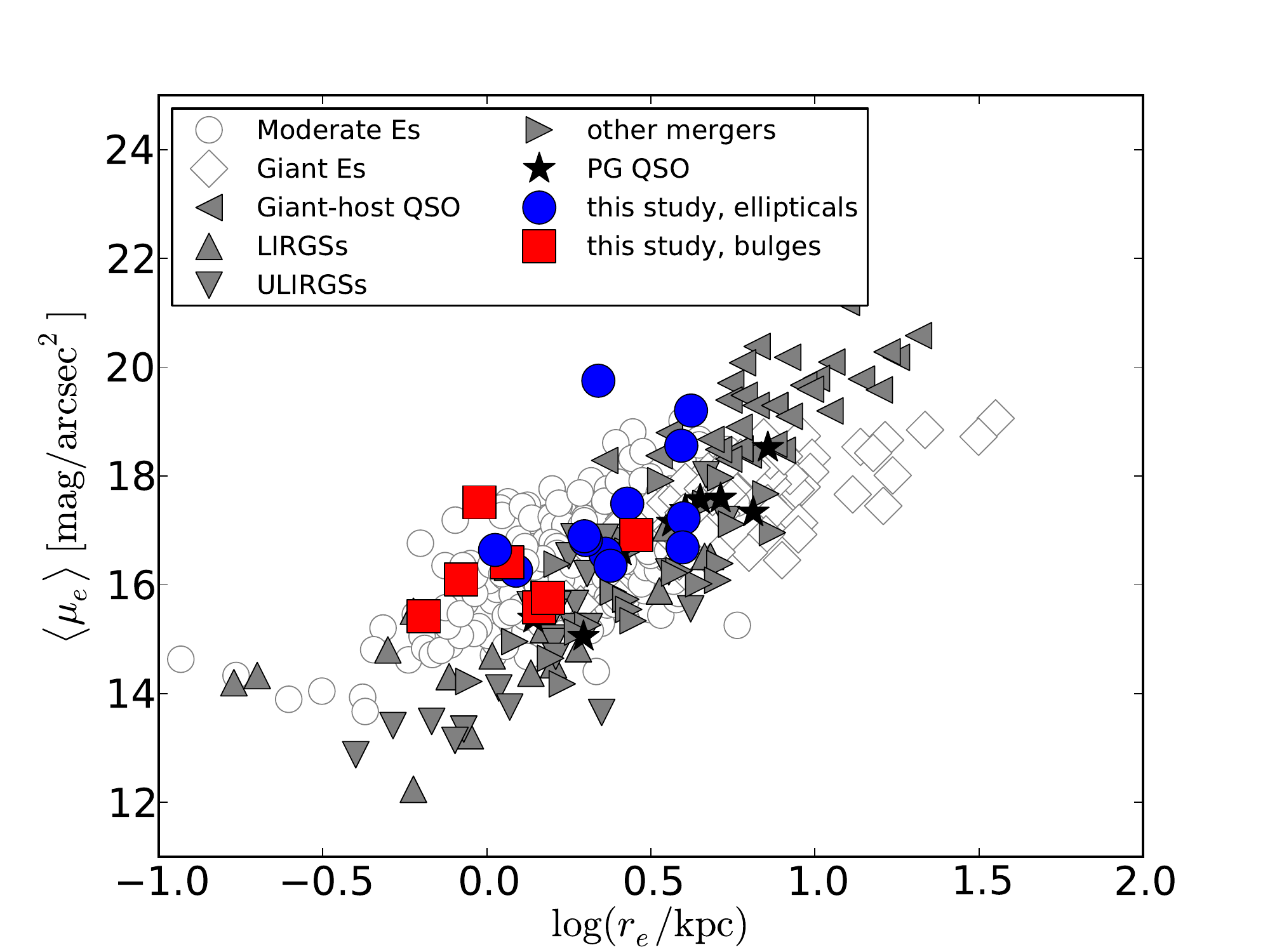}
\caption{Comparison of the observed LLQSOs with data from other studies in a plot of mean surface brightness within the effective radius $\langle \mu_{e} \rangle$ vs. effective radius $r_{e}$(in kpc), as presented in \cite{2007ApJ...657..102D}. This plot is a projection of the fundamental plane of early-type galaxies, often called Kormendy relation.}
\label{fig:fp}
\end{figure*}

In Fig. \ref{fig:fp}, we show our data points (blue circles for elliptical galaxies, red squares for bulges of spiral galaxies) together with data points for cluster-, moderate-, and giant ellipticals, luminous infrared galaxies (LIRGs), ultraluminous infrared galaxies (ULIRGs), other merger remnants, giant host QSOs and PG QSOs, collected by \cite{2007ApJ...657..102D}. In this diagram, the data points of the observed galaxies lie in the region of classically built (i.e. by mergers) bulges and ellipticals and therefore do not show clear evidence for the dominance of secular processes.

\subsection{Stellar masses}
\label{sec:stellmasses}
When photometric measurements of the host galaxies are available, estimates of the stellar mass can be made using the mass-to-light ratio $M_*/L$. Our 2D decompositions result in models for the different components. We considered all components, except the AGN component, to calculate the luminosity of the host galaxy. \cite{2009MNRAS.400.1181Z} expressed the $M_*/L$ ratio in the $H$-band as a function of the two colors $(g-i)$ and $(i-H)$ with a typical accuracy of 30 per cent. \cite{2003ApJS..149..289B} also used SDSS and 2MASS to calculate galaxy luminosity and stellar mass functions in the local Universe. They found a relation between $M/L$ ratio in the NIR and optical $(B-V)$ colors with the general form
\begin{equation}
\log\left(\frac{M_*/M_\odot}{L/L_{\odot,\lambda}}\right) = a_\lambda + b_\lambda (B-V).
\label{eq:mlratio}
\end{equation}
The constants are $a_J=-0.261,a_H=-0.209,a_K=-0.206,b_J=0.433,b_H=0.210$, and $b_K=0.135$. The authors adopted a ``diet'' Salpeter initial mass function (IMF). In case of a Kroupa IMF, for example, the zero points $a_\lambda$ have to be modified by subtracting $0.15$ dex.

Because of the absence of measured optical colors, we decided to apply Eq. \ref{eq:mlratio} and used average colors of $(B-V)=0.9$ and $(B-V)=0.7$ for inactive elliptical/S0 galaxies and spiral galaxies resp. \citep{1995PASP..107..945F}. As we discuss in the following subsection, the $M_*/L$ ratio of type-1 AGN hosts is unknown (\cite{2010ApJ...711..284S} found that hosts of narrow-line AGNs lie in the ``green valley'', while \cite{2013ApJ...763..133T} reported that more luminous broad-line AGN have bluer host galaxies.). Assuming that typical galaxy colors range between $(B-V)=0.5 - 1.0$, we found that the $M_*/L$-ratio ranges between $0.90-1.49$ for the $J$-band, $0.79-1.00$ for the $H$-band, and $0.73-0.85$ for the $K$-band. The color-dependence of the $M_*/L$ ratio in the $K$-band is the weakest. However, as the AGN fraction grows in the $K$-band, the uncertainty introduced by the AGN subtraction typically becomes larger.

\begin{figure}
\centering
\includegraphics[width=\columnwidth]{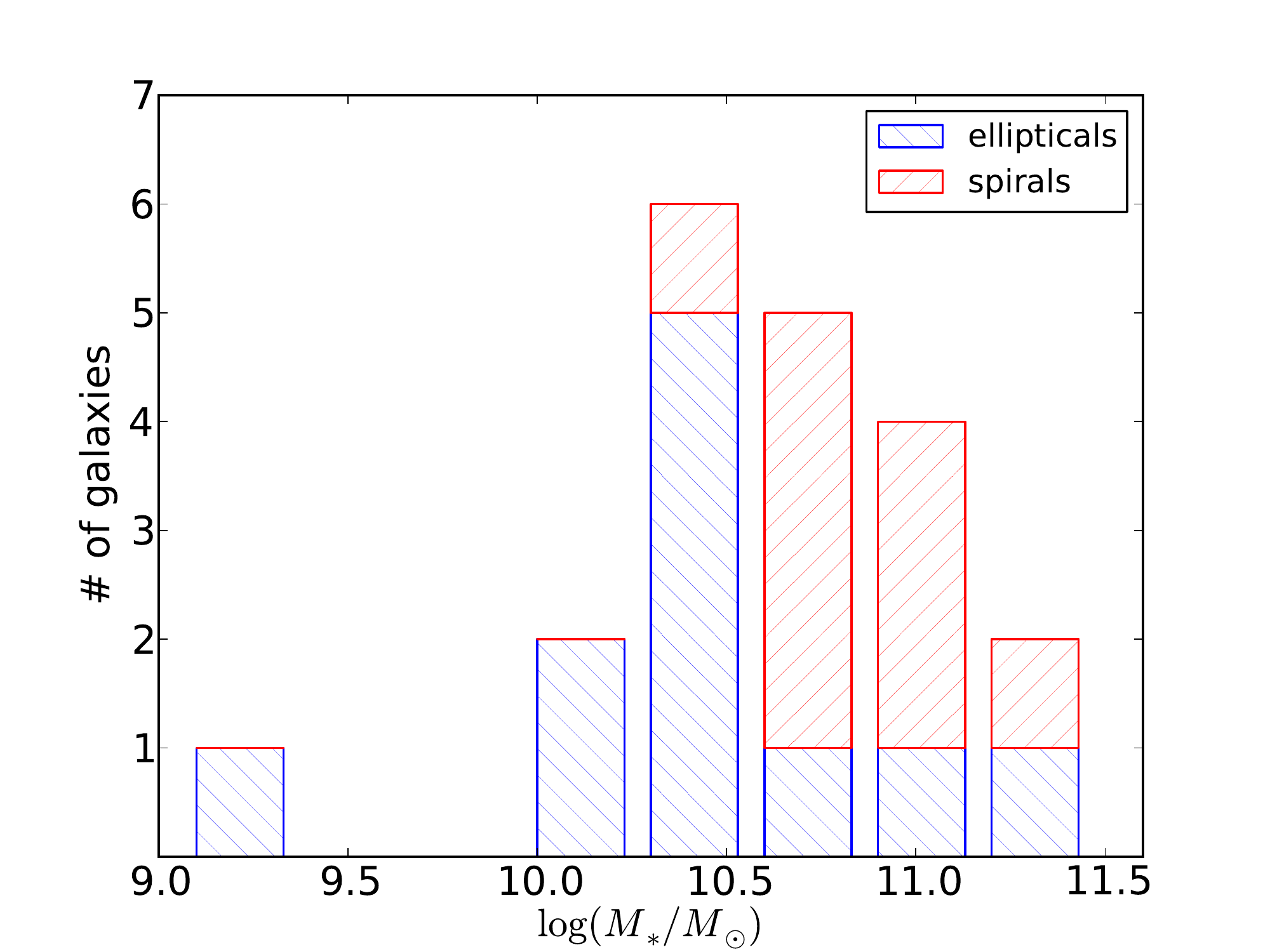}
\caption{Histogram of the stellar masses of the galaxies. The different colors indicate the morphology of the host galaxy.}
\label{fig:stellar_masses}
\end{figure}

Table \ref{tab:stellarmasses} shows the results of our calculations. The mass estimates calculated from the three different bands are consistent despite the mentioned uncertainties. In Fig. \ref{fig:stellar_masses}, we show a histogram of the stellar masses (averages of the mass estimates from the three bands) with different colors indicating the host galaxy's morphology. We found the average mass to be $6.7\times 10^{10} M_\odot$. \cite{2008MNRAS.388..945B} found a stellar mass function in the local Universe with $M^*=4.6\times 10^{10} M_\odot$. However, they assumed a Chabrier IMF \citep{2003PASP..115..763C}. Converting this to the IMF assumed in our calculations, we obtained $M^*\approx 6-8\times 10^{10} M_\odot$, that is, the observed LLQSOs have stellar masses on the order of ${\sim} M^*$.

\cite{2007ApJ...657..102D} found the stellar masses of the spheroids to be $2.1\times 10^{11} M_\odot$ on average, which corresponds to ${\sim} 3 M^*$ (or ${\sim} 1.5 M^*$ in their choice of $M^*$). We conclude that the hosts of our low-luminosity type-1 QSOs are on average less massive than PG quasars and therefore have stellar masses between local galaxies and PG quasars.

\begin{table*}
\centering
\caption{Host luminosities and stellar masses.}
\label{tab:stellarmasses}
\begin{tabular}{c|ccc|c|ccc|c} \hline \hline
Name & \multicolumn{3}{|c|}{$\log(L_{\mathrm{host},\lambda}/L_\odot)$} & type & \multicolumn{3}{|c|}{$\log (M_{*,\lambda}/M_\odot)$} & $\log (M_{*,\mathrm{av}}/M_\odot)$ \\ 
 & J & H & K & & J & H & K & \\ \hline
05 HE0036-5133 & 10.24 & 10.3 & 10.46 & E & 10.37 & 10.28 & 10.38 & 10.35 \\ 
08 HE0045-2145 & 10.45 & 10.59 & 10.77 & S & 10.49 & 10.53 & 10.65 & 10.57 \\ 
11 HE0103-5842 & 10.77 & 10.95 & 11.09 & S & 10.81 & 10.89 & 10.98 & 10.9 \\ 
16 HE0119-0118 & 11.03 & 11.24 & 11.28 & S & 11.07 & 11.17 & 11.16 & 11.14 \\ 
24 HE0224-2834 & 10.85 & 11.03 & 11.15 & S & 10.89 & 10.97 & 11.04 & 10.97 \\ 
29 HE0253-1641 & 10.54 & 10.72 & 10.81 & S & 10.59 & 10.66 & 10.7 & 10.65 \\ 
69 HE1248-1356 & 10.49 & 10.69 & 10.76 & S & 10.53 & 10.63 & 10.65 & 10.61 \\ 
70 HE1256-1805 & 9.19 & 9.27 & 9.54 & E & 9.32 & 9.25 & 9.46 & 9.35 \\ 
71 HE1310-1051 & 10.35 & 10.52 & 10.66 & E & 10.48 & 10.5 & 10.57 & 10.52 \\ 
74 HE1330-1013 & 10.73 & 10.87 & 11.07 & S & 10.77 & 10.81 & 10.96 & 10.86 \\ 
75 HE1338-1423 & 11.12 & 11.35 & 11.52 & S & 11.16 & 11.28 & 11.41 & 11.3 \\ 
77 HE1348-1758 & 9.91 & 10.08 & 10.15 & E & 10.04 & 10.06 & 10.07 & 10.06 \\ 
79 HE1417-0909 & 10.29 & 10.47 & 10.58 & E & 10.42 & 10.45 & 10.49 & 10.45 \\ 
80 HE2112-5926 & 10.61 & 10.79 & 10.89 & E & 10.74 & 10.77 & 10.8 & 10.77 \\ 
81 HE2128-0221 & 10.47 & 10.56 & 10.7 & E & 10.6 & 10.54 & 10.62 & 10.59 \\ 
82 HE2129-3356 & 10.32 & 10.42 & 10.63 & E & 10.45 & 10.4 & 10.54 & 10.47 \\ 
83 HE2204-3249 & 11.08 & 11.22 & 11.36 & E & 11.21 & 11.2 & 11.27 & 11.23 \\ 
84 HE2211-3903 & 11.06 & 11.23 & 11.34 & S & 11.1 & 11.17 & 11.23 & 11.17 \\ 
85 HE2221-0221 & 10.79 & 10.83 & 11.05 & E & 10.92 & 10.81 & 10.96 & 10.9 \\ 
89 HE2236-3621 & 10.1 & 10.16 & 10.4 & E & 10.22 & 10.14 & 10.32 & 10.23 \\ 
\hline
\end{tabular}
\tablefoot{
Host luminosity in the $J$-, $H$-, and $K$-band (columns 2-4), the classification assumed to chose a $M_*/L$ ratio (column 5), stellar masses estimated from the host luminosities in the three bands and the averaged mass (columns 6-9).
}
\end{table*}

\subsection{Black hole masses}
\label{sec:bhmasses}
For 74 galaxies from the LLQSO sample, eleven of which are analyzed in this paper, black hole masses are available from \cite{2009A&A...507..781S}, \cite{2010A&A...516A..87S}, and Schulze (priv. communication). In several observation campaigns, they took optical spectra of most of the type-1 AGN in the HES and computed BH masses using the scaling relation between broad line region (BLR) size and continuum luminosity by \cite{2009ApJ...697..160B} and the scale factor $f=3.85$ from \cite{2006A&A...456...75C},
\begin{equation}
M_\mathrm{BH} = 6.7 f  \left( \frac{L_{5100}}{10^{44}~ \mathrm{erg}~\mathrm{s}^{-1}} \right)^{0.52} \left( \frac{\sigma_{\mathrm{H}\beta}}{\mathrm{km}~\mathrm{s}^{-1}} \right)^2 M_\odot .
\end{equation}

\begin{figure}
\centering
\includegraphics[width=\columnwidth]{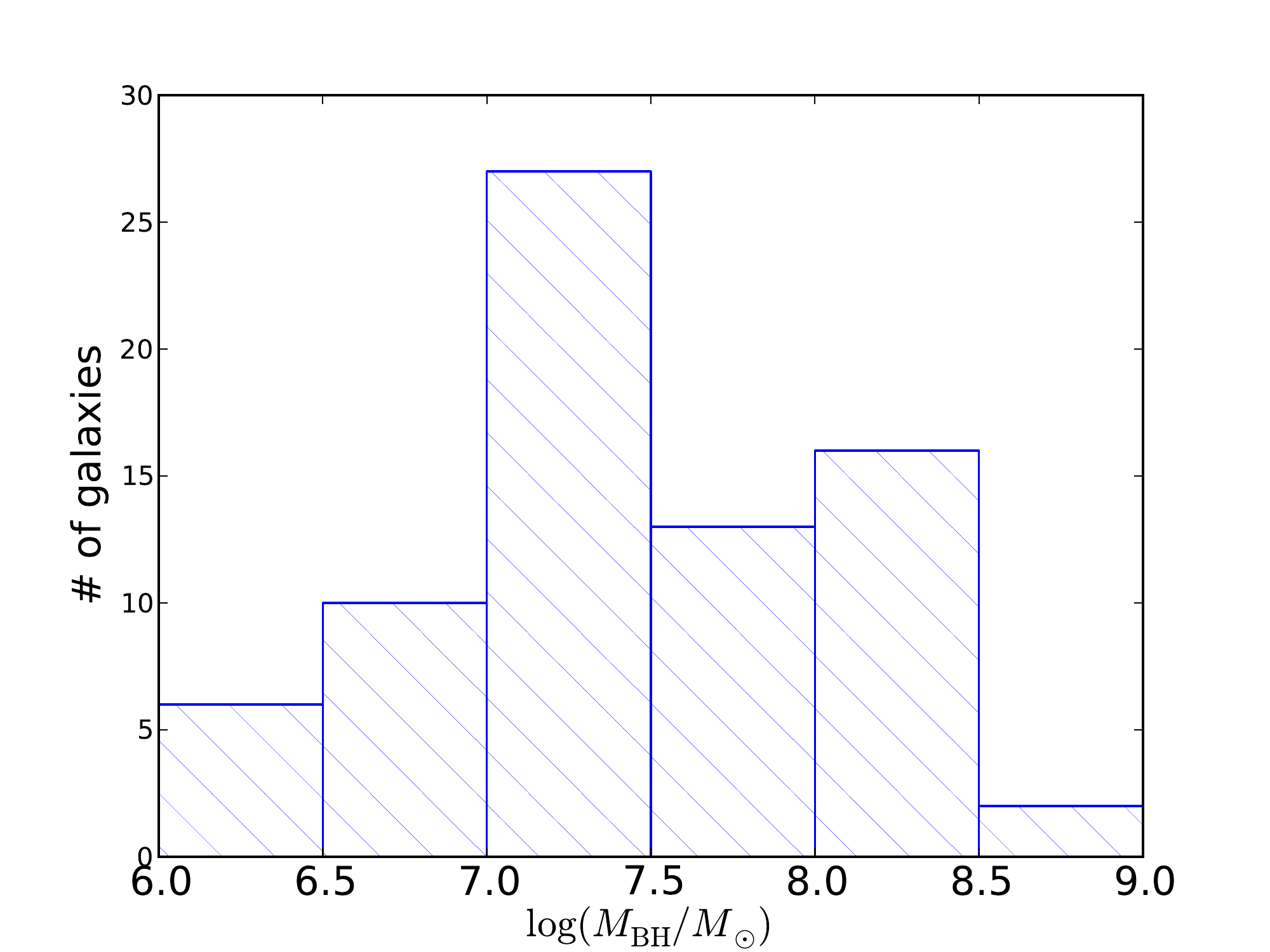}
\caption{Histogram of the SMBH masses in the low-luminosity type-1 QSO sample. BH masses of 74 galaxies have been measured.}
\label{fig:bhmasshist}
\end{figure}

Depending on the choice of slope and scale factor $f$, virial BH-mass estimators can differ by about $0.38$ dex \citep{2008ApJ...673..703M}. Particularly, using a more recent scale factor for active galaxies $f=5.9$ \citep{2013ApJ...772...49W} shifts the BH masses up by $\Delta M_\mathrm{BH} = 0.19\, \mathrm{dex}$. In Fig. \ref{fig:bhmasshist}, we show a histogram of the measured BH masses. In the LLQSO sample, they range from $6.0<\log (M_\mathrm{BH}/M_\odot) <8.7$ with a median of $\log (M_\mathrm{BH}/M_\odot) = 7.4$. \cite{2007ApJ...657..102D} found the BH masses to be typically around $10^8 M_\odot$.  We conclude that the observed low-luminosity type-1 QSOs have less massive central black holes than PG quasars. \cite{2006ApJ...641L..21G} compiled a sample of 88 spectroscopically identified AGN with $z\leq 0.05$. The BH masses typically range from $5.0<\log (M_\mathrm{BH}/M_\odot) <8.5$ with $\log (M_\mathrm{BH}/M_\odot) = 7.0$. This is consistent with our findings. \cite{2011ApJ...726...59B} observed type-1 AGN in a similar redshift range but selected only galaxies with $M_\mathrm{BH} > 10^7 M_\odot$. However, also in this study, only a few (2/25) galaxies with $M_\mathrm{BH} > 10^8 M_\odot$ were found. \cite{2006A&A...455..173P} analyzed 60 Seyfert galaxies from the Palomar optical spectroscpic survey of nearby galaxies (39 type-2, 13 type-1, 8 at boundary between Seyferts and LINERs or \ion{H}{ii}-region classification). Their BH masses range from $4.9<\log (M_\mathrm{BH}/M_\odot) <8.8$ with a median of $\log (M_\mathrm{BH}/M_\odot) = 7.6$ (median mass $\log (M_\mathrm{BH}/M_\odot) = 7.3$ for type-1 only) which is similar to our slightly more distant objects.

In the past decade, tight connections between black hole masses and properties of the central spheroidal component have been found. With high-quality near-infrared imaging data at hand, we are interested in the $M_\mathrm{BH}-L_\mathrm{bulge}$ relation that connects black hole mass and NIR luminosity of the bulge component. \cite{2003ApJ...589L..21M} first found a correlation
\begin{equation}
\log \left( \frac{M_\mathrm{BH}}{M_\odot} \right) = a_\lambda + b_\lambda \cdot \left[ \log \left( \frac{L_{\mathrm{bul},\lambda}}{L_{\odot,\lambda}} \right) - c_\lambda \right] ,
\end{equation}
which, for the BH mass range of their sample, has a scatter similar to that of the $M_\mathrm{BH}-\sigma$ relation. The constants for the three bands used in this study are $a_J=8.10,a_H=8.04,a_K=8.08$, $b_J=1.24,b_H=1.25,b_K=1.21,c_J=10.7,c_H=10.8$, and $c_K=10.9$.

\begin{figure*}
\sidecaption
\includegraphics[width=12cm]{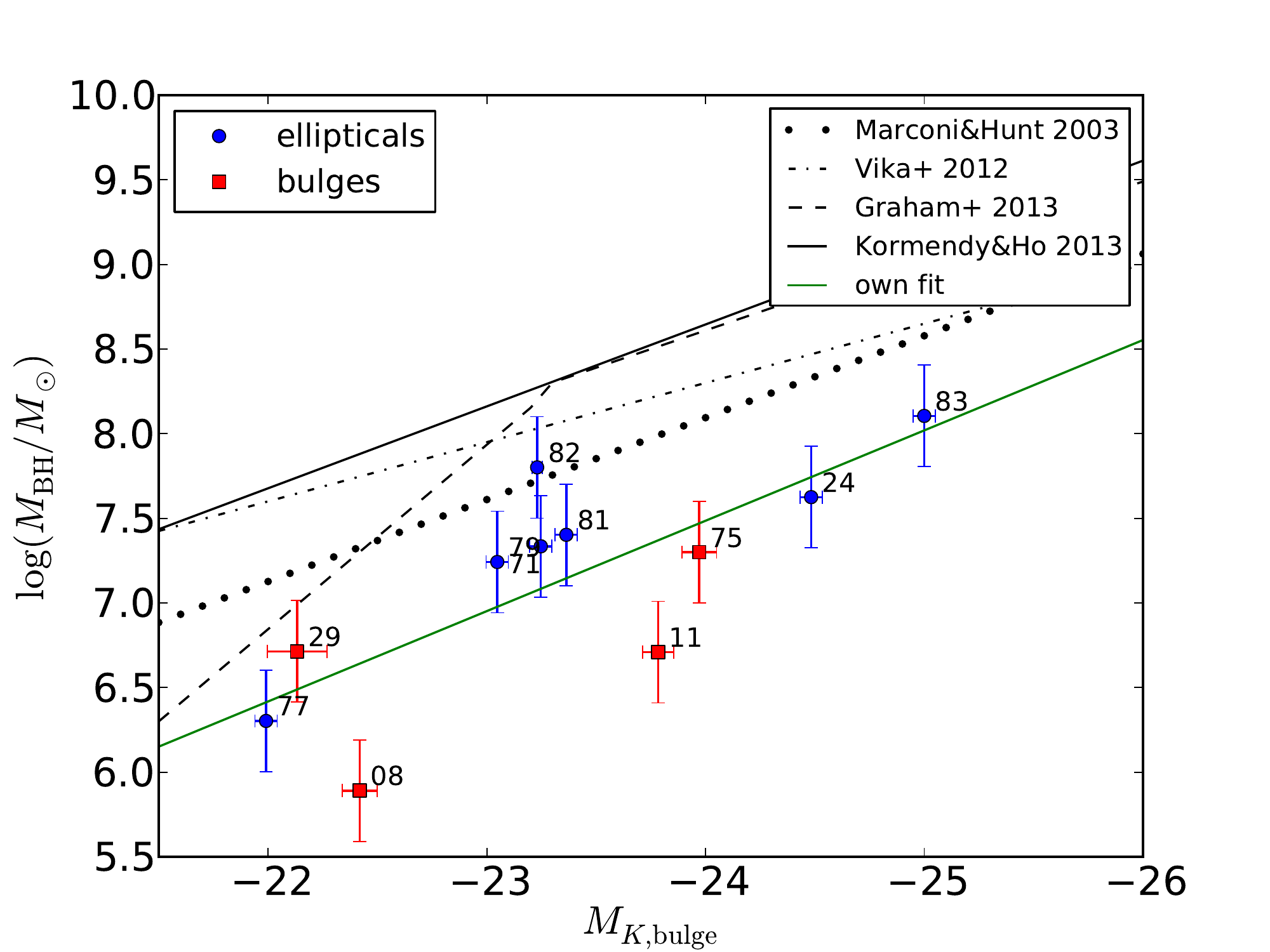}
\caption{Correlation between black hole mass $M_\mathrm{BH}$ and absolute $K$-band magnitude $M_K$ of the bulge or the elliptical galaxy. Our data (red squares indicate disk-dominated galaxies, blue circles bulge-dominated galaxies) plotted together with the $M_\mathrm{BH}-L_\mathrm{bulge}$ relations of \cite{2003ApJ...589L..21M}, \cite{2012MNRAS.419.2264V}, \cite{2013ApJ...764..151G}, and \cite{2013ARA&A..51..511K}. Our best fit is shown as the green solid line for comparison.}
\label{fig:bhmass}
\end{figure*}

\begin{figure*}
\centering
\includegraphics[width=0.45\linewidth]{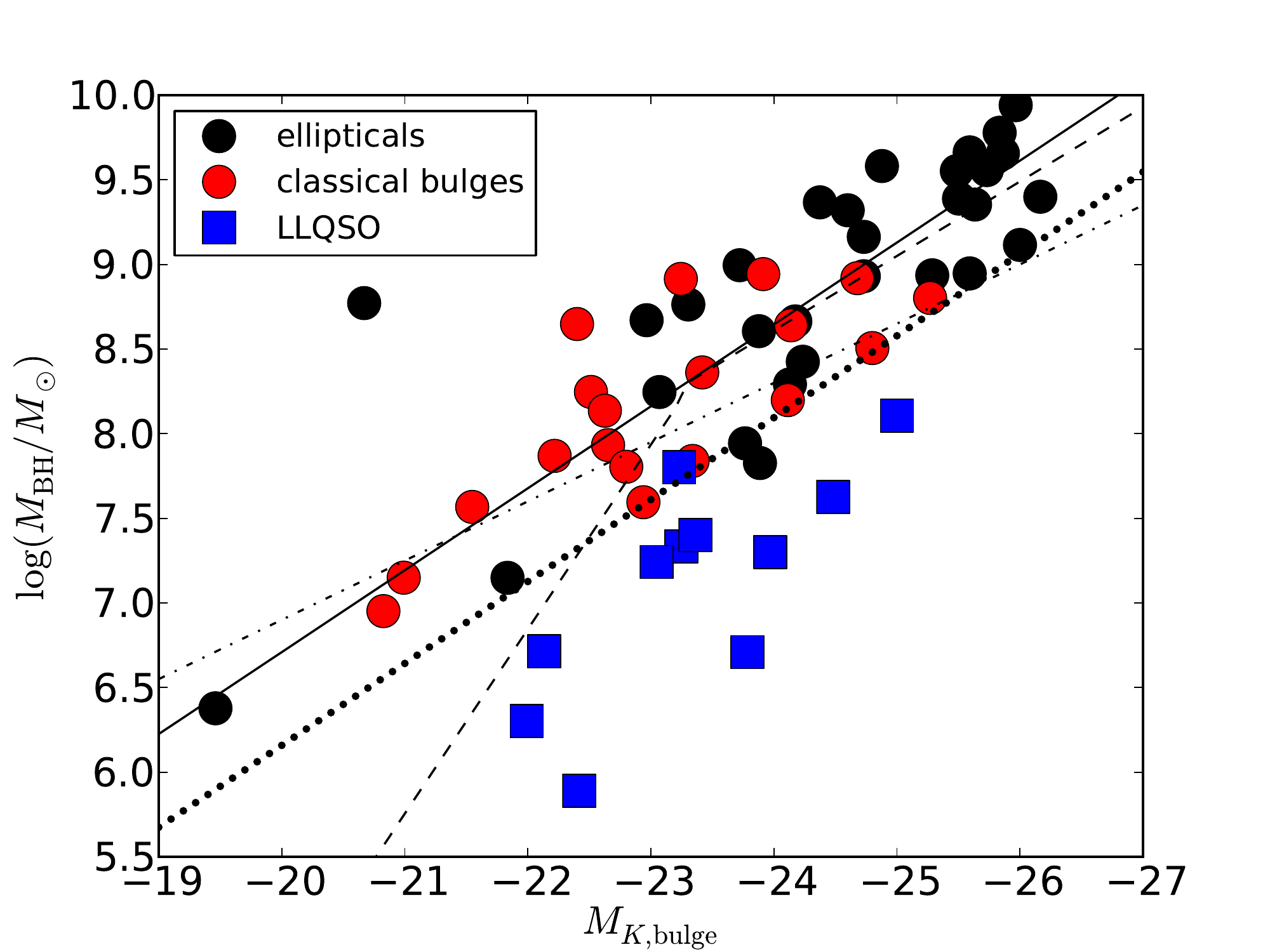}
\includegraphics[width=0.45\linewidth]{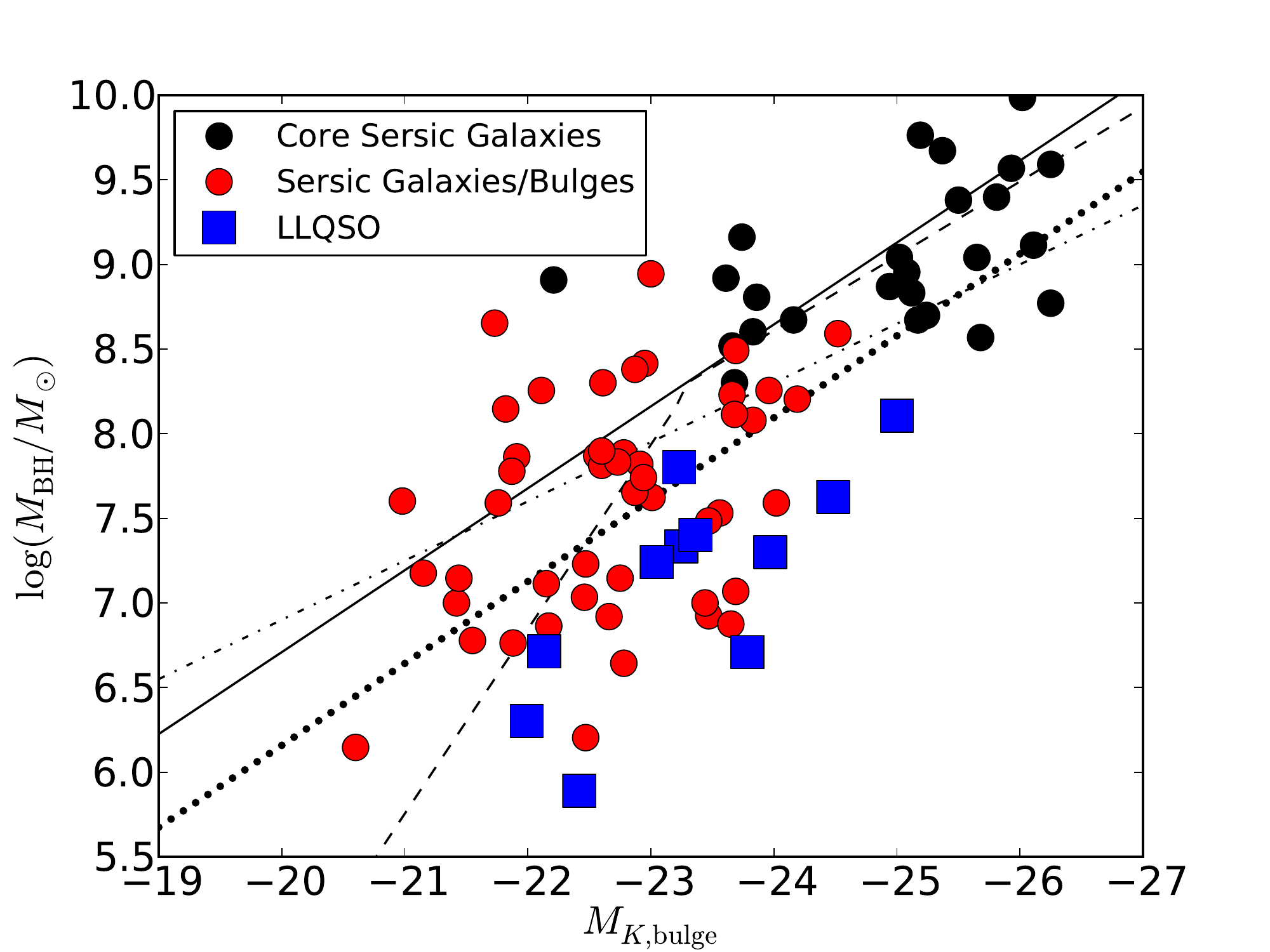} 
\caption{Correlation between black hole mass $M_\mathrm{BH}$ and absolute $K$-band magnitude $M_K$ of the bulge or the elliptical galaxy. Left: LLQSOs (blue squares) together with data collected by \cite{2013ARA&A..51..511K}. Right: LLQSOs together with data collected by \cite{2013ApJ...764..151G}. The lines indicate the same relations as in Fig. \ref{fig:bhmass}.}
\label{fig:bhmass2}
\end{figure*}

The BH masses and bulge magnitudes derived for the LLQSOs in this paper are listed in Table \ref{tab:bhmasses}. We have corrected the magnitudes for the systematic effect seen in Sect. \ref{sec:reliability}. To do this, we determined the mean deviation of the fitted bulge magnitude in bins of 20\% in the bulge fraction and subtracted this value from the measured value. The corrections are lower than 0.15 mag. In Fig. \ref{fig:bhmass}, we plot our data points together with a simple regression line and published $M_\mathrm{BH}-L_\mathrm{bulge}$ relations by \cite{2003ApJ...589L..21M}, \cite{2012MNRAS.419.2264V}, and \cite{2013ApJ...764..151G}. As error of the bulge magnitudes, we adopted the scatter we measured in the corresponding bulge-fraction bin in Sect. \ref{sec:reliability} (Fig. \ref{fig:qualbudda} right). The errors of the black hole masses are assumed to be 0.3 dex. Our data points follow none of these relations. In Fig. \ref{fig:bhmass2} we plot the black hole masses of the galaxies of this study as function of the absolute $K$-band magnitude together with data points collected in previous studies. We clearly see that the LLQSOs observed in this study lie below the location of the inactive classical bulges and ellipticals. 

Basically, we can consider two explanations of this discrepancy: (a) the bulges are overluminous and/or (b) the black holes are undermassive. In Sec. \ref{sec:reliability}, we show that the decomposition with \textsc{Budda} causes errors. However, the results of our tests indicate that the bulge-brightness fit uncertainties (a few tenths of dex in luminosity) are not large enough to explain our finding. \cite{2009A&A...507..781S} reported that they did not correct their continuum luminosities $L_{5100}$ for the host galaxy contribution. However, this effect is expected to be weak and more importantly shift the black hole masses to even lower masses. Thus, the deviation of the observed LLQSOs from the $M_\mathrm{BH}-L_\mathrm{bulge}$ relations of inactive galaxies is too high to be explained by the expected measurement uncertainties. 

An offset of type-1 AGN from the (optical) $M_\mathrm{BH}-L_\mathrm{bulge}$ relation has been reported by \cite{2004ApJ...615..652N} and \cite{2008ApJ...687..767K}. In a pilot study, \cite{2011ApJ...726...59B} connected SDSS images and high-quality Keck/LIRS long-slit spectra of nearby type-1 AGNs to study the relations between central black holes and host galaxies. While they found that the $M_\mathrm{BH}-M_*$ and $M_\mathrm{BH}-\sigma$ relation of active galaxies agrees with those of inactive ones, they also found an offset in the $M_\mathrm{BH}-L_\mathrm{bulge}$ relation (in $V$-band) and reported the host galaxies to be overluminous by 0.4 mag on average. \cite{2004ApJ...615..652N} suggested that brighter bulges can be explained by lower $M/L$ ratios. They used Guy Worthey's Dial-a-Galaxy Website \citep{1994ApJS...95..107W} to show that the mixing of 15\% 1 Gyr and 85\% 12 Gyr population results in a $\approx$0.5 mag brighter bulge than a bulge with old populations only. 

The so-called narrow-line Seyfert-1s (NLS1s) are galaxies with relatively small black holes and high Eddington ratios, indicating a strong black hole growth (e.g., \citealp{2011nlsg.confE...3B}). The observed objects have a mean Eddington ratio\footnote{The Eddington ratio is a measure of the accretion efficiency and is defined as $\eta \equiv \frac{L_\mathrm{bol}}{L_\mathrm{edd}} = 9.47\times 5100 L_{5100}/1.26\times 10^{38} (M_\mathrm{BH}/M_\odot)$ \citep{1994ApJS...95....1E}.} of $\eta_\mathrm{av}=0.18$ (median $\eta_\mathrm{median}=0.082$), indicating efficient accretion but not conspicuously heavily growing black holes as seen in some NLS1s. It has been found that NLS1s also lie below the $M_\mathrm{BH}-L_\mathrm{bulge}$ relation of inactive galaxies \citep{2002ApJ...565..762W,2007ApJ...654..799R}, while they seem to follow the $M_\mathrm{BH}-\sigma_*$ relation \citep{2007ApJ...667L..33K}. Since NLS1s might be the low-$M_\mathrm{BH}$ tail \citep[see discussion in][]{2013arXiv1305.3273V} of the type-1 AGN population, we expect them to show the same trend as our sample of type-1 sources.

\begin{figure*}
\centering
\includegraphics[width=0.45\linewidth]{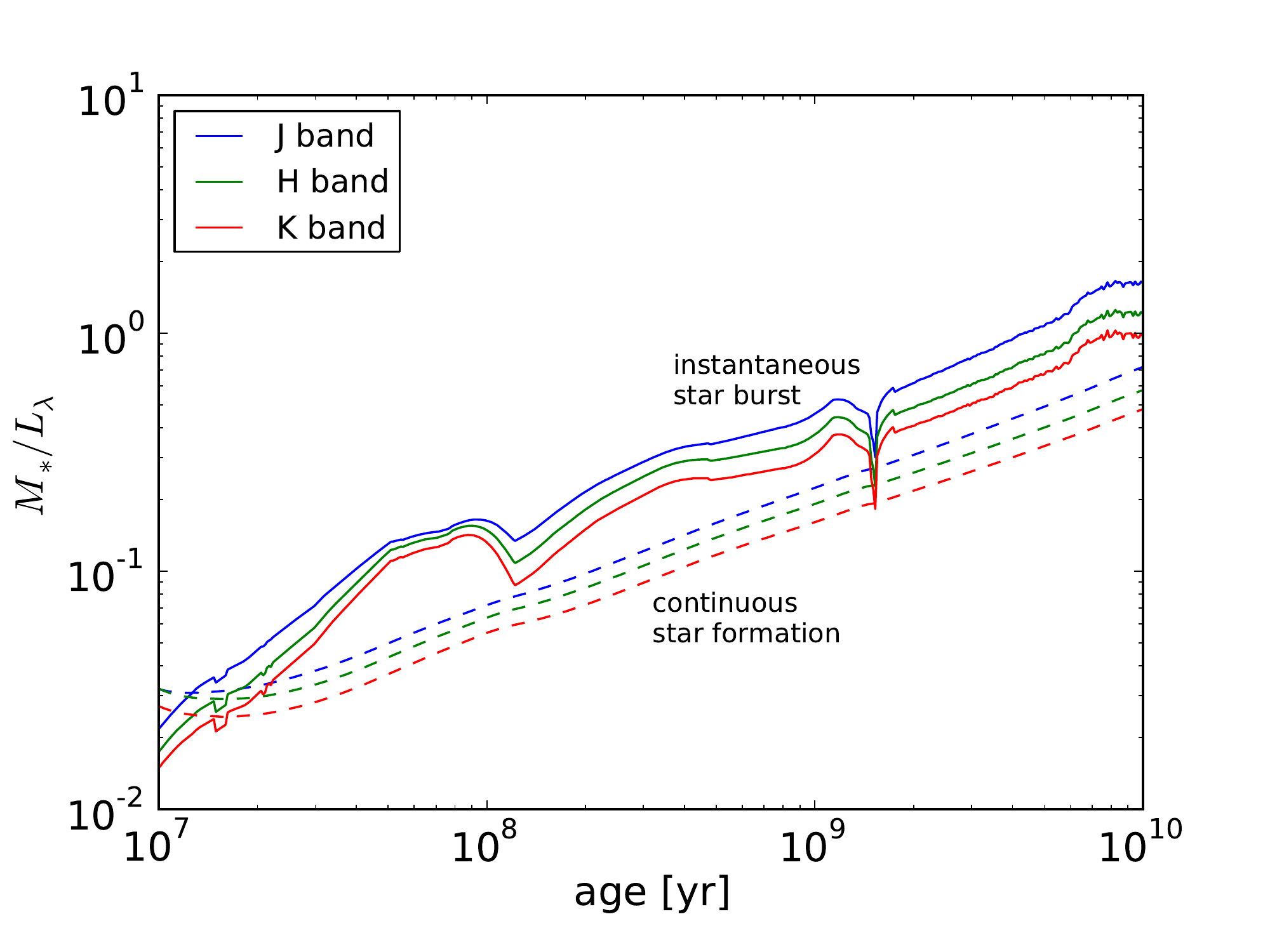}
\includegraphics[width=0.45\linewidth]{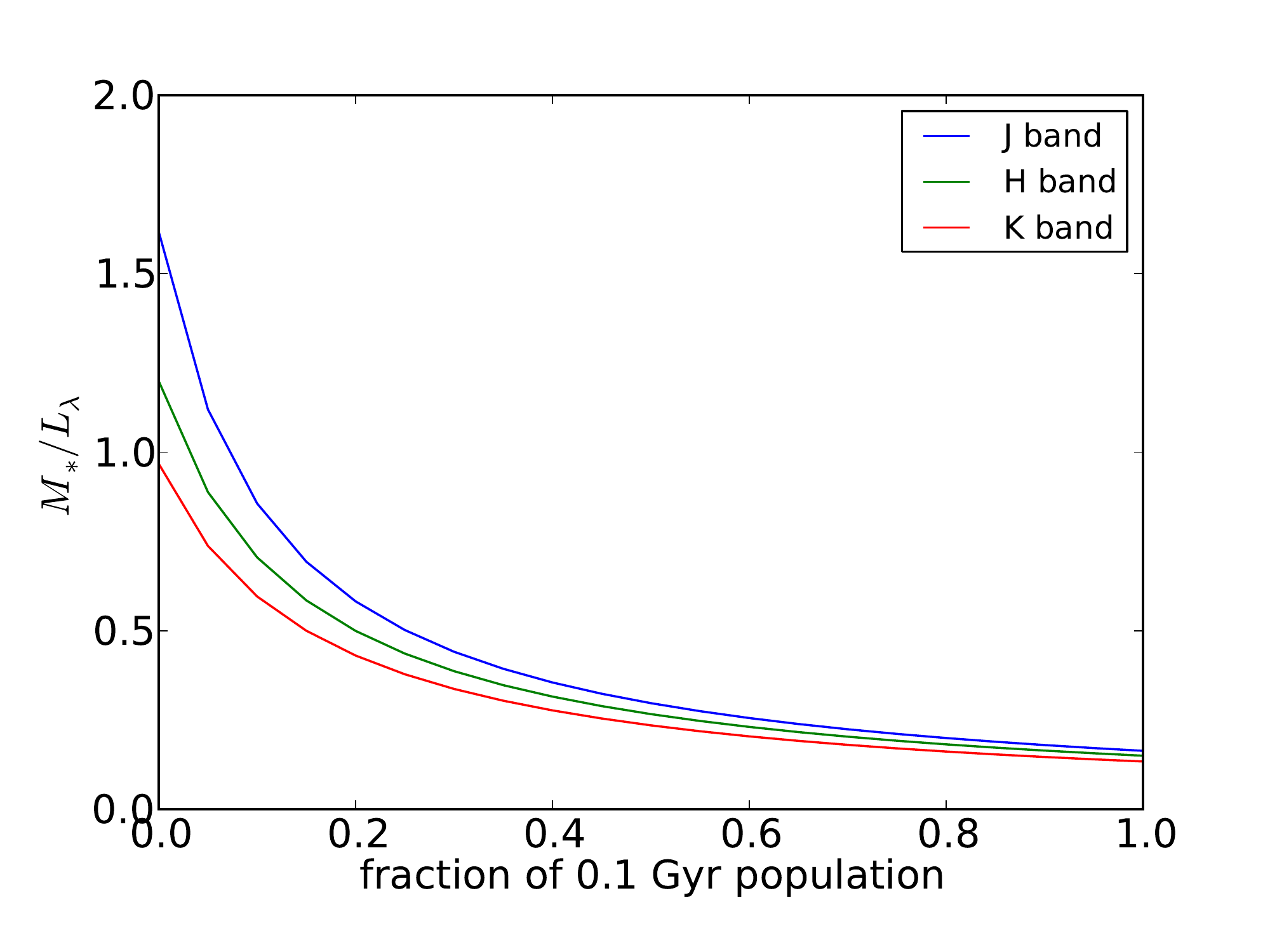}
\caption{Left: $M_*/L_\lambda$ ratio in $J$-, $H$-, and $K$-band of the young stellar population as a function of age. Different colors indicate the three bands. The solid line shows an instantaneous star burst, the dashed line plots continuous star formation ($1\ M_\odot \ \mathrm{yr}^{-1}$). Right: $M_*/L_\lambda$ ratio as function of the mass fraction of a $0.1$ Gyr population added to an old (10 Gyr) population.}
\label{fig:mlratio}
\end{figure*}

Model calculations show that the ratio of stellar mass and $K$-band luminosity, $M_*/L_K$, is a good estimator for the age of stellar populations because it increases monotonically with time. In practice, old and young stellar populations cannot easily be separated. However, the overall $M_*/L_K$ ratio can be used as an upper limit of the age of the young stellar population \citep[see discussion in][]{2007ApJ...671.1388D}. We used \textsc{starburst99} \citep{1999ApJS..123....3L,2005ApJ...621..695V} to estimate the $M_*/L_\lambda$ ratio for the $J$, $H$ and $K$ band as a function of the age of the stellar population (Fig. \ref{fig:mlratio} left). In Fig. \ref{fig:mlratio} (right), we show the $M_*/L_{\lambda}$ ratio that results when one mixes a 0.1 Gyr population with an old 10 Gyr population. The $M_*/L_{\lambda}$ ratio decreases monotonically with increasing fraction of the intermediate-age population.

Given that former studies \citep{2011ApJ...726...59B} indicated that active galaxies follow the same $M_\mathrm{BH}-M_*$ relation as inactive ellipticals, we calculated the $M_*/L_{\lambda}$ ratios that would be necessary to shift the observed galaxies on the $M_\mathrm{BH}-M_*$ relation
\begin{equation}
\log \left( M_\mathrm{BH}/M_\odot \right) = 8.20 + 1.12 \times \log \left( M_\mathrm{bulge}/10^{11} M_\odot \right)
\label{eq:mm-rel}
\end{equation}
of inactive galaxies \citep{2004ApJ...604L..89H}. With the $M_*/L_\lambda$ ratios presented in Table \ref{tab:bhmasses}, we can now estimate the amount of intermediate-age stellar populations that have to be added to an old population to reproduce the observed $M_*/L_\lambda$ ratio. With the exception of galaxy 82, none of the galaxies would only contain old stellar populations. Instead, the galaxies should contain a considerable amount of intermediate-age stellar populations \citep[e.g.][]{2004MNRAS.352..399J}. Under the assumption that these galaxies follow the $M_\mathrm{BH}-M_*$ relation of inactive galaxies, at least four or five galaxies (08, 11, 75, 77, and probably 24) would be dominated by a young 0.1 Gyr population, in contrast to the traditional view of bulge composition. 

Discarding the assumption that active galaxies follow the same $M_\mathrm{BH}-M_*$ relation as inactive galaxies, an alternative interpretation is that undermassive black holes are hosted in normal bulges. At the current accretion rate, the systems would require one typical AGN duty cycle (without additional star formation) to reach the $M_\mathrm{BH}-L_\mathrm{bulge}$ relation of inactive galaxies.

Recently, a bent $M_\mathrm{BH}-L_\mathrm{bulge}$ relation has been suggested \citep{2012ApJ...746..113G,2013ApJ...764..151G}. This relation has different slopes for core-S\'ersic galaxies \citep{2003AJ....125.2951G} and S\'ersic galaxies that are thought to originate from different building mechanisms (dry merging vs. gaseous formation processes). The scatter of the relation for S\'ersic galaxies that show a trend for lower $M_\mathrm{BH}$ is higher than that of the single power-law relations. In our study, we cannot distinguish between core-S\'ersic and S\'ersic galaxies because of the AGN subtraction and the lower resolution (owing to the higher distances of the observed LLQSOs). We note that several galaxies that are classified as S\'ersic galaxies by \cite{2013ApJ...764..151G} are classified as hosts of pseudo-bulges by \cite{2013ARA&A..51..511K} and have therefore been excluded from the $M_\mathrm{BH}-L_\mathrm{bulge}$ relation fit of the latter. 

When comparing the observed LLQSOs with the bent relation by \cite{2013ApJ...764..151G}, more than half of the observed galaxies would lie within the scatter of the relation (see Fig.~\ref{fig:bhmass2}). However, the trend that they are significantly below the predicted relation remains and has to be investigated in more detail.

We are carrying out further studies with multiwavelength spectroscopic techniques to gain deeper insights into the circumnuclear star-forming properties of the observed galaxies and specify the amount of star formation in low-luminosity type-1 QSOs.

\begin{table*}
\centering
\caption{Bulge magnitudes and black hole masses.}
\label{tab:bhmasses}
\begin{tabular}{c|ccc|c|ccc|cc} \hline \hline
Name & \multicolumn{3}{|c|}{abs. magnitude} & $\log (M_{\mathrm{BH},\mathrm{H} \beta}/M_\odot)$ & \multicolumn{3}{|c|}{$M_*/L_\lambda$ ratio} & fraction of \\
 & $M_{J,\mathrm{bul}}$ & $M_{H,\mathrm{bul}}$ & $M_{K,\mathrm{bul}}$ & & J & H & K & 0.1 Gyr population \\ \hline

08 HE0045-2145  & -20.36 & -21.88 & -22.42 & 5.89 & 0.22 & 0.07 & 0.05 & $>70\%$ \\ 
11 HE0103-5842  & -22.50 & -23.37 & -23.78 & 6.71 & 0.16 & 0.10 & 0.07 & $>70\%$ \\ 
24 HE0224-2834  & -23.43 & -24.15 & -24.48 & 7.63 & 0.46 & 0.32 & 0.24 & $40-80\%$ \\ 
29 HE0253-1641  & -20.78 & -21.80 & -22.13 & 6.71 & 0.80 & 0.42 & 0.32 & $10-30\%$ \\ 
71 HE1310-1051  & -22.12 & -22.86 & -23.25 & 7.33 & 0.84 & 0.57 & 0.41 & $10-30\%$ \\ 
75 HE1338-1423  & -23.44 & -23.88 & -23.97 & 7.30 & 0.23 & 0.21 & 0.20 & $>70\%$ \\ 
77 HE1348-1758  & -21.09 & -21.77 & -21.99 & 6.30 & 0.26 & 0.19 & 0.16 & $>70\%$ \\ 
79 HE1417-0909  & -21.97 & -22.74 & -23.05 & 7.24 & 0.80 & 0.52 & 0.41 & $10-30\%$ \\ 
81 HE2128-0221  & -22.49 & -22.97 & -23.36 & 7.40 & 0.69 & 0.59 & 0.43 & $10-30\%$ \\ 
82 HE2129-3356  & -22.10 & -22.68 & -23.23 & 7.80 & 2.23 & 1.75 & 1.10 & $0\%$ \\ 
83 HE2204-3249  & -24.01 & -24.68 & -25.00 & 8.11 & 0.72 & 0.52 & 0.40 & $10-30\%$ \\

\hline
\end{tabular}
\tablefoot{
Measured $J$-, $H$-, and $K$-band magnitudes of the spheroidal component (columns 2-4) and the BH masses (column 5). Columns 6-8 show $M_*/L_\lambda$ ratios, which are necessary to shift galaxies on the $M_\mathrm{BH}-M_\mathrm{bulge}$ relation of inactive galaxies. Column 9 shows an estimate of the mass fraction of the 0.1 Gyr population that has to be added to an old 10 Gyr population to reproduce the observed $M_*/L_\lambda$ ratios.
}
\end{table*}

\section{Summary and conclusions}
\label{sec:summary}

The \emph{low-luminosity type-1 QSO sample} is suitable for detailed spatially resolved studies of nearby QSOs. We investigated 20 galaxies with high-quality NIR imaging data. The results of our analysis can be summarized as follows:
\begin{enumerate}
\item We presented the morphological classification and supported the statistics with additional data that we inspected by eye only. We found that 86\% of the spiral galaxies are barred and many of them have other peculiarities such as inner rings.

\item We measured NIR colors in different apertures and found that the galaxies are broadly distributed across the whole range between inactive galaxies and zero-redshift quasars. The colors of the AGN-subtracted galaxies are consistent with previous studies.

\item We found \textsc{Budda} to be suitable and reliable for decompositions of nearby active galaxies. The limits of this method have been probed in Sect. \ref{sec:reliability}. We used \textsc{Budda} to extract single components or subtract the AGN component to investigate host galaxy properties. 

\item From the models obtained by decomposition with \textsc{Budda} we estimated stellar masses that range from $2\times 10^9 M_\odot$ to $2\times 10^{11} M_\odot$ with an average mass of $7\times 10^{10} M_\odot$ by applying a typical mass-to-light ratio.

\item In comparison with other samples that contain more luminous QSOs, we found our galaxies to have lower stellar and black hole masses. They are less luminous, considering both nuclear and host galaxy luminosity. In the fundamental plane, the observed LLQSOs lie between luminous QSOs and inactive local galaxies. Along with the findings of \cite{2007A&A...470..571B} and \cite{2013arXiv1309.6921M}, this implies that LLQSOs populate an intermediate region between nearby Seyfert galaxies, (ultra-) luminous infrared galaxies and luminous QSOs in terms of many parameters such as $z$, $M_\mathrm{BH}$, $L_\mathrm{bol}$, $L_\mathrm{host}$, $\mathrm{SFR}$.

\item We found that the low-luminosity type-1 QSOs observed in this study do not follow the $M_\mathrm{BH}-L_\mathrm{bulge}$ relations for inactive galaxies. Previous studies at optical wavelengths have often explained this by star formation in the spheroids of active galaxies.
An in-depth investigation of the amount of star formation in the observed sources in order to constrain the account of star formation in shifting active galaxies off the $M_\mathrm{BH}-L_\mathrm{bulge}$ relations is needed.

\end{enumerate}

\begin{acknowledgements}
The authors kindly thank Andreas Schulze for providing black-hole mass estimates and emission-line characteristics of the LLQSOs. Furthermore, the authors thank the anonymous referee, Alister Graham, Dongchan Kim, and Davide Lena for comments that helped to improve the manuscript.
GB is member of the \emph{Bonn-Cologne Graduate School of Physics and Astronomy} and acknowledges support from the \emph{Konrad-Adenauer-Stiftung}.
JZ acknowledges support from the European project EuroVO DCA under the Research e-Infrastructures area (RI031675). JS, SF and JZ acknowledge support by the German Academic Exchange Service (DAAD) under project number 50753527. JS acknowledges the European Research Council for the Advanced Grant Program Num 267399-Momentum.
MV-S thanks for the funding from the European Union Seventh Framework Programme (FP7/2007-2013) under grant agreement No.312789.
We acknowledge fruitful discussions with members of the European Union funded COST Action MP0905: Black Holes in a violent Universe and the COST Action MP1104: Polarization as a tool to study the Solar System and beyond. 
Part of this work was supported by the German \emph{Deut\-sche For\-schungs\-ge\-mein\-schaft, DFG\/} project numbers SFB~494 and SFB~956. This work is supported in part by the German Federal Department for Education and Research (BMBF) under the grants Verbundforschung 05AL5PKA/0 and 05A08PKA.
Parts of the observations were conducted at the LBT Observatory, a joint facility of the Smithsonian Institution and the University of Arizona. LBT observations were obtained as part of the Rat Deutscher Sternwarten guaranteed time on Steward Observatory facilities through the LBTB coorperation.
This research has made use of the NASA/IPAC Extragalactic Database (NED) which is operated by the Jet Propulsion Laboratory, California Institute of Technology, under contract with the National Aeronautics and Space Administration.
\end{acknowledgements}

\bibliographystyle{aa}
\bibliography{llqso2}

\begin{figure*}
\centering
\includegraphics{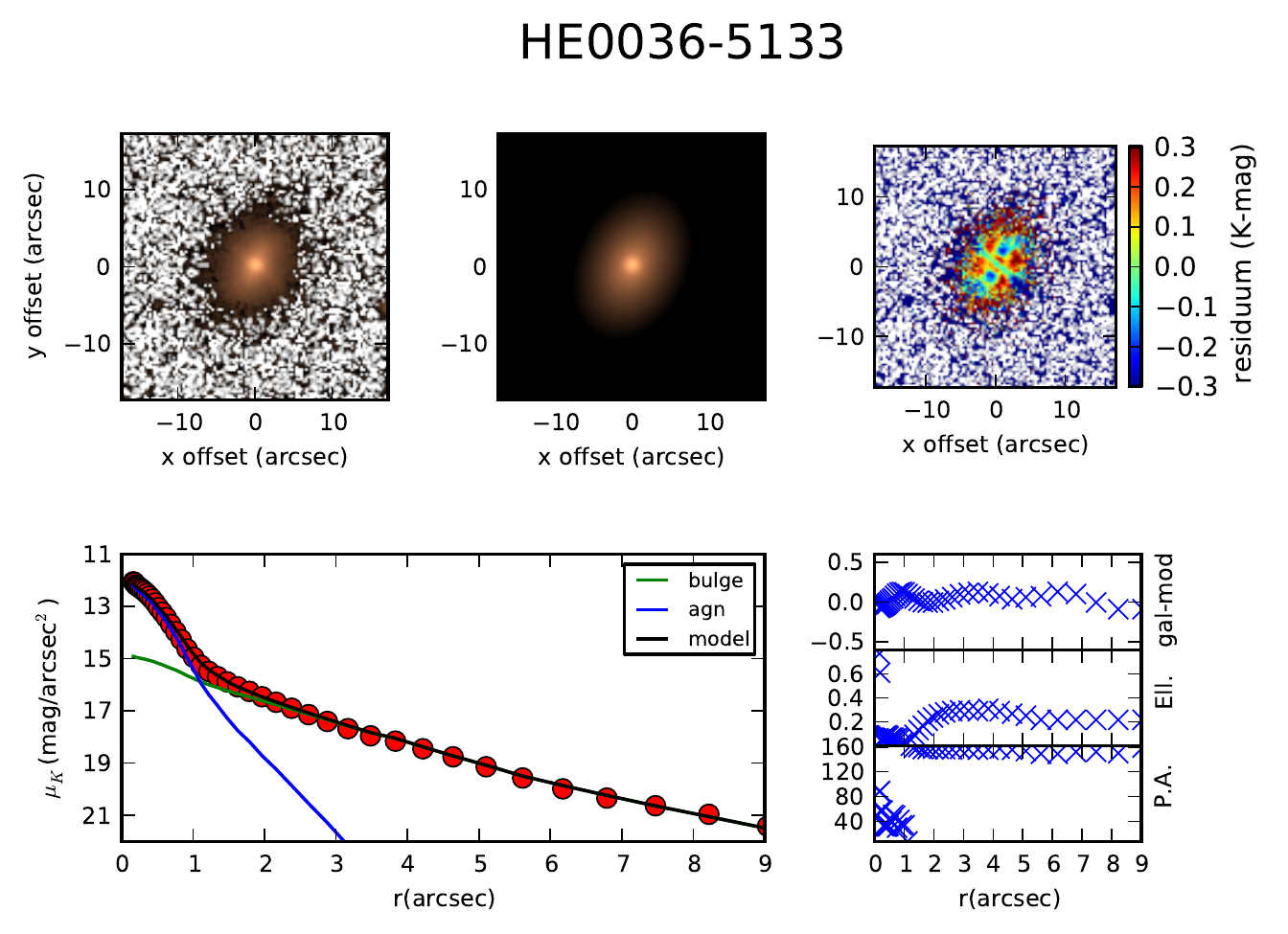}
\includegraphics{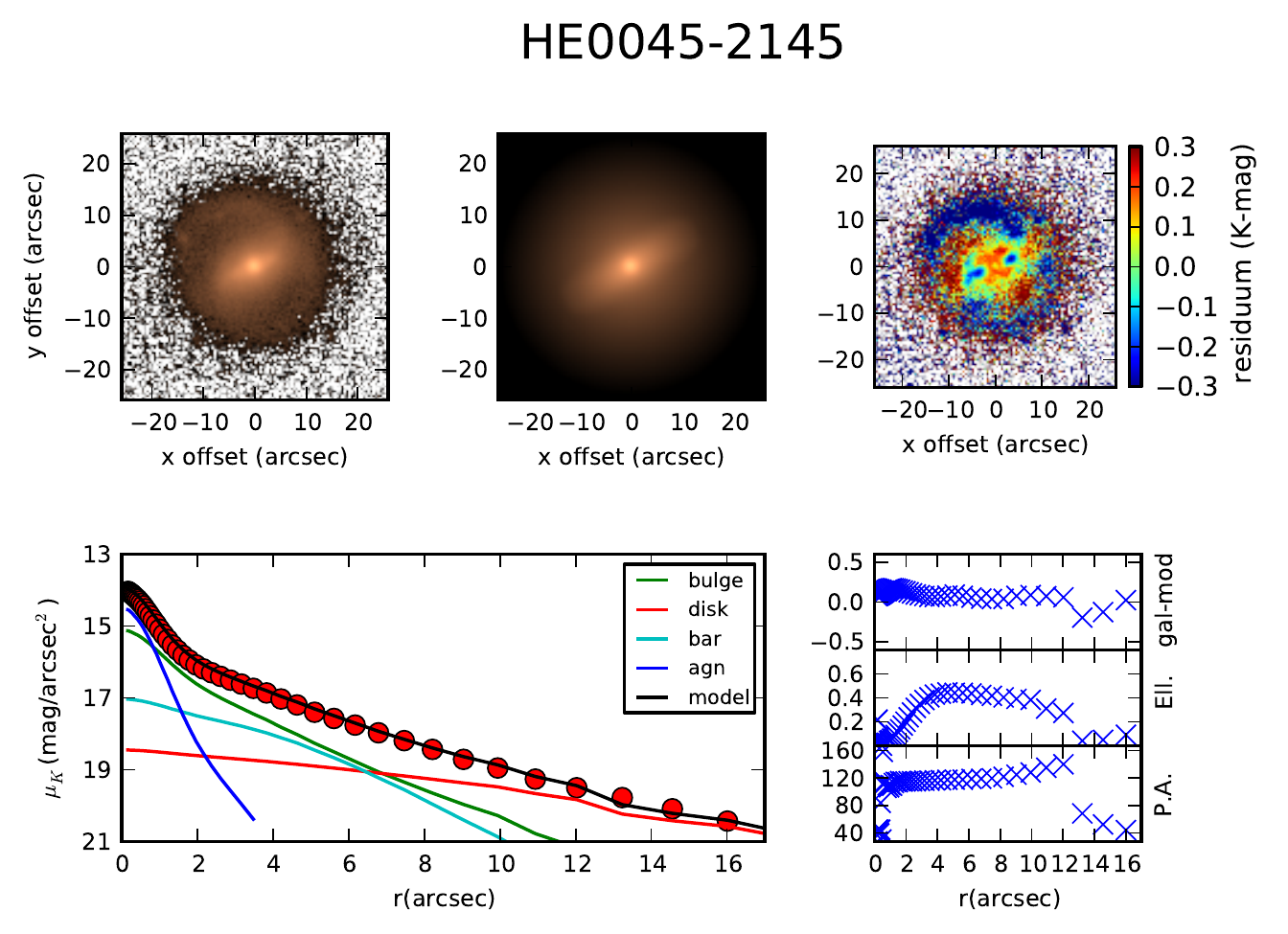}
\caption{Same as Fig. \ref{fig:decomp_ex} for the remaining galaxies.}
\label{fig:decomp}
\end{figure*}

\begin{figure*}
\ContinuedFloat
\centering
\includegraphics{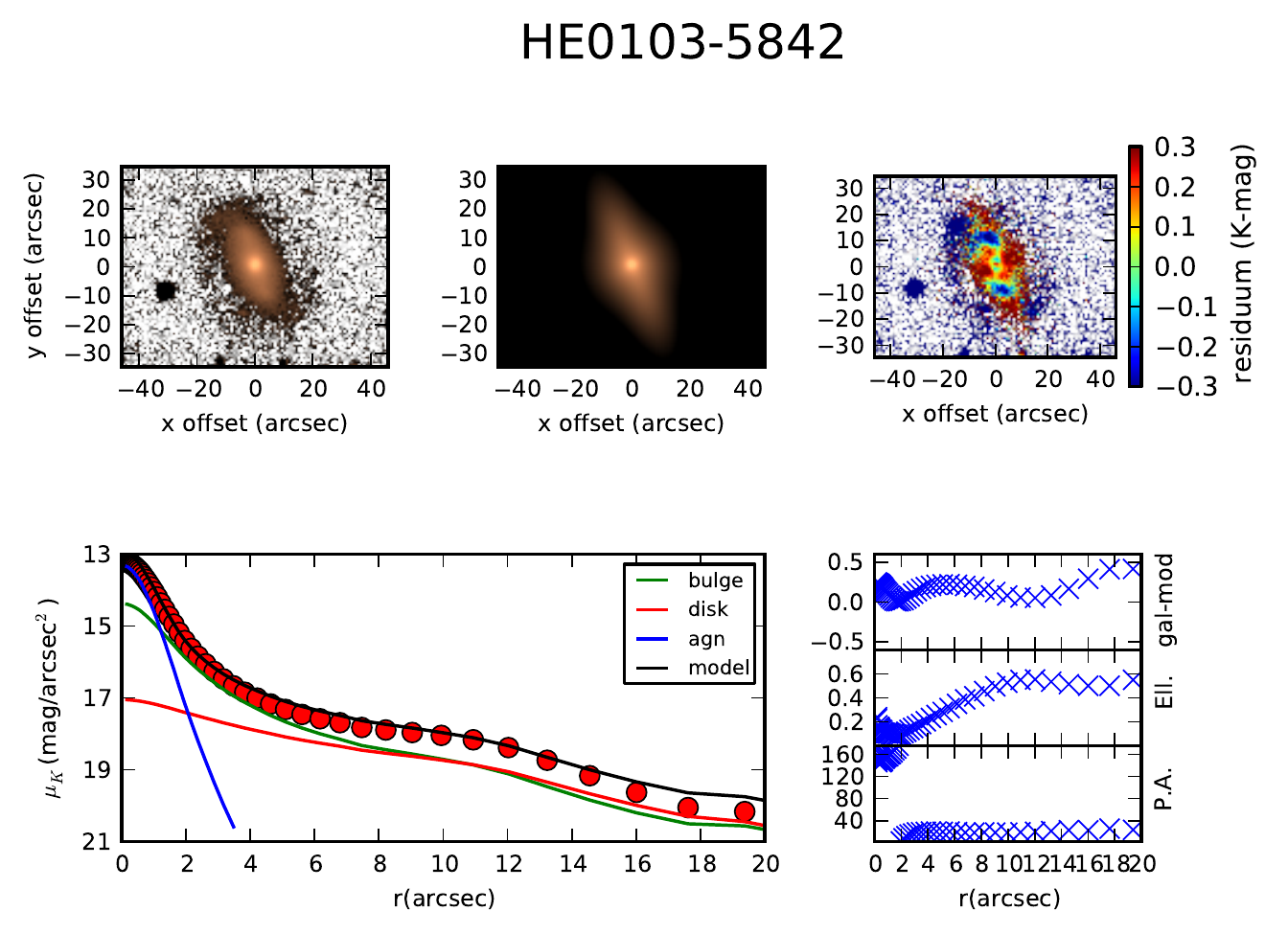}
\includegraphics{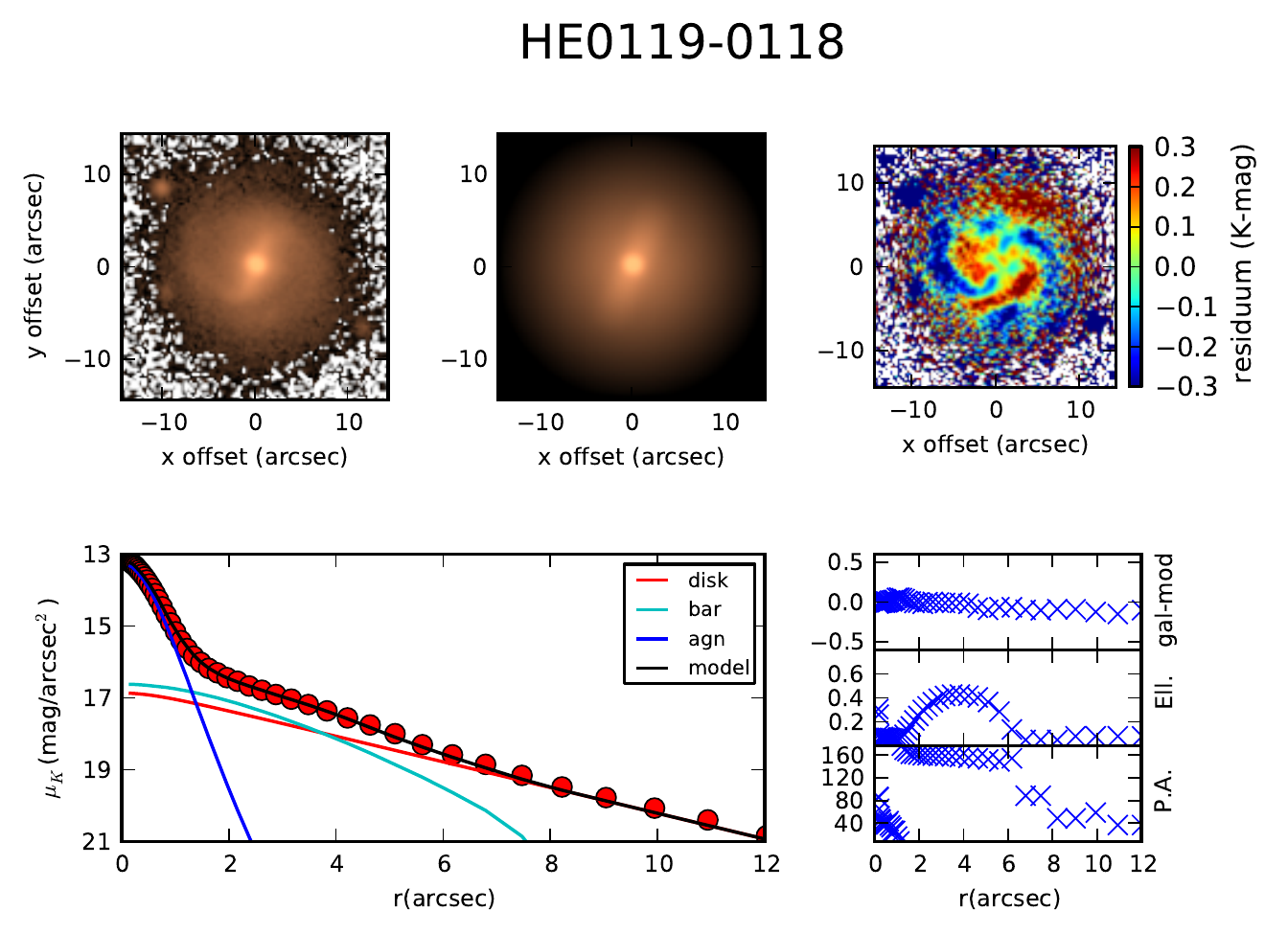}
\caption{continued}
\end{figure*}

\begin{figure*}
\ContinuedFloat
\centering
\includegraphics{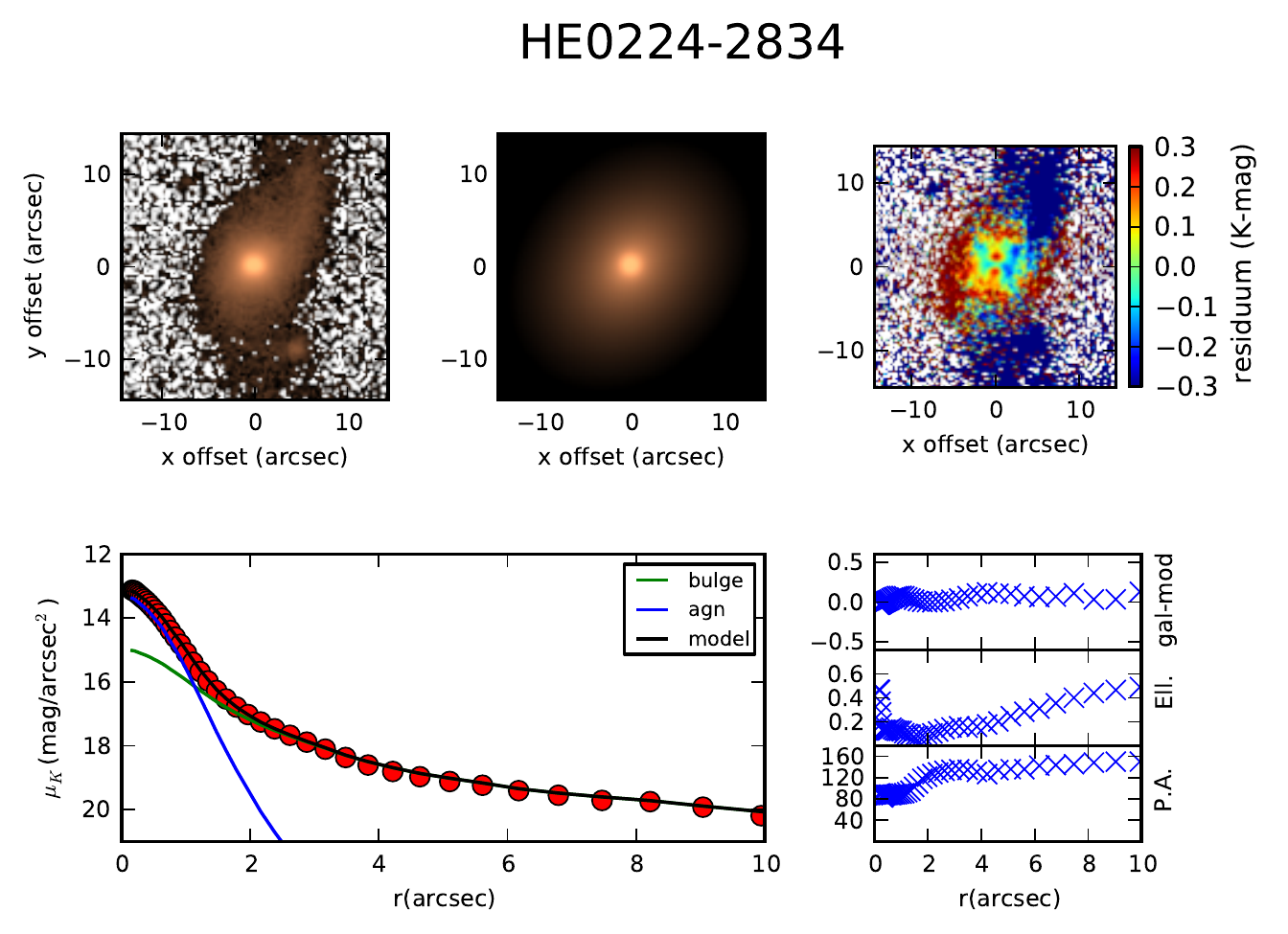}
\includegraphics{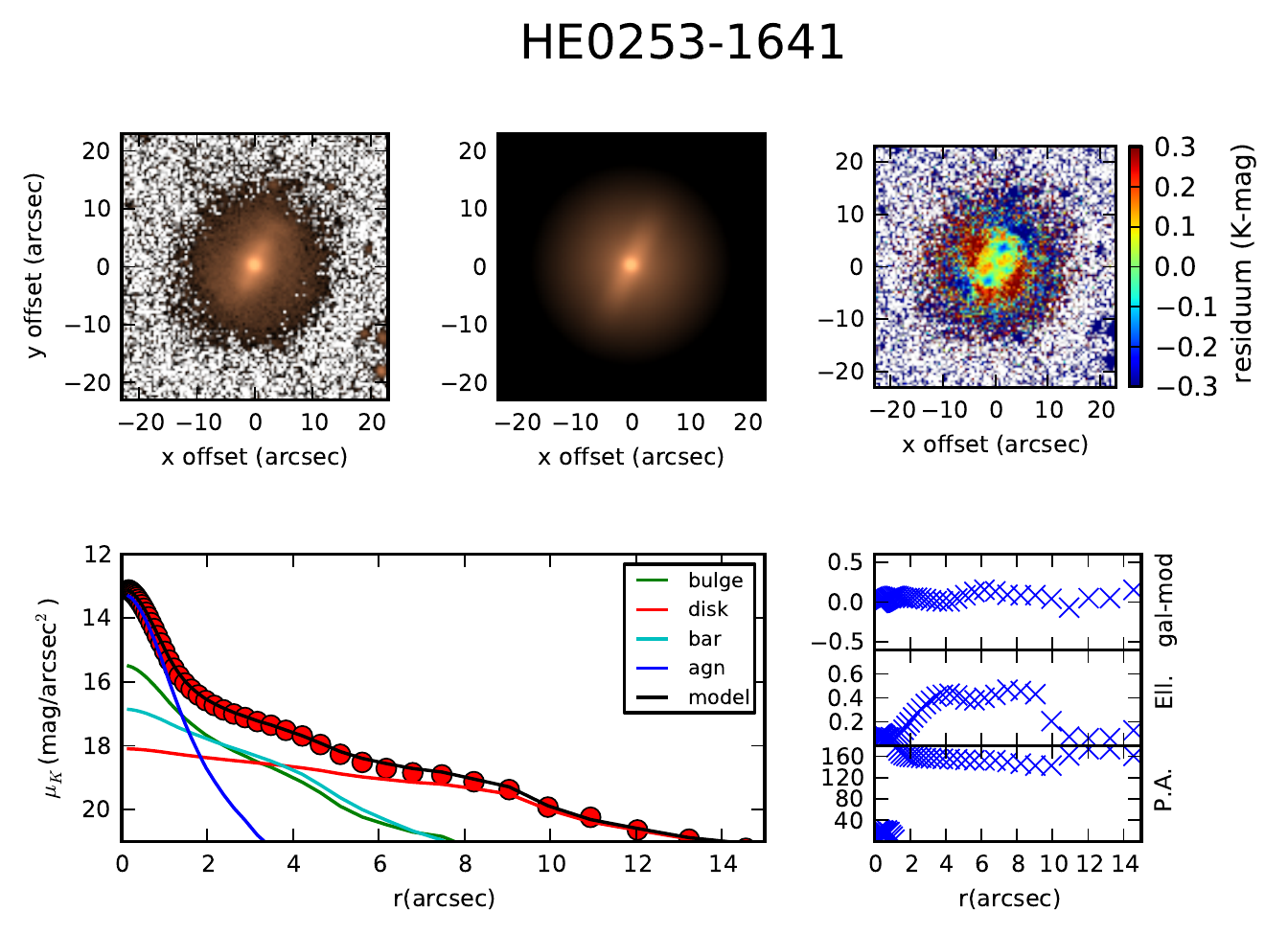}
\caption{continued}
\end{figure*}

\begin{figure*}
\ContinuedFloat
\centering
\includegraphics{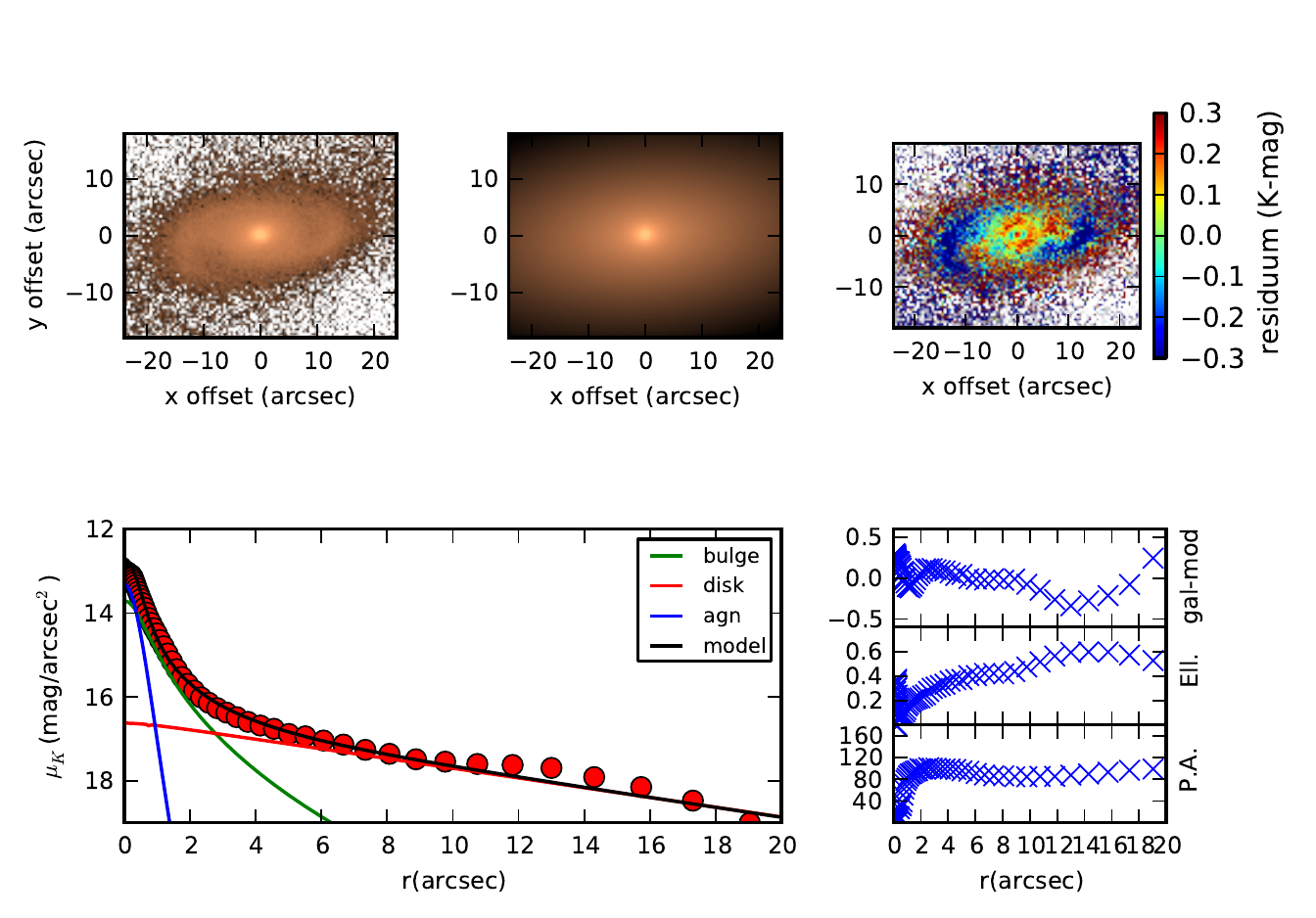}
\includegraphics{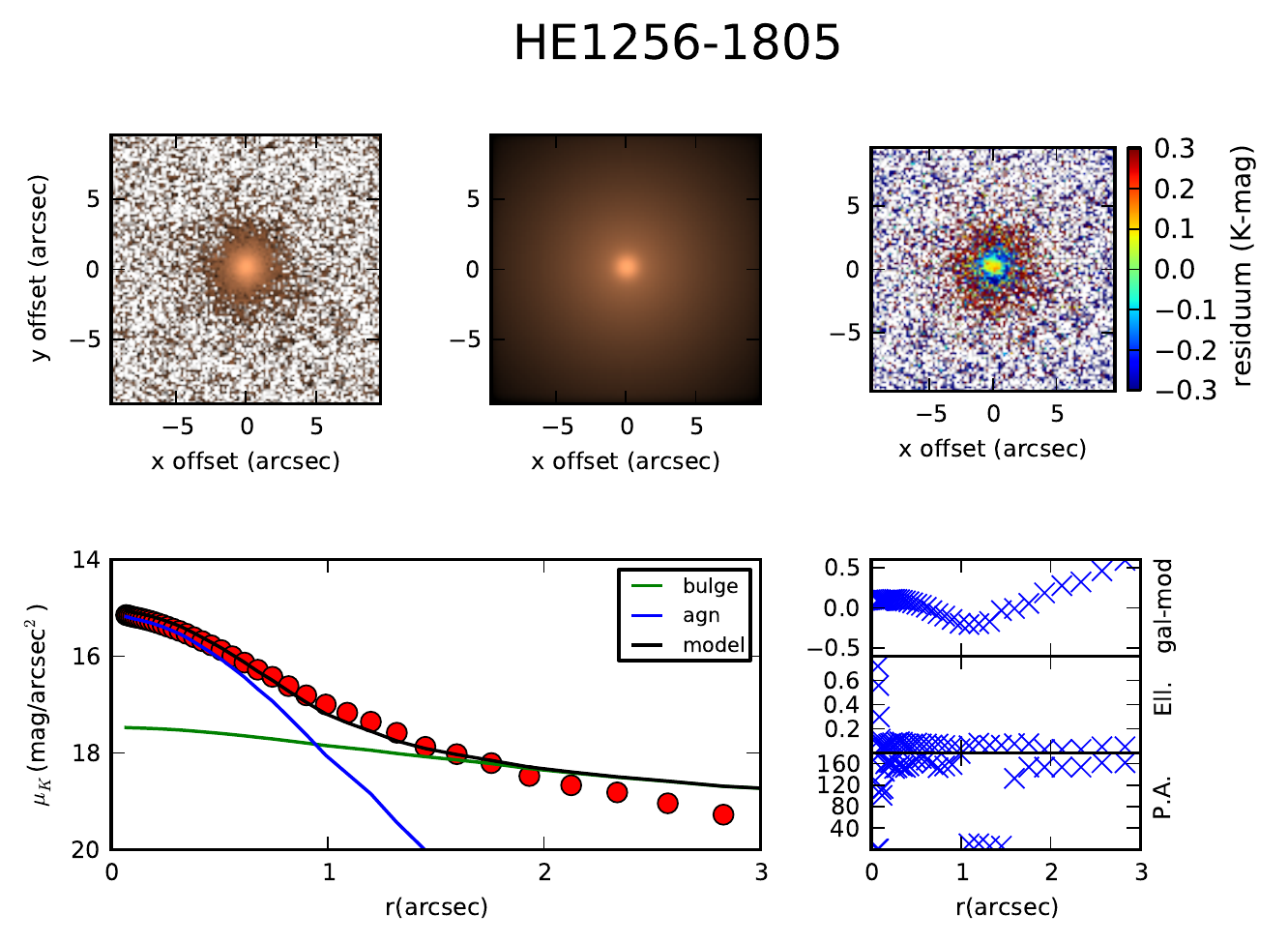}
\caption{continued}
\end{figure*}

\begin{figure*}
\ContinuedFloat
\centering
\includegraphics{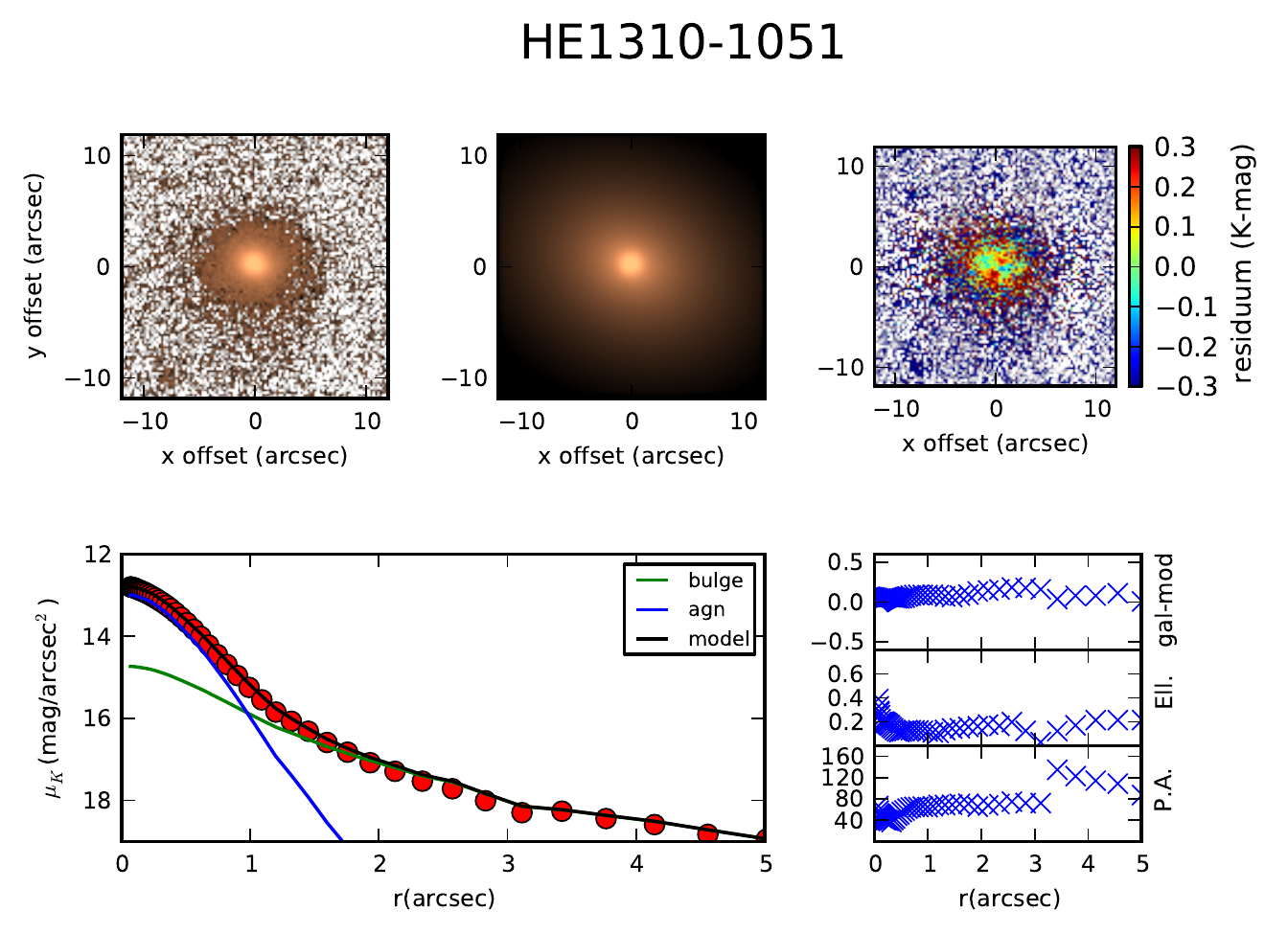}
\includegraphics{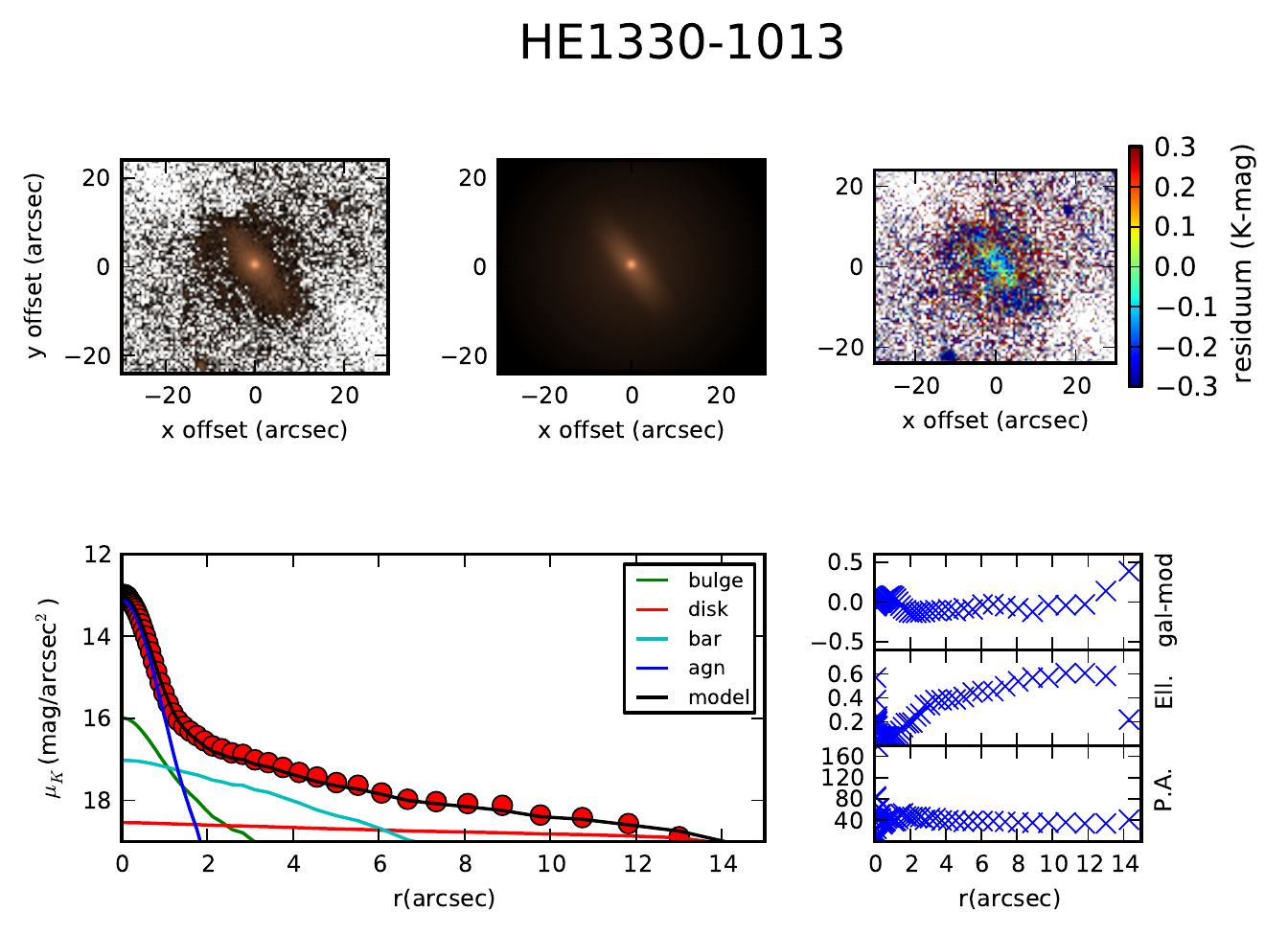}
\caption{continued}
\end{figure*}

\begin{figure*}
\ContinuedFloat
\centering
\includegraphics{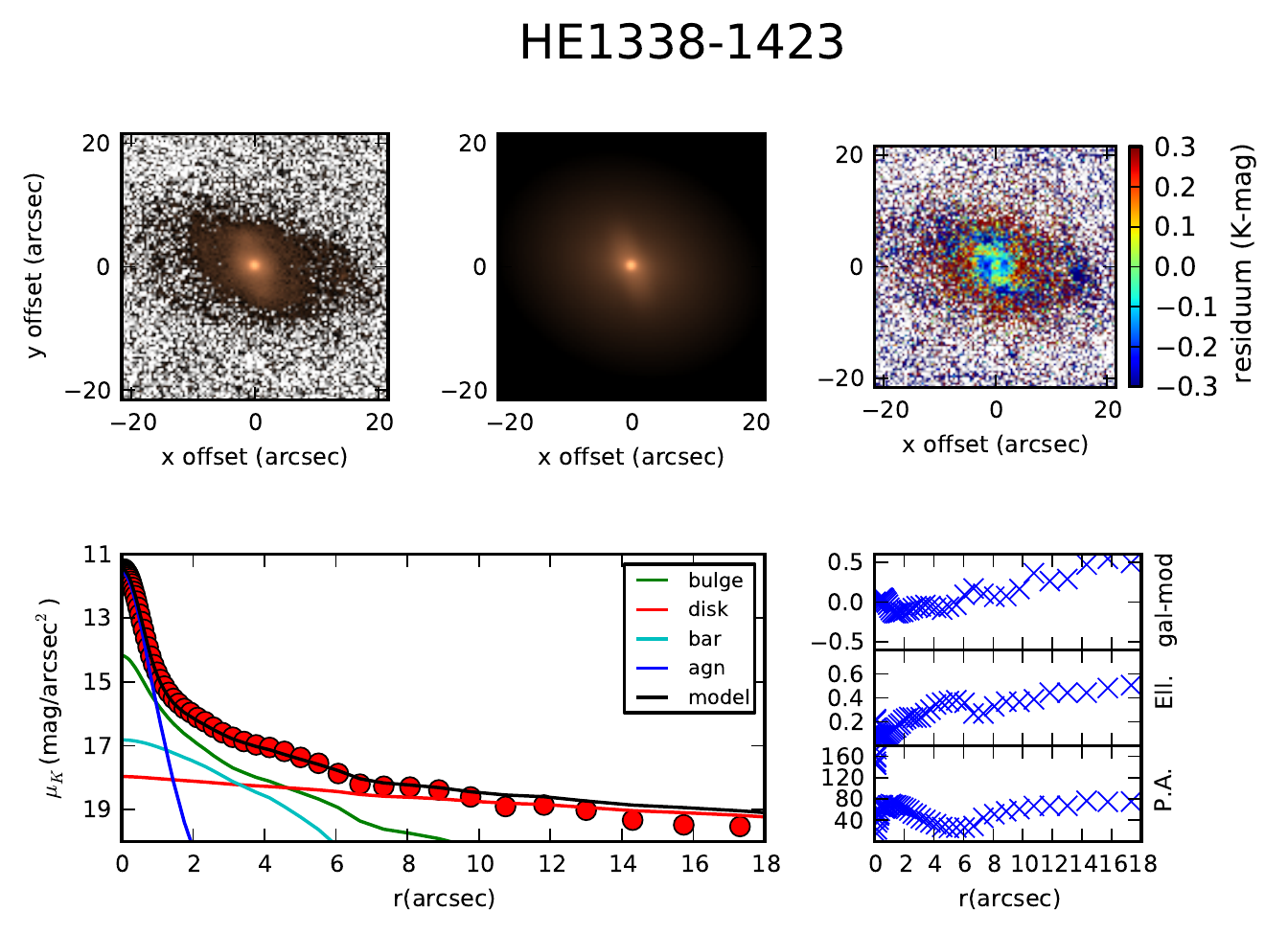}
\includegraphics{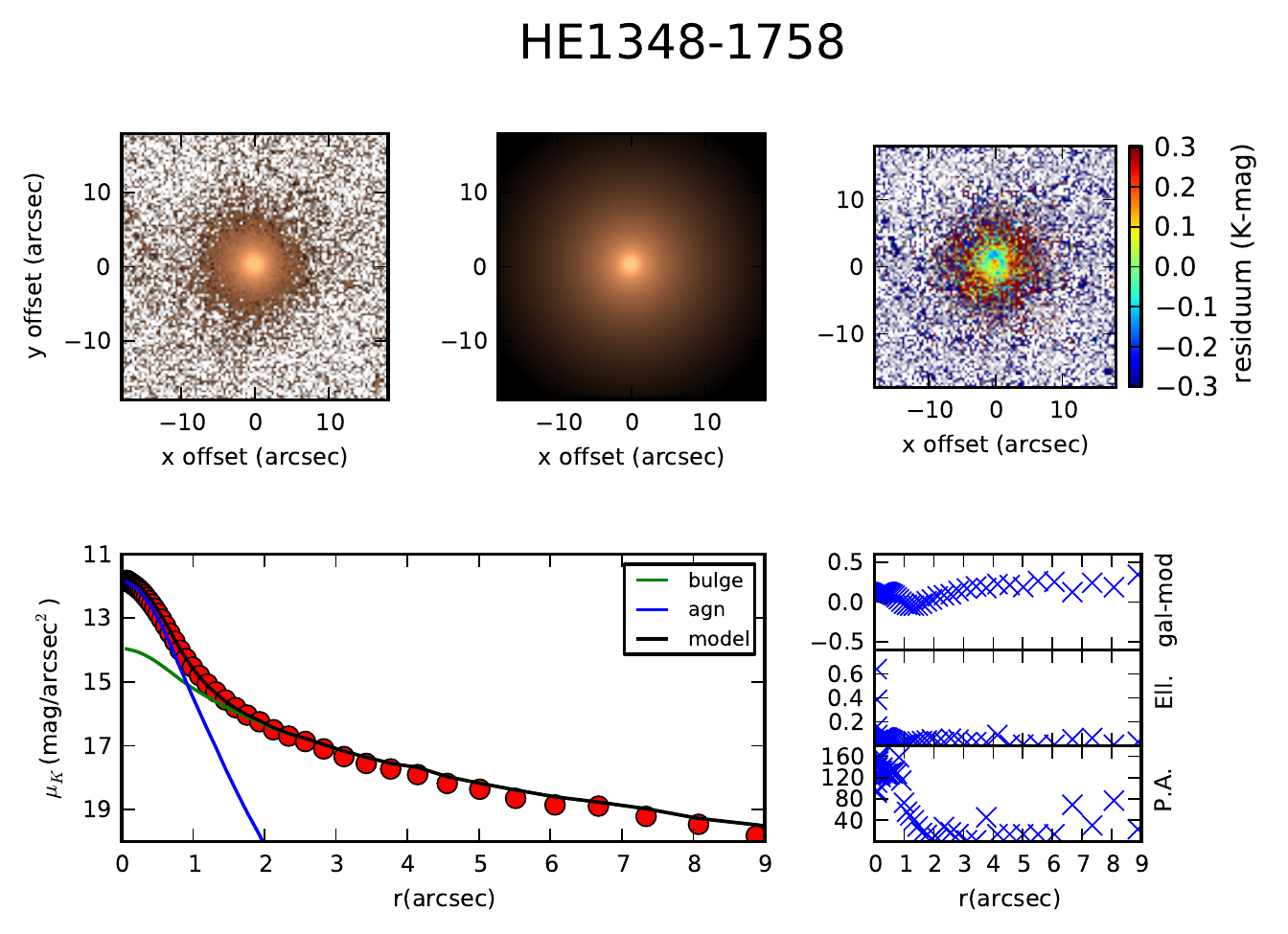}
\caption{continued}
\end{figure*}

\begin{figure*}
\ContinuedFloat
\centering
\includegraphics{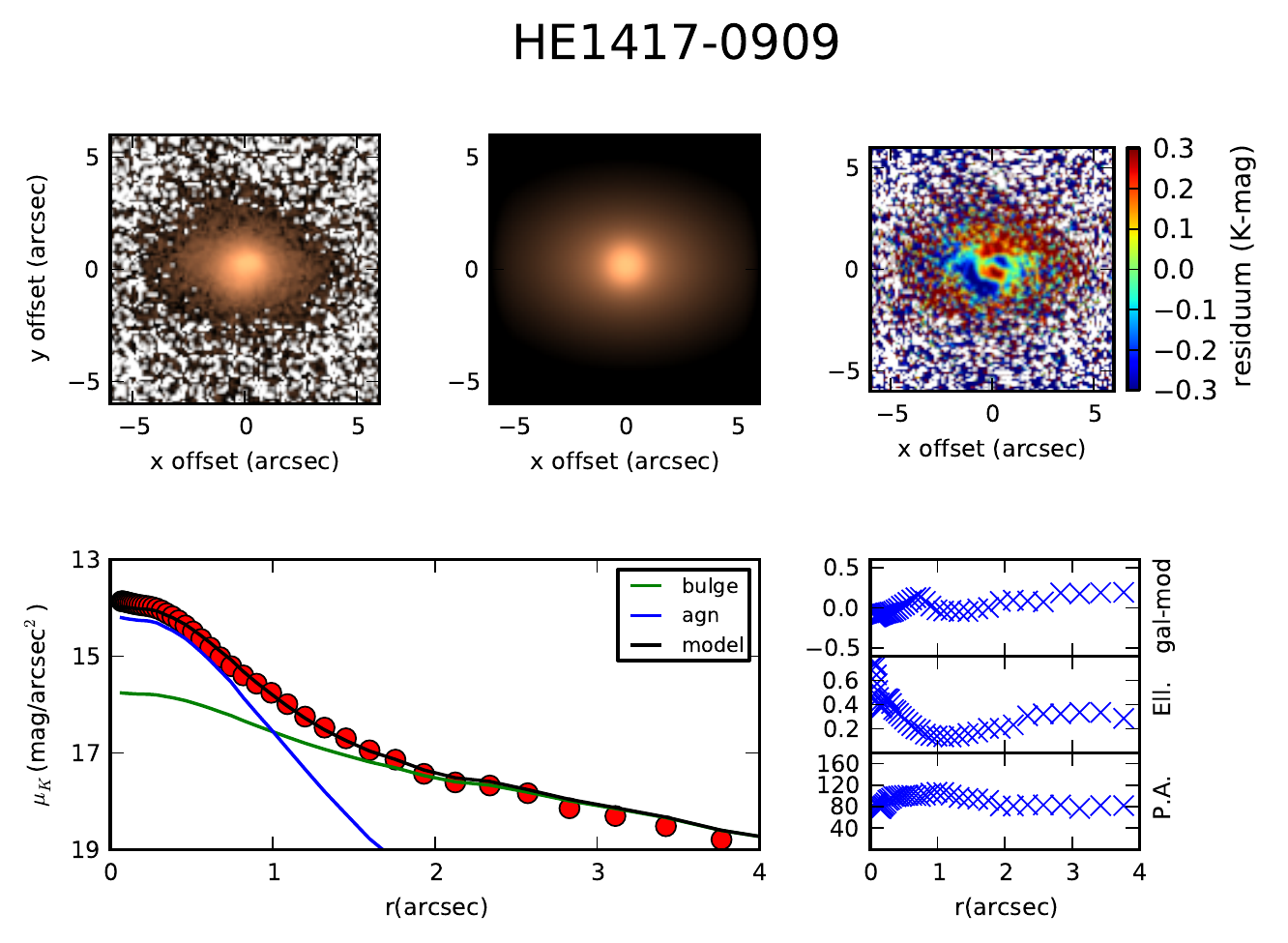}
\includegraphics{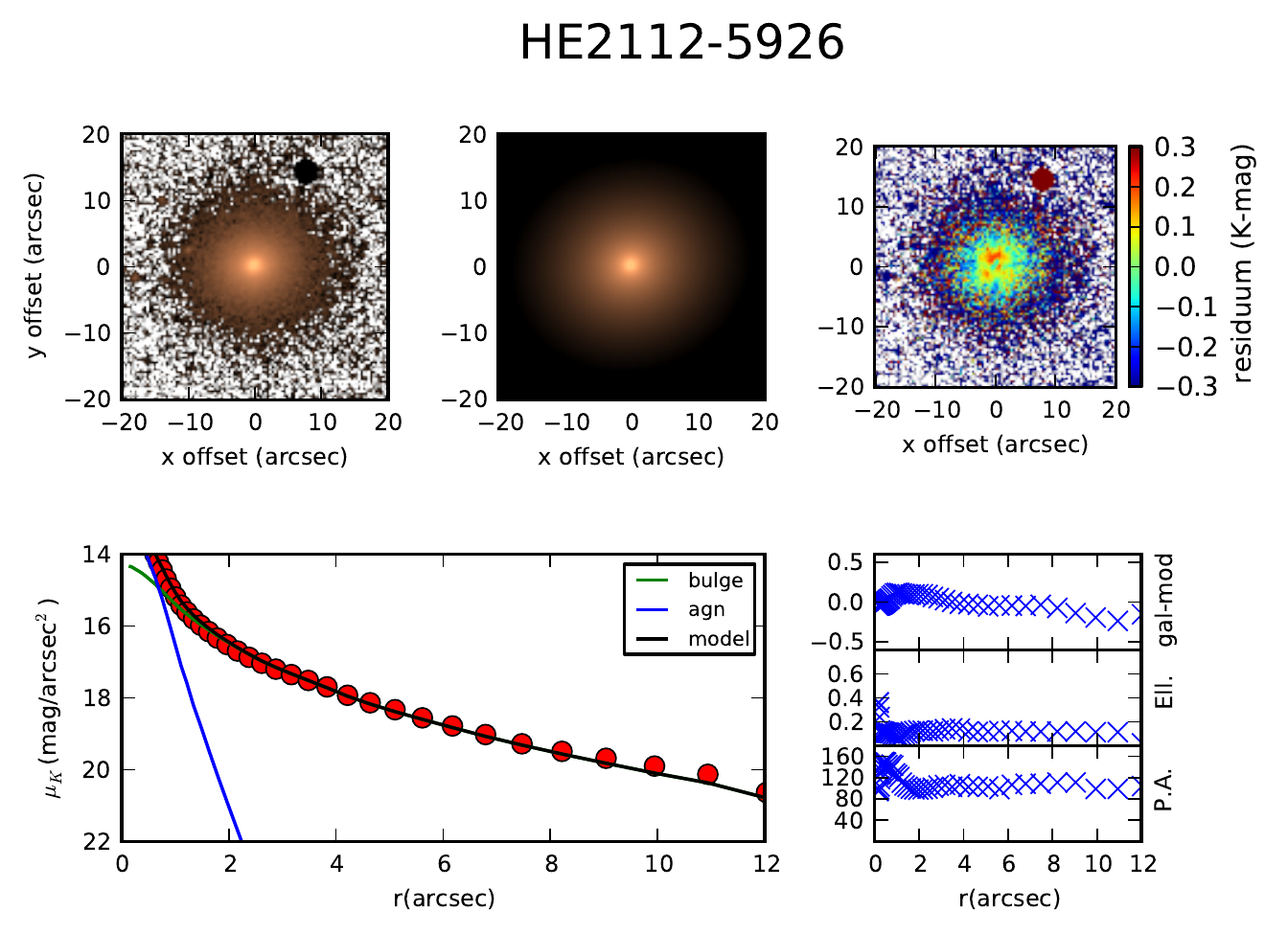}
\caption{continued}
\end{figure*}

\begin{figure*}
\ContinuedFloat
\centering
\includegraphics{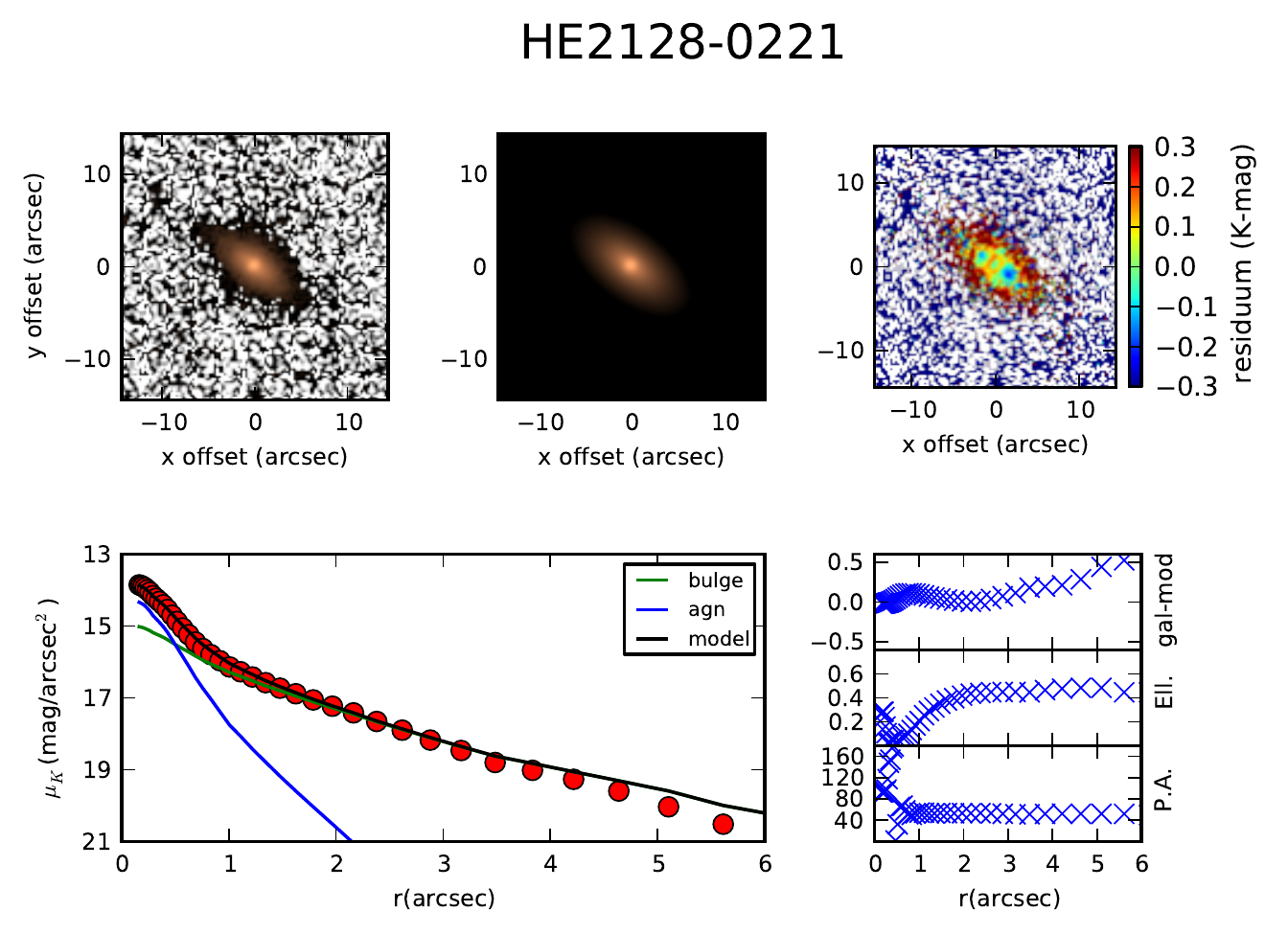}
\includegraphics{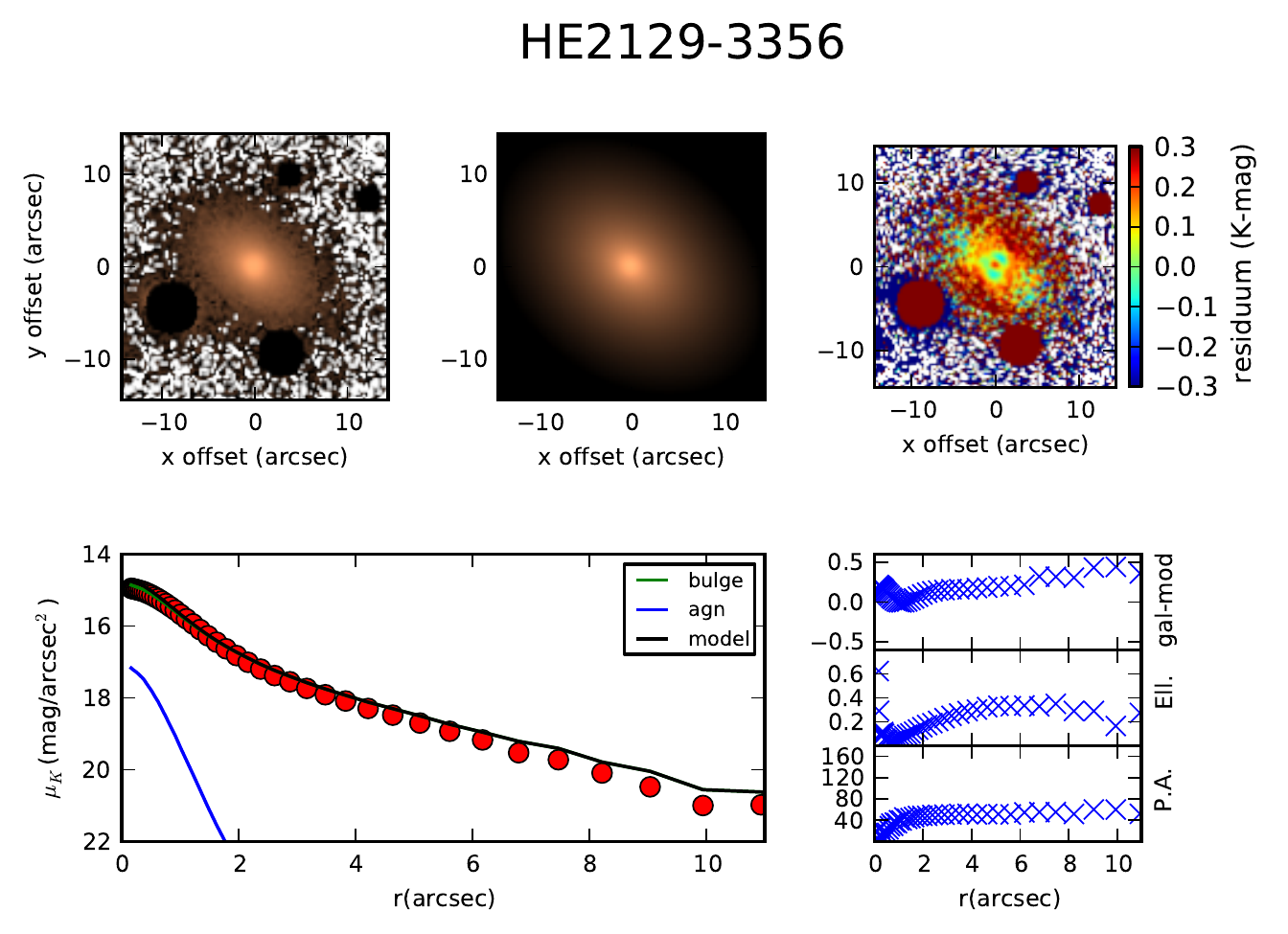}
\caption{continued}
\end{figure*}

\begin{figure*}
\ContinuedFloat
\centering
\includegraphics{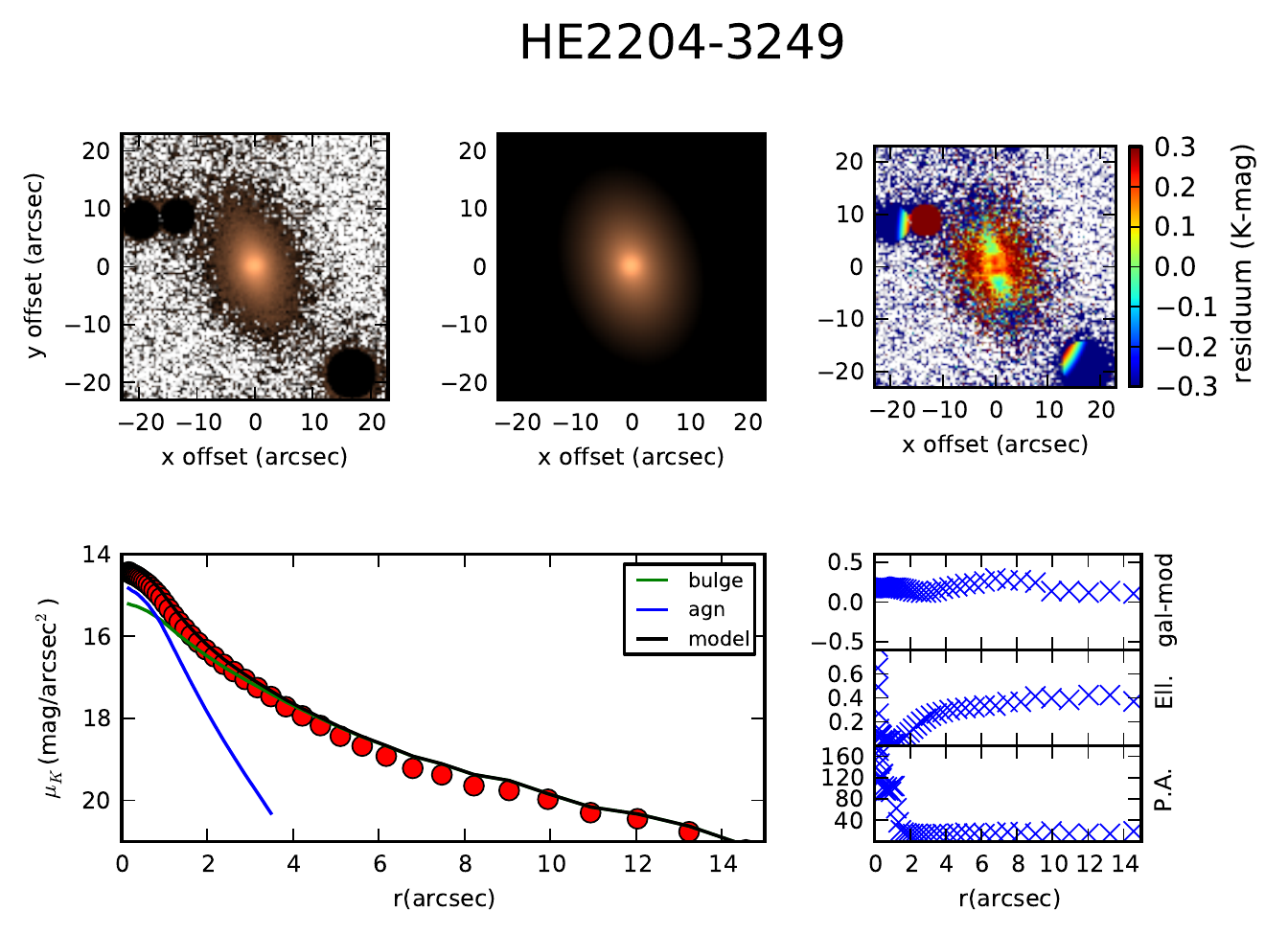}
\includegraphics{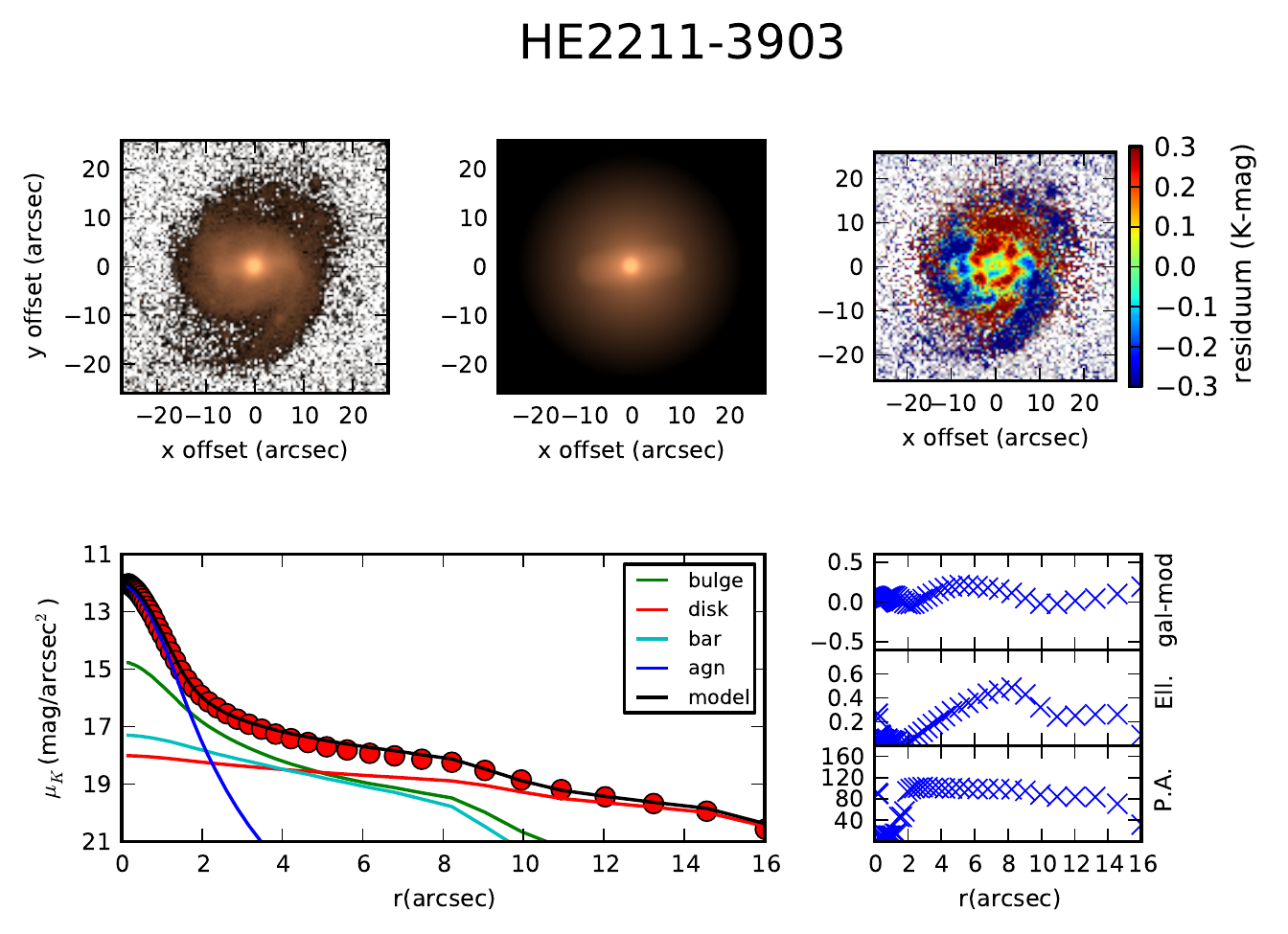}
\caption{continued}
\end{figure*}

\begin{figure*}
\ContinuedFloat
\centering
\includegraphics{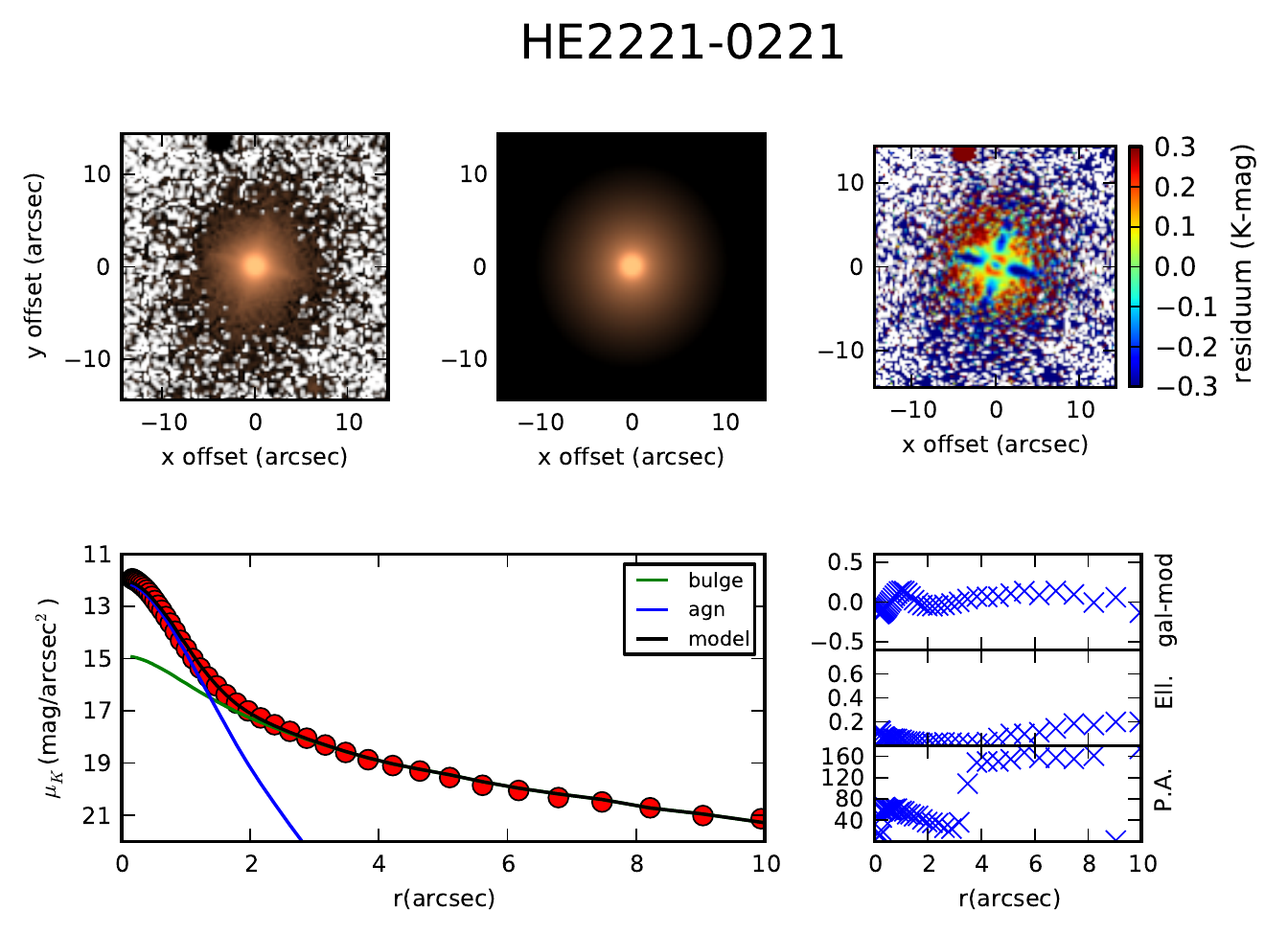}
\includegraphics{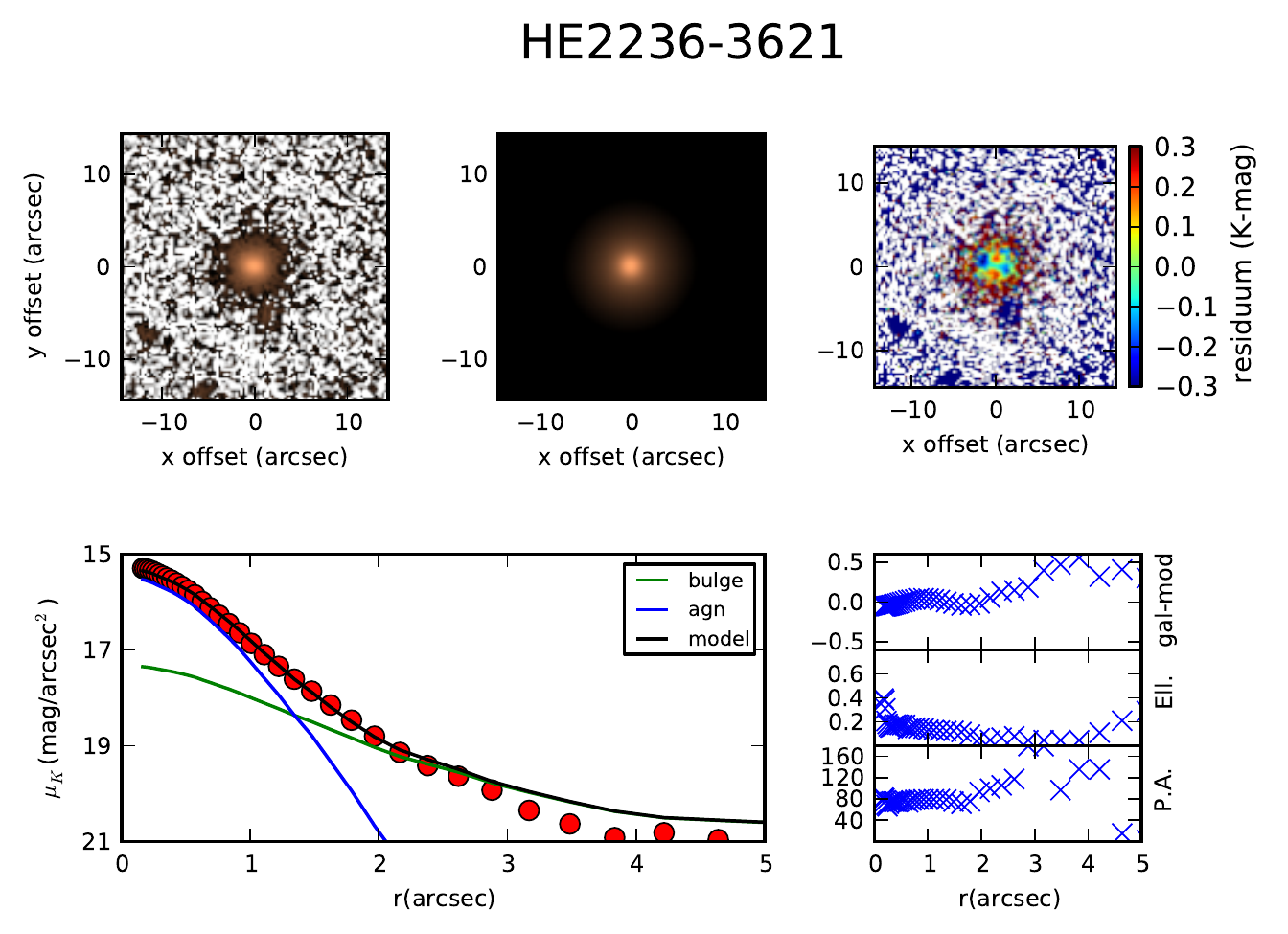}
\caption{continued}
\end{figure*}

\end{document}